\documentclass[aps,prd,preprint,preprintnumbers,groupedaddress,nofootinbib]{revtex4} 
\usepackage{graphicx}
\usepackage{graphics}
\usepackage{amsmath, amsthm, amssymb}
\usepackage{feynmf}
\usepackage{verbatim}
\usepackage{subfig}
\usepackage{breqn}
\numberwithin{equation}{section}

\def\SLAC{SLAC National Accelerator Center\\
  2575 Sand Hill Rd., Menlo Park, CA 94025}
\def\O{\mathcal{O}}
\def\Re{{\rm Re}}
\def\vv{\mathcal{V}}
\def\Im{{\rm Im}}
\def\psih{\hat{\psi}}
\def\Ah{\hat{A}}
\def\lambdah{\hat{\lambda}}
\def\coms{\stackrel{\star}{,}}
\def\O{\mathcal{O}}
\def\tev{\,{\ifmmode\mathrm {TeV}\else TeV\fi}}
\def\gev{\,{\ifmmode\mathrm {GeV}\else GeV\fi}}
\def\mev{\,{\ifmmode\mathrm {MeV}\else MeV\fi}}
\def\ipb{\,{\ifmmode\mathrm{pb}^{-1}\else pb$^{-1}$\fi}}
\def\ifb{\,{\ifmmode\mathrm{fb}^{-1}\else fb$^{-1}$\fi}}
\def\mz{\ifmmode M_Z\else $M_Z$\fi}
\def\mw{\ifmmode M_W\else $M_W$\fi}
\def\mws{\ifmmode M_W^2 \else $M_W^2$\fi}
\def\mhs{\ifmmode m_H^2 \else $m_H^2$\fi}   
\def\mzs{\ifmmode M_Z^2 \else $M_Z^2$\fi}

\def\to{\ifmmode \rightarrow\else $\rightarrow$\fi}
\def\xw{\ifmmode x\subw\else $x\subw$\fi}
\def\epem{\ifmmode e^+e^-\else $e^+e^-$\fi}
\def\ppb{\ifmmode \bar pp\else $\bar pp$\fi}
\def\ffb{\ifmmode \bar ff\else $\bar ff$\fi}
\def\qqb{\ifmmode \bar qq\else $\bar qq$\fi}
\def\uub{\ifmmode \bar uu\else $\bar uu$\fi}
\def\ddb{\ifmmode \bar dd\else $\bar dd$\fi}
\def\wpwm{\ifmmode W^+ W^- \else $W^+ W^-$\fi}
\def\shat{\ifmmode \hat s\else $\hat s$\fi}
\def\that{\ifmmode \hat t\else $\hat t$\fi}
\def\uhat{\ifmmode \hat u\else $\hat u$\fi}

\def\Th{\mathit{\Theta}}

\begin{document}

\begin{fmffile}{ncsmfmf}

  \preprint{SLAC-PUB-13471}

  \title{Effects of the Noncommutative Standard Model in\\ WW Scattering}
  \author{John A. Conley}
  \email{conley@stanford.edu}
  \author{JoAnne L. Hewett}
  \email{hewett@slac.stanford.edu}
  \affiliation{\SLAC}
  \thanks{Work supported by the US Department of Energy,
                     contract DE--AC02--76SF00515.}

  \date{November 24, 2008}
  \begin{abstract}
   We examine $W$ pair production in the Noncommutative Standard Model 
   constructed with the Seiberg-Witten map.  Consideration of partial wave 
   unitarity in the reactions $WW\to WW$ and $e^+e^-\to WW$ shows that the 
   latter process is more sensitive and that tree-level unitarity is violated 
   when scattering energies are of order a TeV and the noncommutative scale is below about a TeV. 
   We find that $WW$ production at the LHC is not sensitive to scales above the unitarity bounds. 
   $WW$ production in $e^+e^-$ annihilation, however, provides a good probe of such effects with
   noncommutative scales below 300-400 GeV being excluded at LEP-II, and the ILC being sensitive
   to scales up to 10-20 TeV.  In addition, we find that the ability to measure the helicity states 
   of the final state $W$ bosons at the ILC provides a diagnostic tool to determine and disentangle
   the different possible noncommutative contributions.
  \end{abstract}
  \maketitle

  \section{Introduction}

  With motivation coming from connections to string theory and possible relevance to quantum gravity, the possibility of noncommutative spacetime has 
  received a great deal of attention in many different contexts in the last decade.
  The idea that spacetime coordinates might not commute, however, dates to Heisenberg \cite{Wess:2002dd}, who originally hoped that quantum field
  theory on a noncommutative spacetime (NCQFT) could be free of the ultraviolet divergences that typically plague QFT.  This motivated Snyder in 1947 to furnish the
  first explicit construction of a noncommutative spacetime \cite{Snyder:1946qz}. 
  The idea that NCQFT could cut off divergences dwindled since the success of renormalization, though it has received some attention in recent years
  \cite{Rivasseau:2007ab}.
  The theory and phenomenology of 
  NC quantum mechanics has been well studied \cite{Nair:2000ii,Gamboa:2000yq,Ho:2001aa,Bellucci:2001xp}.
  Connes and his collaborators
  have sought to derive the Standard Model gauge groups and representations as a unique consequence of NC geometry
  \cite{Chamseddine:2007hz}.  There is a strong relationship between NCQFT and
  string theory which has been studied by many authors \cite{Douglas:2001ba,Szabo:2001kg}.  
  
  In particular, interest in noncommutative extensions to the Standard Model increased when Seiberg and Witten \cite{Seiberg:1999vs}
  described how NC gauge theory could arise as the low-energy limit of string theory with a background field.  This leads to a
  commutation relation between spacetime operators,
  \begin{equation}
    \label{canon}
    [\hat{x}^\mu,\hat{x}^\nu]=i\theta^{\mu\nu},
  \end{equation}
  where here, and throughout the rest of the paper, hatted quantities are in the algebra of operators generated by the non-commuting coordinates, 
  and $\theta^{\mu\nu}$ is a constant, antisymmetric matrix. 
  Only ``space-space'' noncommutativity ($\theta_{0i}=0$) can arise as a consistent limit of string theory in Seiberg and Witten's
  construction.  One can, however, try to formulate a ``space-time'' ($\theta_{0i}\neq 0$) NCQFT nonetheless; though this leads to potential problems
  as we will discuss below.

  There are several approaches to constructing a NCQFT. 
  Starting with the canonical commutation relation, one can use the analogue of Weyl quantization to associate 
  a function $f(x)$ to an operator $\hat{f}$ by 
  \begin{equation}
    \label{weyl}
	\hat{f}=\int\frac{d^4k}{(2\pi)^4}\tilde{f}(k)e^{ik_\mu\hat{x}^\mu},
  \end{equation}
  where $\tilde{f}$ is the Fourier transform of $f$.  One can then show \cite{Szabo:2001kg} that the product of two
  operators $\hat{f}\hat{g}=\widehat{f\star g}$, with
  \begin{equation}
    \label{Moyal}
    f(x)\star g(x)\equiv f(x)\exp(\frac{i}{2}\theta^{\mu\nu}\overleftarrow{\partial}_\mu\overrightarrow{\partial}_\nu)g(x)
  \end{equation}
  being known as the Moyal star product.  The commutator likewise has a noncommutative version
  \begin{equation}
    \label{MoyalBracket}
    [f(x)\stackrel{\star}{,}g(x)]\equiv f(x)\star g(x)-g(x)\star f(x)\;,
  \end{equation}
  known as the Moyal bracket.

  Using this, one can establish a map between any NC Lagrangian built from operators (hatted fields) and an ordinary Lagrangian
  built from conventional fields.  The NC Lagrangian is formulated with star products replacing ordinary products and Moyal brackets substituting for commutators.
  It is easy to show that $\int d^4xf\star g=\int d^4xfg$,
  so it is clear that this procedure leaves kinetic and mass terms unchanged from the commutative case, but introduces momentum-dependent phase
  factors in the interaction terms.  In a gauge theory, the Moyal bracket also leads to new interaction terms, such as three- and four-photon vertices in NC
  quantum electrodynamics.  The primary impediment to formulating a noncommutative Standard Model (NCSM) in this way is that $SU(N)$ groups
  do not close under star multiplication.  $U(N)$ groups do, however, and an NCSM has been constructed using a $U(3)\times U(2)\times U(1)$ gauge
  group with some new fields added to implement symmetry breaking effectively \cite{Chaichian:2001py}.  This NCSM, however, was shown to 
  violate unitarity in gauge boson scattering at high energies \cite{Hewett:2001im}.

  In a different approach, Seiberg and Witten \cite{Seiberg:1999vs} showed how to solve NC and normal gauge equivalence conditions and derived expressions for
  noncommuting gauge fields in terms of normal gauge fields, order by order in $\theta^{\mu\nu}$.  Called the Seiberg-Witten map (SWM), this construction
  is viable for $SU(N)$ gauge groups. The SWM formulation of the NCSM 
  thus uses the SM gauge group and particle content, but adds new interactions
  \cite{Calmet:2001na,Melic:2005fm,Melic:2005am}.  This is the model we will study in this paper, and we will simply refer to it as the 
  NCSM from now on.

  Because of the nonlocal nature of Eq.~\ref{Moyal}, the quantization of NC field theory is fraught with subtlety.  Many authors have pointed out
  potential issues with unitarity, especially in the space-time NC case 
  \cite{Gomis:2000zz,Seiberg:2000gc,Chaichian:2000ia,Chu:2002fe,Ohl:2003zd,Bassetto:2001vf,AlvarezGaume:2001ka}.  
  On the other hand, with suitable modifications to
  the formulation of perturbation theory, others have claimed that unitary space-time NCQFTs can be constructed 
  \cite{Doplicher:1994tu,Bahns:2002vm,Rim:2002if,Liao:2002xc,Liao:2002pj,Berrino:2002ss,Chaichian:2003yu}.  We will not review
  this discussion in any detail, but instead will take the point of view espoused by Calmet \cite{Calmet:2006iy}, that the NCSM formulated
  using the SWM is a valid effective theory
  with a Hermitian Hamiltonian which does not intrinsically violate unitarity, but may violate partial-wave unitarity 
  at some energy scale.\footnote{We note that the this model is in fact renormalizeable with a slightly modified Lagrangian. \cite{Buric:2006wm}
  It would be interesting to see how that modification affects the phenomenology.  See \cite{Alboteanu:2007zz}, p. 42 for a more detailed discussion of
  this issue.}  
  Inspired by the above study of unitarity in the $U(3)\times U(2)\times U(1)$ model \cite{Hewett:2001im}, 
  here, we will determine that scale by studying partial-wave unitarity
  in \epem\to\wpwm in the NCSM.  We will find that the terms in this amplitude that grow with the center-of-mass energy
  do not completely cancel as in the SM.  At a certain energy, therefore, partial-wave unitarity is indeed violated.  We will then discuss
  the ensuing implications for collider phenomenology of this model.

  There has been a great deal of work on the phenomenology of the NCSM.  Some very strong bounds have been placed on NC physics in general,
  but whether they apply to the NCSM formulated using the SWM has not, in every case, been established \cite{Calmet:2004dn}.  
  Defining $\theta_{\mu\nu}\equiv c_{\mu\nu}/\Lambda^2$, where $c_{\mu\nu}$ is an antisymmetric matrix
  of $\O(1)$ numbers and $\Lambda$ is the scale where NC effects become relevant, 
  the strongest bound on the NCSM comes from Lorentz violation \cite{Carroll:2001ws}.
  Many possible phenomenological signatures of this model have been examined 
  \cite{Ohl:2004tn,Schupp:2002up,Haghighat:2005jy,MohammadiNajafabadi:2006iu,Mahajan:2003ze,Iltan:2002ip,Deshpande:2001mu,Najafabadi:2008sa,
    Behr:2002wx,Trampetic:2008bk,Buric:2006nr,Das:2007dn,Abbiendi:2003wva,Melic:2005hb,Melic:2005su,Alboteanu:2007bp}.  Here, in a similar vein to 
  work by Alboteanu, Ohl, and R\"uckl 
  \cite{Alboteanu:2006hh,Alboteanu:2007by}, we study the phenomenology of the NCSM at LEP, the LHC, and the ILC.  We concentrate
  on~\wpwm~production, expecting that the terms in the amplitude that grow large at high energy and threaten unitarity will also
  lead to strong experimental signatures of the NCSM.  We find that this is the case, and that by studying this process one can discover
  the NCSM and measure its parameters at future colliders.

  \section{The Noncommutative Standard Model}

  In the Noncommutative Standard Model, the action is expanded in powers of the 
  antisymmetric noncommutativity tensor $\theta^{\mu\nu}$ 
  using the Seiberg-Witten map.  No new particles are present in the model, but there are new interactions 
  and modifications to existing interactions.

  The Seiberg-Witten map relates each NC field to a nonlinear function of ordinary fields and $\theta^{\mu\nu}$.  The form of this function, for each field, is obtained by 
  solving the so-called gauge equivalence condition, the requirement that an NC gauge transformation of NC fields 
  induces an ordinary gauge transformation of the associated ordinary fields.  For example, for an NC gauge theory with gauge field $\hat{A}$, matter field 
  $\hat{\psi}$, and gauge parameter $\lambdah$, the SWM maps these NC objects to the corresponding functions $\Ah(A,\theta)$, $\psih(\psi,A,\theta)$, and
  $\lambdah(\lambda,A,\theta)$, respectively.
  The NC gauge transformations have the same form as ordinary gauge transformations, with products replaced by star products as expected,
  \begin{align}
    \psih\to\psih'&=\exp(ig\lambdah\star)\psih=\psih+ig\lambdah\star\psih+\frac{(ig)^2}{2!}\lambdah\star\lambdah\star\psih+\O(\lambdah^3)\;,\nonumber\\
    \Ah_\mu\to\Ah_\mu'&=\exp(ig\lambdah\star)\Ah_\mu\exp(-ig\lambdah\star)+\frac{i}{g}\exp(ig\lambdah\star)(\partial_\mu\exp(-ig\lambdah\star))\\\nonumber
    &=\Ah_\mu+ig[\lambdah\coms\Ah_\mu]+\frac{(ig)^2}{2!}[\lambdah\coms[\lambdah\coms\Ah_\mu]]+\partial_\mu\lambdah+ig[\lambdah\coms\partial_\mu\lambdah]+\O(\lambdah^3)\;.
  \end{align}
  The gauge equivalence conditions then state that   
  \begin{align}
    \Ah'(A,\theta)&=\Ah(A',\theta)\;,\nonumber\\
    \psih'(\psi,A,\theta)&=\psih(\psi',A',\theta)\\\nonumber
    \lambdah'(\lambda,A,\theta)&=\lambdah(\lambda',A',\theta)\;.
  \end{align}
  Solving these conditions to first order in $\theta^{\mu\nu}$ yields
  \begin{align}
    \hat{A}_{\mu}(x)&=A_{\mu}(x)+\frac{1}{4}\theta^{\rho\sigma}\{A_\sigma(x),\partial_\rho A_\mu(x)+F_{\rho\mu}(x)\}+\mathcal{O}(\theta^2)\;,\nonumber\\
    \hat{\psi}(x)&=\psi(x)+\frac{1}{2}\theta^{\rho\sigma}A_\sigma(x)\partial_\rho\psi(x)+\frac{i}{8}
    \theta^{\rho\sigma}[A_\rho(x),A_\sigma(x)]\psi(x)+\mathcal{O}(\theta^2)\;,\label{swm1}\\\nonumber
    \lambdah(x)&=\lambda(x)+\frac{1}{4}\theta^{\rho\sigma}\{A_\sigma(x),\partial_\rho\lambda(x)\}+\O(\theta^2)\;.
  \end{align}
  The NCSM Lagrangian can then be formed, essentially, by starting with the SM Lagrangian and replacing all fields with their NC counterparts.  Then, by using
  the SWM map given in the above expressions and Eq.~\ref{Moyal}, the Lagrangian can be expanded order-by-order in $\theta^{\mu\nu}$.

  There are two caveats to this simple procedure.  The first is that in order to solve the charge quantization problem that plagues
  NC gauge theories \cite{Hayakawa:1999yt}, one must introduce a separate NC gauge field for each $U(1)$ eigenvalue.  Since each
  of these is mapped to the same SM gauge field under the SWM, however, there are no new fields in the effective theory \cite{Calmet:2001na}.

  The second is that, as can be seen by the anticommutator on the right-hand side of the first line of Eq.~\ref{swm1},
  the NC gauge field $\hat{A}$ takes values in the enveloping algebra of the Lie algebra of the SM gauge group.  
  This means that the SWM does not fully determine the gauge kinetic terms.  The resulting
  freedom in the gauge sector is parametrized \cite{Behr:2002wx} with the constants $\kappa_1$, $\kappa_2$, and $\kappa_3$.  Since only
  $\kappa_2$ contributes to the interactions of the $W$ boson, it is the only parameter relevant here.  The choice
  $\kappa_1=\kappa_2=\kappa_3=0$ is called the minimal NCSM (mNCSM).  Throughout this paper we will generally work with the non-minimal NCSM, 
  considering various values of $\kappa_2$ and determining
  bounds on it and $\Lambda$ from collider data.  There are also bounds on the $\kappa$ parameters from consistency conditions on the theory
  \cite{Deshpande:2001mu}.

  As mentioned above, the scale of noncommutativity $\Lambda$ appears as
  $\theta_{\mu\nu}=c_{\mu\nu}/\Lambda^2$, where in this paper we will 
  take $c_{\mu\nu}$ to be an antisymmetric matrix with each element being either 0, 1, or -1.
  The space-time case $c_{0i}\neq0$, as stated earlier, has been shown to be problematic
  for unitarity, and we will see below that it is only this case that leads to dangerous unitarity-violating
  terms in $WW$ scattering.  We will also find that the space-space elements of $\theta^{\mu\nu}$ give significant contributions
  to specific $W$ polarizations in $WW$ production at colliders.

  Here we present the Feynman rules relevant to our discussion
  as derived in the NCSM to $\O(\theta)$.
  In the diagrams displayed below,
  a vertex with a square refers to the $\O(\theta)$ contribution to that Feynman
  rule.  Most of the rules relevant to this analysis are derived in Melic et al. \cite{Melic:2005fm}
  All gauge boson momenta are incoming.

  First, there are modifications to the SM fermion-gauge boson 3-point couplings,
  
  \allowdisplaybreaks \begin{align}
    \fmfframe(20,20)(20,20){
      \begin{fmfgraph*}(60,40)
	\fmflabel{$\bar{f}_u$}{i1}
	\fmflabel{$f_d$}{i2}
	\fmflabel{$W^+_\mu$}{o1} 
	\fmfleft{i2,i1}
	\fmfright{o1}
	\fmfv{decoration.shape=square,decoration.filled=empty,decoration.size=40.}{v1}
	\fmf{fermion,label=$p$}{i2,v1}
	\fmf{fermion}{v1,i1}
	\fmf{boson,label=$k$}{v1,o1}
      \end{fmfgraph*}
    }
    &=\frac{e}{4\sqrt{2}\sin \theta_W}\theta_{\mu\nu\rho}k^\nu p^\rho
    (1-\gamma_5)\;,\nonumber\\
    \fmfframe(20,20)(20,20){
      \begin{fmfgraph*}(60,40)
	\fmflabel{$\bar{f}$}{i1}
	\fmflabel{$f$}{i2}
	\fmflabel{$A_\mu$}{o1} 
	\fmfleft{i2,i1}
	\fmfright{o1}
	\fmfv{decoration.shape=square,decoration.filled=empty,decoration.size=40.}{v1}
	\fmf{fermion,label=$p$}{i2,v1}
	\fmf{fermion}{v1,i1}
	\fmf{boson,label=$k$}{v1,o1}
      \end{fmfgraph*}
    }
    &=\frac{eQ_f}{2}\theta_{\mu\nu\rho}k^\nu p^\rho\;,\\\nonumber
    \fmfframe(20,20)(20,20){
      \begin{fmfgraph*}(60,40)
	\fmflabel{$\bar{f}$}{i1}
	\fmflabel{$f$}{i2}
	\fmflabel{$Z_\mu$}{o1} 
	\fmfleft{i2,i1}
	\fmfright{o1}
	\fmfv{decoration.shape=square,decoration.filled=empty,decoration.size=40.}{v1}
	\fmf{fermion,label=$p$}{i2,v1}
	\fmf{fermion}{v1,i1}
	\fmf{boson,label=$k$}{v1,o1}
      \end{fmfgraph*}
    }
    &=\frac{eQ_f}{2}\theta_{\mu\nu\rho}k^\nu p^\rho
    (g_V-g_A\gamma_5)\;,
  \end{align}
  
  where $\theta_{\mu\nu\rho}=\theta_{\mu\nu}\gamma_\rho+\theta_{\nu\rho}\gamma_\mu
  +\theta_{\rho\mu}\gamma_\nu$\;.

  Then there are modifications to the SM three-gauge-boson vertices,
  \begin{align}
    \fmfframe(20,20)(20,20){
      \begin{fmfgraph*}(60,40)
	\fmflabel{$W^+_\rho$}{i1}
	  \fmflabel{$A_\mu$}{i2}
	  \fmflabel{$W^-_\nu$}{o1} 
	  \fmfleft{i2,i1}
	  \fmfright{o1}
	  \fmfv{decoration.shape=square,decoration.filled=empty,decoration.size=40.}{v1}
	  \fmf{boson,label=$p$}{i2,v1}
	  \fmf{boson,label=$r$}{v1,i1}
	  \fmf{boson,label=$q$}{v1,o1}
      \end{fmfgraph*}
    }&=-\frac{em_W^2}{2}f^A_{\mu\nu\rho}(p)\nonumber\\
    &\quad +2e\sin2\theta_WK_{WW\gamma}\Th_{\mu\nu\rho}(p,q,r)\;,\\\nonumber
    \fmfframe(20,20)(20,20){
      \begin{fmfgraph*}(60,40)
	\fmflabel{$W^+_\rho$}{i1}
	  \fmflabel{$Z_\mu$}{i2}
	  \fmflabel{$W^-_\nu$}{o1} 
	  \fmfleft{i2,i1}
	  \fmfright{o1}
	  \fmfv{decoration.shape=square,decoration.filled=empty,decoration.size=40.}{v1}
	  \fmf{boson,label=$p$}{i2,v1}
	  \fmf{boson,label=$r$}{v1,i1}
	  \fmf{boson,label=$q$}{v1,o1}
      \end{fmfgraph*}
    }&=-\frac{em_W^2}{2}f^A_{\mu\nu\rho}(p)+\frac{em_Z^2}{4}f^Z_{\mu\nu\rho}(p,q,r)\\\nonumber
    &\quad +2e\sin2\theta_WK_{WWZ}\Th_{\mu\nu\rho}(p,q,r)\;,
  \end{align}
  where 
  \begin{align}
    f^A_{\mu\nu\rho}(p)\equiv&\,\theta_{\mu\nu}p_\rho+\theta_{\mu\rho}p_\nu\nonumber\\
                            +&\, g_{\mu\nu}(\theta\cdot p)_\rho-g_{\nu\rho}(\theta\cdot p)_\mu+g_{\rho\mu}(\theta\cdot p)_\nu\; ,\nonumber\\
    f^Z_{\mu\nu\rho}(p,q,r)\equiv&\;\theta_{\mu\nu}(p-q)_\rho+\theta_{\nu\rho}(q-r)_\mu+\theta_{\rho\mu}(r-p)_\nu \nonumber\\
                                 -&2g_{\mu\nu}(\theta\cdot r)_\rho-2g_{\nu\rho}(\theta\cdot p)_\mu-2g_{\rho\mu}(\theta\cdot q)_\nu\; ,\\
    \Th_{\mu\nu\rho}(p,q,r)\equiv&\,\theta_{\mu\nu}\left(p\cdot r\; q_\rho-q\cdot r\; p_\rho\right)
                                  +(\theta\cdot p)_\mu\left(q\cdot r\; g_{\nu\rho}-q_\rho r_\nu\right)\nonumber\\
                                -&\,(\theta\cdot p)_\nu\left(q\cdot r\; g_{\rho\mu}-q_\rho r_\mu\right)
                                  -(\theta\cdot p)_\rho\left(q\cdot r\; g_{\mu\nu}-q_\mu r_\nu\right)\nonumber\\
				+&\,p\times q\left(r_\mu g_{\nu\rho}-r_\nu g_{\rho\mu}\right)\nonumber\\
				+&\,(\text{cyclic permutations of }\{p,q,r\}\text{ and }\{\mu,\nu,\rho\}\text{ simultaneously})\,,\nonumber
  \end{align}
  $K_{WW\gamma}=-g^2\kappa_2/2c_ws_w,$ $K_{WWZ}=g^2\kappa_2/2c_w^2$, and $p\times q\equiv p^\mu\theta_{\mu\nu}q^\nu$.  Here, $\kappa_2$ is the parameter
  capturing the freedom in the gauge sector discussed above.

  There is also a 4-point fermion-gauge boson interaction,
  \begin{equation}
    \fmfframe(20,20)(20,20){
      \begin{fmfgraph*}(60,40)
        \fmflabel{$\bar{f}$}{i1}
        \fmflabel{$f$}{i2}
        \fmflabel{$W^+_\mu$}{o2}
        \fmflabel{$W^-_\nu$}{o1}
        \fmfv{decoration.shape=square,decoration.filled=empty,decoration.size=40.}{v1}
        \fmfleft{i1,i2}
        \fmfright{o1,o2}
        \fmf{fermion,label=$p$,label.dist=30}{i1,v1}
        \fmf{fermion}{v1,i2}
        \fmf{boson}{o2,v1}
        \fmf{boson,label=$k_1$,label.side=right,label.dist=30}{o1,v1}
      \end{fmfgraph*}
    }=-\frac{g^2}{8}(\theta_{\mu\nu\rho}(p^\rho+k_1^\rho)(1-\gamma_5)+2\theta_{\mu\nu}m)\;.
  \end{equation}

  Finally there is a modification to the 4-W vertex, which was not previously presented in the literature,
  \begin{equation}
    \fmfframe(20,20)(20,20){
      \begin{fmfgraph*}(60,40)
	\fmflabel{$W^+_\mu$}{i1}
	\fmflabel{$W^-_\nu$}{i2}
	\fmflabel{$W^+_\rho$}{o1} 
	\fmflabel{$W^-_\sigma$}{o2}
	\fmfv{decoration.shape=square,decoration.filled=empty,decoration.size=40.}{v1}
	\fmfleft{i1,i2}
	\fmfright{o1,o2}
	\fmf{boson}{i2,v1,i1}
	\fmf{boson}{o1,v1,o2}
      \end{fmfgraph*}
    }=-\frac{g^2}{4}m_W^2\left(g_{\rho\sigma}\theta_{\mu\nu}+g_{\nu\sigma}\theta_{\mu\rho}+g_{\nu\rho}\theta_{\mu\sigma}+g_{\mu\sigma}\theta_{\nu\rho}\right)\;.
  \end{equation}
  
  Note that all of the $\O(\theta)$ contributions to the Feynman
  rules are opposite in phase from the SM contributions ({\it i.e.} one is real, while the other is imaginary).  
  This fact determines the lowest order in $\theta$ at which the NCSM corrections to collider observables occur.
  To see why this is, we can examine the phases of various contributions to a scattering amplitude.
  For example, consider $\wpwm\to\wpwm$ scattering with $s$-channel photon exchange.  Up to an overall phase, the amplitude can be written as
  \begin{equation}
    A=\epsilon_\mu\epsilon_\nu (V_{SM}^{\mu\nu\lambda}+iV_{NC}^{\mu\nu\lambda})(\frac{-ig_{\lambda\eta}}{s})
    (V_{SM}^{\rho\sigma\eta}+iV_{NC}^{\rho\sigma\eta})\epsilon^*_\rho\epsilon^*_\sigma\;,
  \end{equation}
  where $\epsilon$ is a polarization vector for one of the external $W$ bosons, and $V$, which
  represents the $\gamma W^+W^-$ Feynman rule, can taken to be real.  $V_{SM}$
  is the Standard Model Feynman rule, while $V_{NC}$ is the NC correction at first order in $\theta$.
  Truncating the amplitude at first order in $\theta$ yields
  \begin{equation}
    A=\frac{-ig_{\lambda\eta}}{s}\epsilon_\mu\epsilon_\nu((V_{SM}^{\mu\nu\lambda} V_{SM}^{\rho\sigma\eta}+
    i(V_{SM}^{\mu\nu\lambda}V_{NC}^{\rho\sigma\eta}+V_{SM}^{\rho\sigma\lambda}V_{NC}^{\mu\nu\eta}))
    \epsilon^*_\rho\epsilon^*_\sigma\;.
  \end{equation}
  Defining $\vv^\lambda\equiv\epsilon_\mu\epsilon_\nu V^{\mu\nu\lambda}$ yields
  $\vv^{\lambda *}=\epsilon_\mu^*\epsilon_\nu^* V^{\mu\nu\lambda}$, since V is real.  The amplitude 
  can now be written as
  \begin{equation}
    A=-\frac{i}{s}(\vv_{SM}\cdot\vv_{SM}^*+i(\vv_{SM}\cdot\vv_{NC}^*+\vv_{SM}^*\cdot\vv_{NC})).
  \end{equation}
  If the $\epsilon_\mu$ vectors are real, then $\vv$ is real, in which case the squared amplitude is
  \begin{equation}
    |A|^2=\frac{1}{s^2}((\vv_{SM}\cdot\vv_{SM})^2+(\vv_{SM}\cdot\vv_{NC}+\vv_{SM}\cdot\vv_{NC})^2),
  \end{equation}
  which has no terms that are first order in $\theta$.  In fact, so long as each $\epsilon_\mu$ has only an overall phase, that is,
  each Lorentz component of a given polarization four-vector has the same phase, then the phase can be factored out
  into an overall phase multiplying the whole amplitude and the argument 
  still goes through.  To obtain a first-order contribution, we require that at least one of the polarization vectors
  has a relative phase difference among its components.  For $\wpwm\to\wpwm$, this means that at least one of the $W$ bosons must be transversely 
  polarized.  In the sum over all polarizations, however, $\O(\theta)$ contributions to the amplitude must vanish.  

  With fermion external states, however, every helicity choice has a spinor with relative phase differences among its components.  In this case
  there are nonzero $\O(\theta)$ contributions to the squared amplitude for any helicity choice and for the sum over helicities.
  
  \section{Partial Wave Unitarity}
  
  It is expected that any perturbative model of new physics satisfy tree-level unitarity.  It is well-known that
  any amplitude that exhibits azimuthal rotational invariance
  can be decomposed into partial waves $a_l$ according to
  \begin{equation}
    A=16\pi\sum_{l=0}^{\infty}a_l(2l+1)P_l(\cos\theta),
  \end{equation}
  where $\theta$ is the scattering angle in the center-of-mass frame, $P_l$ is the $l$th Legendre polynomial, 
  and $a_l$ is called the $l$th partial wave amplitude.  The orthogonality of the Legendre polynomials then gives
  \begin{equation}
    \label{al}
    a_l=\frac{1}{32\pi}\int_{-1}^{1}A(\cos\theta)P_l(\cos\theta)d\cos\theta.
  \end{equation}
  It is a consequence of probability conservation that 
  \begin{equation}
    \label{unitbounds}
    \Re(a_l)\leq 1/2\,,\quad0\le\Im(a_l)\le1\,,\quad\text{and}\quad|a_l|^2\le1\,.  
  \end{equation}

  In the Standard Model without a Higgs boson, the~\wpwm\to\wpwm~amplitude contains terms that grow linearly with $s$.  At high enough scattering energy, these terms
  grow large enough to violate the above partial-wave unitarity bounds.  
  The inclusion of Higgs exchange diagrams cancels these growing terms, and (for large Higgs mass $m_H$)
  the leading terms that remain are proportional to $m_H^2$.  Partial-wave unitarity then gives an upper bound to the Higgs mass $m_H\lesssim 800\gev$.
  One can also check unitarity in other processes in the SM, such as $\epem\to\wpwm$.  In this case the terms in the amplitude that
  go as $s$ once again cancel, though they do so in the absence of diagrams involving the Higgs, as they should, since the Higgs diagrams vanish
  in the limit of small electron mass.

  If, in a model of new physics, this cancellation of terms that grow with scattering energy does not occur, unitarity will be violated at some scale.  
  Above this scale, the model is invalid or incomplete.  As we will see, this is the case in the NCSM, though since a partial cancellation occurs 
  for~\wpwm\to\wpwm,~a stronger effect occurs in~\epem\to\wpwm.

  The partial wave unitarity analysis in the NCSM is complicated by the fact that the $\theta_{\mu\nu}$ tensor
  gives preferred directions in space.  This breaks the azimuthal rotational invariance.
  The simple partial-wave analysis discussed above is thus not necessarily valid.  The appropriate form for the partial-wave unitarity bounds 
  for amplitudes without azimuthal symmetry was given by Chaichian, Montonen, and Tureanu 
  \cite{Chaichian:2003yu}.  These bounds apply to a partial-wave expansion of the amplitude in all of the independent angular variables.  Evaluating these bounds
  requires calculating the amplitudes with an arbitrary crossing angle, which is beyond the scope of this paper.

  To get an estimate of the unitarity constraints, we will instead remove the $\phi$ dependence from the amplitude by fixing $\phi$ at a value that
  maximizes the NC contribution and use the
  bounds for azimuthally symmetric systems given in Eq.~\ref{unitbounds}.  While not strictly accurate, if we fix $\phi$ so as to maximize the 
  NC contribution, this procedure should provide a ``worst-case
  scenario.'' If the unitarity bounds are weak in this case we can conclude that a more thorough analysis will not give significantly stronger limits.
  
  We will first examine \wpwm\to\wpwm, the process that is most dangerous in the SM.
  In the NCSM, there are twenty diagrams contributing to this process at tree level.  There are six s-channel gauge boson exchange diagrams,
  \begin{equation}
    \fmfframe(20,20)(20,20){
      \begin{fmfgraph*}(60,40)
	\fmflabel{$W^+$}{i1}
	\fmflabel{$W^-$}{i2}
	\fmflabel{$W^+$}{o1} 
	\fmflabel{$W^-$}{o2}
	\fmfleft{i1,i2}
	\fmfright{o1,o2}
	\fmf{boson}{i2,v1,i1}
	\fmf{boson,label=$\gamma$,,$Z$}{v1,v2}
	\fmf{boson}{o1,v2,o2}
      \end{fmfgraph*}
    }
    \fmfframe(20,20)(20,20){
      \begin{fmfgraph*}(60,40)
	\fmflabel{$W^+$}{i1}
	\fmflabel{$W^-$}{i2}
	\fmflabel{$W^+$}{o1} 
	\fmflabel{$W^-$}{o2}
	\fmfleft{i1,i2}
	\fmfright{o1,o2}
	\fmf{boson}{i2,v1,i1}
	\fmfv{decoration.shape=square,decoration.filled=empty,decoration.size=40.}{v1}
	\fmf{boson,label=$\gamma$,,$Z$}{v1,v2}
	\fmf{boson}{o1,v2,o2}
      \end{fmfgraph*}
    } 
    \fmfframe(20,20)(20,20){
      \begin{fmfgraph*}(60,40)
	\fmflabel{$W^+$}{i1}
	\fmflabel{$W^-$}{i2}
	\fmflabel{$W^+$}{o1} 
	\fmflabel{$W^-$}{o2}
	\fmfleft{i1,i2}
	\fmfright{o1,o2}
	\fmf{boson}{i2,v1,i1}
	\fmfv{decoration.shape=square,decoration.filled=empty,decoration.size=40.}{v2}
	\fmf{boson,label=$\gamma$,,$Z$}{v1,v2}
	\fmf{boson}{o1,v2,o2}
      \end{fmfgraph*}
    } 
  \end{equation}
  six t-channel gauge boson exchange diagrams,
  \begin{equation}
    \fmfframe(20,20)(20,20){
      \begin{fmfgraph*}(60,40)
	\fmflabel{$W^+$}{i1}
	\fmflabel{$W^-$}{i2}
	\fmflabel{$W^+$}{o1} 
	\fmflabel{$W^-$}{o2}
	\fmfleft{i1,i2}
	\fmfright{o1,o2}
	\fmf{boson}{i2,v2,o2}
	\fmf{boson,label=$\gamma$,,$Z$}{v1,v2}
	\fmf{boson}{i1,v1,o1}
      \end{fmfgraph*}
    }
    \fmfframe(20,20)(20,20){
      \begin{fmfgraph*}(60,40)
	\fmflabel{$W^+$}{i1}
	\fmflabel{$W^-$}{i2}
	\fmflabel{$W^+$}{o1} 
	\fmflabel{$W^-$}{o2}
	\fmfleft{i1,i2}
	\fmfright{o1,o2}
	\fmf{boson}{i1,v1,o1}
	\fmfv{decoration.shape=square,decoration.filled=empty,decoration.size=40.}{v1}
	\fmf{boson,label=$\gamma$,,$Z$}{v1,v2}
	\fmf{boson}{i2,v2,o2}
      \end{fmfgraph*}
    } 
    \fmfframe(20,20)(20,20){
      \begin{fmfgraph*}(60,40)
	\fmflabel{$W^+$}{i1}
	\fmflabel{$W^-$}{i2}
	\fmflabel{$W^+$}{o1} 
	\fmflabel{$W^-$}{o2}
	\fmfleft{i1,i2}
	\fmfright{o1,o2}
	\fmf{boson}{i1,v1,o1}
	\fmfv{decoration.shape=square,decoration.filled=empty,decoration.size=40.}{v2}
	\fmf{boson,label=$\gamma$,,$Z$}{v1,v2}
	\fmf{boson}{i2,v2,o2}
      \end{fmfgraph*}
    } 
  \end{equation}
  three s-channel Higgs exchange diagrams,
  \begin{equation}
    \fmfframe(20,20)(20,20){
      \begin{fmfgraph*}(60,40)
	\fmflabel{$W^+$}{i1}
	\fmflabel{$W^-$}{i2}
	\fmflabel{$W^+$}{o1} 
	\fmflabel{$W^-$}{o2}
	\fmfleft{i1,i2}
	\fmfright{o1,o2}
	\fmf{boson}{i2,v1,i1}
	\fmf{dashes,label=$H$}{v1,v2}
	\fmf{boson}{o1,v2,o2}
      \end{fmfgraph*}
    }
    \fmfframe(20,20)(20,20){
      \begin{fmfgraph*}(60,40)
	\fmflabel{$W^+$}{i1}
	\fmflabel{$W^-$}{i2}
	\fmflabel{$W^+$}{o1} 
	\fmflabel{$W^-$}{o2}
	\fmfleft{i1,i2}
	\fmfright{o1,o2}
	\fmf{boson}{i2,v1,i1}
	\fmfv{decoration.shape=square,decoration.filled=empty,decoration.size=40.}{v1}
	\fmf{dashes,label=$H$}{v1,v2}
	\fmf{boson}{o1,v2,o2}
      \end{fmfgraph*}
    } 
    \fmfframe(20,20)(20,20){
      \begin{fmfgraph*}(60,40)
	\fmflabel{$W^+$}{i1}
	\fmflabel{$W^-$}{i2}
	\fmflabel{$W^+$}{o1} 
	\fmflabel{$W^-$}{o2}
	\fmfleft{i1,i2}
	\fmfright{o1,o2}
	\fmf{boson}{i2,v1,i1}
	\fmfv{decoration.shape=square,decoration.filled=empty,decoration.size=40.}{v2}
	\fmf{dashes,label=$H$}{v1,v2}
	\fmf{boson}{o1,v2,o2}
      \end{fmfgraph*}
    } 
  \end{equation}
  three t-channel Higgs exchange diagrams,
  \begin{equation}
    \fmfframe(20,20)(20,20){
      \begin{fmfgraph*}(60,40)
	\fmflabel{$W^+$}{i1}
	\fmflabel{$W^-$}{i2}
	\fmflabel{$W^+$}{o1} 
	\fmflabel{$W^-$}{o2}
	\fmfleft{i1,i2}
	\fmfright{o1,o2}
	\fmf{boson}{i2,v2,o2}
	\fmf{dashes,label=$H$}{v1,v2}
	\fmf{boson}{i1,v1,o1}
      \end{fmfgraph*}
    }
    \fmfframe(20,20)(20,20){
      \begin{fmfgraph*}(60,40)
	\fmflabel{$W^+$}{i1}
	\fmflabel{$W^-$}{i2}
	\fmflabel{$W^+$}{o1} 
	\fmflabel{$W^-$}{o2}
	\fmfleft{i1,i2}
	\fmfright{o1,o2}
	\fmf{boson}{i1,v1,o1}
	\fmfv{decoration.shape=square,decoration.filled=empty,decoration.size=40.}{v1}
	\fmf{dashes,label=$H$}{v1,v2}
	\fmf{boson}{i2,v2,o2}
      \end{fmfgraph*}
    } 
    \fmfframe(20,20)(20,20){
      \begin{fmfgraph*}(60,40)
	\fmflabel{$W^+$}{i1}
	\fmflabel{$W^-$}{i2}
	\fmflabel{$W^+$}{o1} 
	\fmflabel{$W^-$}{o2}
	\fmfleft{i1,i2}
	\fmfright{o1,o2}
	\fmf{boson}{i1,v1,o1}
	\fmfv{decoration.shape=square,decoration.filled=empty,decoration.size=40.}{v2}
	\fmf{dashes,label=$H$}{v1,v2}
	\fmf{boson}{i2,v2,o2}
      \end{fmfgraph*}
    } 
  \end{equation}
  and, finally, two contact interaction diagrams.
  \begin{equation}
    \fmfframe(20,20)(20,20){
      \begin{fmfgraph*}(60,40)
	\fmflabel{$W^+$}{i1}
	\fmflabel{$W^-$}{i2}
	\fmflabel{$W^+$}{o1} 
	\fmflabel{$W^-$}{o2}
	\fmfleft{i1,i2}
	\fmfright{o1,o2}
	\fmf{boson}{i2,v1,i1}
	\fmf{boson}{o1,v1,o2}
      \end{fmfgraph*}
    }
    \fmfframe(20,20)(20,20){
      \begin{fmfgraph*}(60,40)
	\fmflabel{$W^+$}{i1}
	\fmflabel{$W^-$}{i2}
	\fmflabel{$W^+$}{o1} 
	\fmflabel{$W^-$}{o2}
	\fmfv{decoration.shape=square,decoration.filled=empty,decoration.size=40.}{v1}
	\fmfleft{i1,i2}
	\fmfright{o1,o2}
	\fmf{boson}{i2,v1,i1}
	\fmf{boson}{o1,v1,o2}
      \end{fmfgraph*}
    }
  \end{equation}
  Here again, a vertex with a square refers to the $\O(\theta)$ NCSM contribution.

  We will consider the case in which all four $W$ bosons are longitudinally polarized, as (like in the SM) this gives the worst high-energy behavior.
  We also assume a coordinate system where the beam line is aligned along the $\hat{z}$- or $\hat{3}$-axis.  We note that in fact, if the orientation of $\theta$
  were fixed with respect to, say, the cosmic microwave background, then the earth's motion would 
  continually alter the orientation of the beamline with respect to $\theta$.
  It is a straightforward though involved procedure to account for this (it was carried out for a study of noncommutative QED at LEP 
  \cite{Kawamoto:2004mc}), and will be neglected here.
  
\begin{figure}[htb!]
    \begin{center}
      \includegraphics[width=10cm]{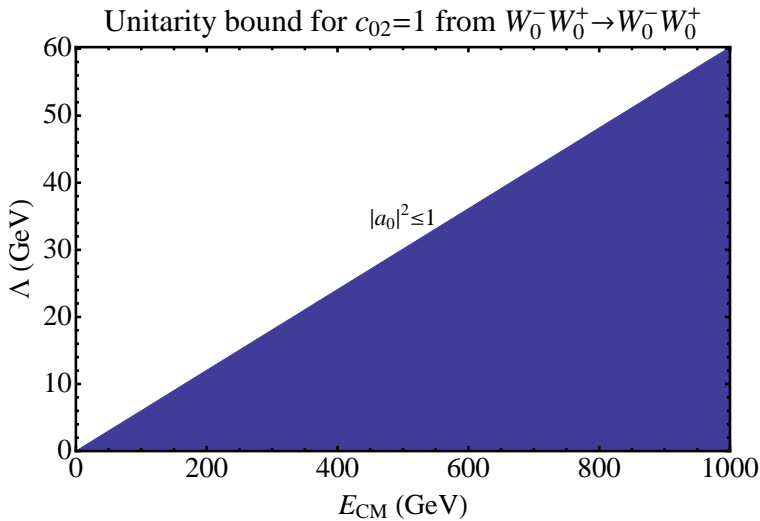}
    \end{center}
    \caption{The bound on the noncommutativity scale $\Lambda$ as a function of $\sqrt{s}$ from partial-wave unitarity violation  
      in WW scattering.  In the shaded area, unitarity is violated.}
    \label{fig:pu1}
  \end{figure}

  For the SM contribution, as mentioned above, the leading terms are 
  independent of  $s$, and in the limit $m_H\gg m_W$ simplify to
  \begin{equation}
    \mathcal{M}_{SM}(W^+_0W^-_0\to W^+_0W^-_0)_{\text{leading terms in}~s}=\frac{g^2m_H^2}{2m_W^2}\,,
  \end{equation}
  Where the subscript ``0'' indicates longitudinal polarization.
  For the NCSM contribution, the leading terms for each diagram go as $s^2$, but these cancel in the sum over diagrams.  There is an incomplete cancellation among
  the terms that go as $s$, however, and what remains is
  \begin{multline}
    \mathcal{M}_{NC}(W^+_0W^-_0\to W^+_0W^-_0)_{\text{leading terms in}~s}=\\
    i\frac{g^2}{16c_w^2\Lambda^2}\frac{(\cos\theta+1)(\cos\theta-3)}{\cos\theta-1}(\sin\theta(\cos\phi\,c_{01}+\sin\phi\,c_{02})+(\cos\theta-1)c_{03})s\,.
  \end{multline}
  Setting $\phi=\pi/2$ and using Eq.~\ref{al} we get
  \begin{equation}
    a_0(W^+_0W^-_0\to W^+_0W^-_0)=\frac{g^2m_H^2}{32\pi m_W^2}+i\frac{g^2}{3072\pi c_w^2\Lambda^2}(32 c_{03}-21\pi c_{02})s\,,
  \end{equation}
  to which we can now apply the bounds of Eq.~\ref{unitbounds}.  Because $\mathcal{M}_{NC}$ is pure imaginary, 
  the bounds on the magnitude and on the imaginary part of $a_0$ are the relevant ones, and they give numerically equivalent results.
  We choose $c_{02}=1$ and $c_{\mu\nu}=0$ otherwise to get the strongest bound
  on $\Lambda$, shown in Fig.~\ref{fig:pu1}.  Only very small values (compared to $\sqrt{s}$) of $\Lambda$ are excluded.

  Since partial-wave unitarity in~\wpwm\to\wpwm~does not give a meaningful bound on $\Lambda$, we now 
  turn to the process~\epem\to\wpwm.
  There are ten diagrams that contribute to this process at tree level in the NCSM.  
  First, there are s-channel gauge boson exchange diagrams.
  \begin{equation}
    \label{bosons}
    \fmfframe(20,20)(20,20){
      \begin{fmfgraph*}(60,40)
	\fmflabel{$e^+$}{i1}
	\fmflabel{$e^-$}{i2}
	\fmflabel{$W^+$}{o1} 
	\fmflabel{$W^-$}{o2}
	\fmfleft{i1,i2}
	\fmfright{o1,o2}
	\fmf{fermion}{i2,v1,i1}
	\fmf{boson,label=$\gamma$,,$Z$}{v1,v2}
	\fmf{boson}{o1,v2,o2}
      \end{fmfgraph*}
    }
    \fmfframe(20,20)(20,20){
      \begin{fmfgraph*}(60,40)
	\fmflabel{$e^+$}{i1}
	\fmflabel{$e^-$}{i2}
	\fmflabel{$W^+$}{o1} 
	\fmflabel{$W^-$}{o2}
	\fmfleft{i1,i2}
	\fmfright{o1,o2}
	\fmf{fermion}{i2,v1,i1}
	\fmfv{decoration.shape=square,decoration.filled=empty,decoration.size=40.}{v1}
	\fmf{boson,label=$\gamma$,,$Z$}{v1,v2}
	\fmf{boson}{o1,v2,o2}
      \end{fmfgraph*}
    } 
    \fmfframe(20,20)(20,20){
      \begin{fmfgraph*}(60,40)
	\fmflabel{$e^+$}{i1}
	\fmflabel{$e^-$}{i2}
	\fmflabel{$W^+$}{o1} 
	\fmflabel{$W^-$}{o2}
	\fmfleft{i1,i2}
	\fmfright{o1,o2}
	\fmf{fermion}{i2,v1,i1}
	\fmfv{decoration.shape=square,decoration.filled=empty,decoration.size=40.}{v2}
	\fmf{boson,label=$\gamma$,,$Z$}{v1,v2}
	\fmf{boson}{o1,v2,o2}
      \end{fmfgraph*}
    } 
  \end{equation}
  Second, there are neutrino exchange diagrams.  Note that for up-type fermions these will be $u$-channel diagrams, while we 
  show the $t$-channel diagrams appropriate for down-type fermions.
  \begin{equation}
    \label{neutrinos}
    \fmfframe(20,20)(20,20){
      \begin{fmfgraph*}(60,40)
	\fmflabel{$e^+$}{i1}
	\fmflabel{$e^-$}{i2}
	\fmflabel{$W^+$}{o1} 
	\fmflabel{$W^-$}{o2}
	\fmfleft{i1,i2}
	\fmfright{o1,o2}
	\fmf{fermion}{i2,v2}
	\fmf{fermion}{v1,i1}
	\fmf{fermion,label=$\nu$}{v2,v1}
	\fmf{boson}{o1,v1}
	\fmf{boson}{o2,v2}
      \end{fmfgraph*} 
    } 
    \fmfframe(20,20)(20,20){
      \begin{fmfgraph*}(60,40)
	\fmflabel{$e^+$}{i1}
	\fmflabel{$e^-$}{i2}
	\fmflabel{$W^+$}{o1} 
	\fmflabel{$W^-$}{o2}
	\fmfleft{i1,i2}
	\fmfright{o1,o2}
	\fmfv{decoration.shape=square,decoration.filled=empty,decoration.size=40.}{v2}
	\fmf{fermion}{i2,v2}
	\fmf{fermion}{v1,i1}
	\fmf{fermion,label=$\nu$}{v2,v1}
	\fmf{boson}{o1,v1}
	\fmf{boson}{o2,v2}
      \end{fmfgraph*} 
    }
    \fmfframe(20,20)(20,20){
      \begin{fmfgraph*}(60,40)
	\fmflabel{$e^+$}{i1}
	\fmflabel{$e^-$}{i2}
	\fmflabel{$W^+$}{o1} 
	\fmflabel{$W^-$}{o2}
	\fmfleft{i1,i2}
	\fmfright{o1,o2}
	\fmfv{decoration.shape=square,decoration.filled=empty,decoration.size=40.}{v1}
	\fmf{fermion}{i2,v2}
	\fmf{fermion}{v1,i1}
	\fmf{fermion,label=$\nu$}{v2,v1}
	\fmf{boson}{o1,v1}
	\fmf{boson}{o2,v2}
      \end{fmfgraph*} 
    } 
  \end{equation}
  Finally, there is the contact term.
  \begin{equation}
    \label{contact}
    \fmfframe(20,20)(20,20){
      \begin{fmfgraph*}(60,40)
	\fmflabel{$e^+$}{i1}
	\fmflabel{$e^-$}{i2}
	\fmflabel{$W^+$}{o1} 
	\fmflabel{$W^-$}{o2}
	\fmfv{decoration.shape=square,decoration.filled=empty,decoration.size=40.}{v1}
	\fmfleft{i1,i2}
	\fmfright{o1,o2}
	\fmf{fermion}{i2,v1,i1}
	\fmf{boson}{o1,v1,o2}
      \end{fmfgraph*}
    }
  \end{equation}
  
  We will examine the helicity choice $e^+_Le^-_R\to W^-_0W^+_0$, which has the worst high-energy behavior.  Adding the diagrams, we find, 
  as mentioned above, that the terms which grow with $s$ cancel in the SM, and the remaining constant term is numerically negligible.
  Unlike for~\wpwm\to\wpwm, however, in this process the NCSM terms that go as $s^2$ do not completely cancel, and we are 
  left with
  \begin{equation}
    \label{a0}
    \mathcal{M}(e^+_Le^-_R\to W^-_0W^+_0)_{\text{leading terms in}~s}=-i\frac{g^2}{8m_W^2\Lambda^2}s^2c_{03}\sin\theta\, e^{-i\phi}.
  \end{equation}
  
\begin{figure}[htb!]
    \begin{center}
      \includegraphics[width=10cm]{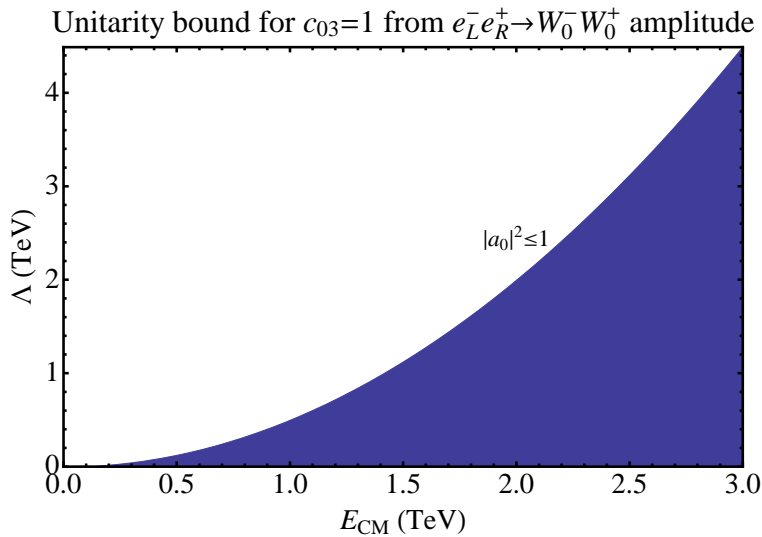}
    \end{center}
    \caption{The bound on the noncommutativity scale $\Lambda$ as a function of $\sqrt{s}$ from partial-wave unitarity violation in \epem\to\wpwm.
      In the shaded area, unitarity is violated.}
    \label{fig:pu1epem}
  \end{figure}

  We once again employ our procedure regarding $\phi$, in this case we set $\phi=0$ and use
  $\Im(a_0)\leq 1/2$.
  Because of the $s^2$ dependence, we find strong bounds on $\Lambda$ as $s$ increases, as shown in Fig.~\ref{fig:pu1epem}.
  This unitarity limit should be compared to the collider search reach in NCSM parameter space, which we discuss below.

  \section{Collider Phenomenology}
  Since unitarity does not exclude the NCSM as a valid or complete theory at LHC and ILC energies, it is natural to investigate its phenomenology at
  these colliders.  Here we examine its effects in $WW$ production.  We will show that this process gives unique signatures that are easily discernible
  for large regions of parameter space, and that by measuring the $W$ polarization one can disentangle and measure the different parameters of the model.

  As mentioned above, our calculations of collider observables assume that the $\hat{3}$ direction ({\it i.e.} the direction corresponding to 
  $\mu=3$ or $\nu=3$ in $\theta_{\mu\nu}$) is aligned with the collider's beam axis, and we do not undertake the procedure \cite{Kawamoto:2004mc}
  of taking into account the earth's motion with respect to the frame in which $\theta_{\mu\nu}$ is fixed.

  \subsection{LEP phenomenology: $\epem\to\wpwm$}

  The process $\epem\to\wpwm$, besides giving unitarity constraints on the NCSM as detailed above, also provides observable signatures at high-energy
  $\epem$ colliders.  The NCSM predicts deviations in the $\epem\to\wpwm$ differential cross section that, for certain
  regions of parameter space, would have already been observable at the high-energy running of the Large Electron-Positron Collider.

  Each $W$ boson can decay either hadronically or leptonically.  There are then three distinct final states for $\epem\to\wpwm$, hadronic
  semileptonic, and leptonic, which occur 45\%, 45\%, and 10\% of the time, respectively.  At lepton colliders, including LEP
  and ILC, each of these final states is easily observable with high efficiency, and in the hadronic and semileptonic cases the full event can be reconstructed.
  For the LEP analysis, we can therefore simply calculate the $\epem\to\wpwm$ differential cross section and multiply by 90\% to exclude the fully leptonic 
  events where reconstruction is impossible and therefore the $W^-$ direction cannot be determined.

  We calculate the amplitude for this process using the diagrams displayed above.  After squaring the amplitude and summing over helicities, we find that
  for the $\O (\theta)$ correction to the differential cross section, the
  dependence on the noncommutativity tensor $\theta_{\mu\nu}$ and on the azimuthal angle $\phi$ takes the simple form 
  \begin{equation}
    \delta d\sigma/d\cos\theta\,d\phi\propto\frac{1}{\Lambda^2}(c_{01}\sin\phi-c_{02}\cos\phi)\;. 
  \end{equation}
  Thus LEP-II is only sensitive to space-time noncommutativity.  Also, any
  observables that are integrated over the full range in $\phi$ will not be sensitive to the NC contributions, since $\cos\phi$ and $\sin\phi$ both
  integrate to zero.  Also note that the leading term in the amplitude
  given in Eq.~\ref{a0}~does not contribute to the {\it squared} amplitude at leading order in $\theta$. 

  To determine the region of parameter space excluded by LEP-II, we take the double differential cross section $\delta d\sigma/d\cos\theta\,d\phi$ 
  binned in $20\times 20$ equally sized bins.  We calculate the $\chi^2$ 
  for this distribution with respect to the SM, assuming a total integrated luminosity of 700~\ipb~and taking into account statistical error 
  plus a $0.1\%$ luminosity uncertainty.
  Choosing $c_{01}=c_{02}=1$, we vary $\Lambda$ and $\kappa_2$ to find the exclusion contours.

  LEP-II operated at a number of center-of-mass 
  energies ranging from 130~\gev~to 209~\gev.  Running at multiple energies is essential to rule out
  acceptance effects as an explanation for any putative azimuthal dependence.  Azimuthal dependence from new physics should change (typically, increase)
  with increasing energy, while azimuthal dependence from acceptance effects should not change with energy.  To simplify our analysis, however, we just
  take all of the integrated luminosity at $E_{CM}=200~\gev$, which is close to the luminosity-weighted average scattering energy.

\begin{figure}[htb!]
    \begin{center}
      \includegraphics[width=10cm]{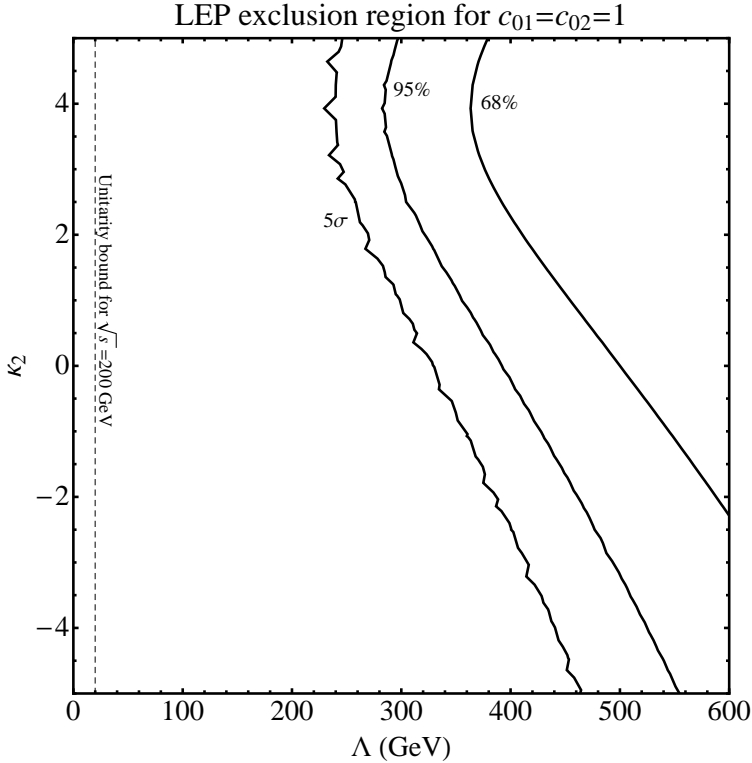}
    \end{center}
    \caption{From left to right, The five-$\sigma$, 95\%, and 68\% confidence level exclusion contours from the process $\epem\to\wpwm$ 
      for the NCSM at LEP-II.  We also display the bound obtained previously from partial-wave unitarity.}
    \label{fig:LEPex}
  \end{figure}

  Fig.~\ref{fig:LEPex}~shows the resulting exclusion region from $W$ pair production at LEP-II.  The reach in $\Lambda$ is above the center-of-mass energy
  for all values of $\kappa_2$.

\begin{figure}[htb!]
    \begin{center}
      \includegraphics[width=10cm]{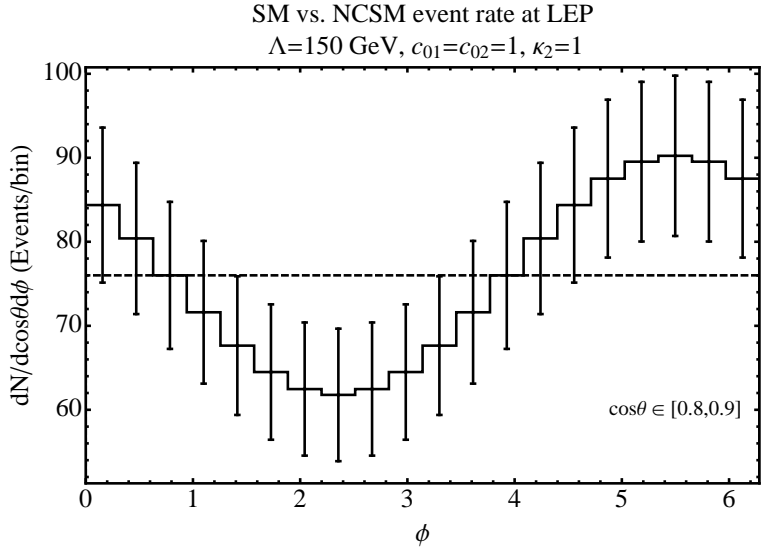}
    \end{center}
    \caption{A comparison of LEP-II data for the SM (dotted) \textit{vs.} the NCSM (solid).  Shown is the the $0.8<\cos\theta<0.9$ bin of 
      the double differential cross section $dN/d\cos\theta d\phi$ for \epem\to\wpwm.  The error bars are statistical plus certain systematics (see text).}
    \label{fig:LEPcomp}
  \end{figure}

  A typical example of the signal that would be observed in the excluded region is given in Fig.~\ref{fig:LEPcomp}.  
  Here, as throughout the paper, we plot the differential event rate $dN/d\cos\theta\,d\phi$, normalized such that the value in each bin
  is the actual number of events that would fall in that bin.  Shown is the $\phi$ distribution for the $0.8<\cos\theta<0.9$ bin.
  The error bars depict the same errors used in the $\chi^2$ analysis--statistical error plus a 0.1\% luminosity uncertainty.
  For $\Lambda=150~\gev$ and $\kappa_2=1$, the $\phi$ distribution deviates from the SM well beyond the errors, and the 
  characteristic sinusoidal shape is visible.

  \subsection{LHC phenomenology: pp\to\wpwm}

  At the LHC, $W$ pairs are copiously produced (in both the SM and NCSM) via the process $\qqb\to\wpwm$, with $q=u,d$ giving the dominant contribution.
  The NCSM diagrams for 
  these processes are the same as were given above for $\epem\to\wpwm$, except that for $\uub$ initial states,
  the $t$-channel diagram is instead a $u$-channel diagram.
  The squared amplitude, once summed over helicities, exhibits the same $\theta_{\mu\nu}$ and $\phi$ dependence as in $\epem\to\wpwm$, given above.

  The large QCD backgrounds at the LHC swamp the hadronic final state, and, as at LEP, for the leptonic final state the event cannot be 
  reconstructed.  Because the relative momentum of the center-of-mass frame is unknown, the full event reconstruction 
  in the semileptonic case is also not possible at the LHC (unlike at an \epem collider).  If one makes the approximation that the leptonically-decaying
  $W$ is always on shell, however,
  then there is enough information to reconstruct the event. 
  Quantifying the validity of this approximation would require calculating the full 4-fermion production amplitudes and using these
  to generate events.  This is beyond the scope of this paper. Here, in order to determine the LHC sensitivity to NCSM parameter space,
  we calculate the production rate of stable, on-shell $W$ bosons and 
  assume that reconstruction is possible in the semileptonic 45\% of events.  Because the $W$ width is relatively small, this should
  give a fairly good approximation to the true search reach.

  To obtain the differential cross section, we use the Mathematica implementation \cite{olness} of the CTEQ5 parton distribution functions \cite{Lai:1999wy}.  
  Because, for a hadron collider,
  the longitudinal-boost-invariant $WW$ invariant mass $m_{WW}$ is a more convenient variable, we will
  consider the $WW$ invariant mass distribution $dN/dm_{WW}d\phi$.  This also takes advantage of the spread of scattering energies available at the LHC.  The NCSM
  effects become easier to observe at higher invariant masses where the the SM event rate falls off.  We choose 20 equal-sized bins 
  spanning from threshold, $m_{WW}=2m_W$, up to $m_{WW}=5.5~\tev$.

  Once again we calculate the $\chi^2$
  for the NCSM distribution relative to its SM counterpart.  As we did for LEP, we include statistical error and a luminosity uncertainty,
  which we take to be 5\% for the LHC.
  Taking $c_{01}=c_{02}=1$, we use the $\chi^2$ to determine the LHC discovery reach
  in the $\Lambda$-$\kappa_2$ plane.  We assume an integrated luminosity of $100~\ifb$.  The result is shown in Fig.~\ref{fig:LHCex}.
  \begin{figure}[htb!]
    \begin{center}
      \includegraphics[width=10cm]{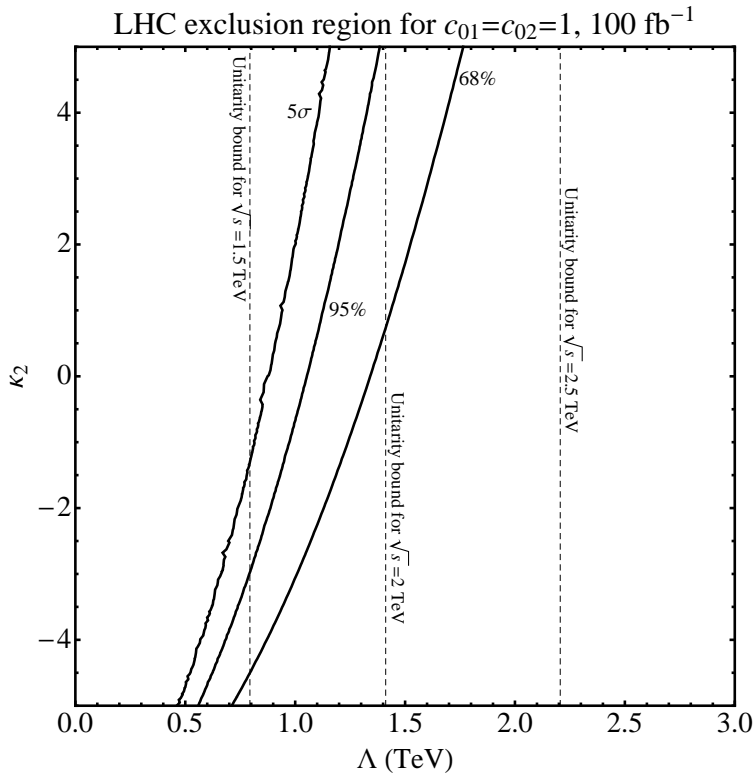}
    \end{center}
    \caption{The search reach for the NCSM from $pp\to \wpwm$ at the LHC for $100~\ifb$ integrated luminosity.  Superimposed are the perturbative unitarity bounds
      for specific parton scattering energies.}
    \label{fig:LHCex}
  \end{figure} 

  Superimposed on these exclusion contours are the unitarity constraints on $\Lambda$ for various values of $\sqrt{s}$, obtained from Fig.~\ref{fig:pu1epem}.
  For example, in order for the model to satisfy tree-level unitarity at scattering energies of $E_{CM}=1~\tev$, $\Lambda$ must be greater than about $200~\gev$.  The LHC
  search reach does not extend very far into the region of parameter space where the NCSM is unitary at typical LHC partonic scattering energies, so we conclude that 
  the process $pp\to WW$ is not an optimal tool to probe the NCSM.

  A typical example of the signal that would be observed in the discovery region is displayed in Fig.~\ref{fig:LHCcomp}.  Shown is the $\phi$ distribution in the
   $1.4\lesssim m_{WW}\lesssim 1.6~\tev$ bin of the double differential event rate $dN/dm_{WW}d\phi$.  The error bars depict the statistical errors plus luminosity
  uncertainty used in the $\chi^2$.  For a low enough value of $\Lambda$, the sinusoidal deviation from the flat SM prediction is observable.
  \begin{figure}[htb!]
    \begin{center}
      \includegraphics[width=10cm]{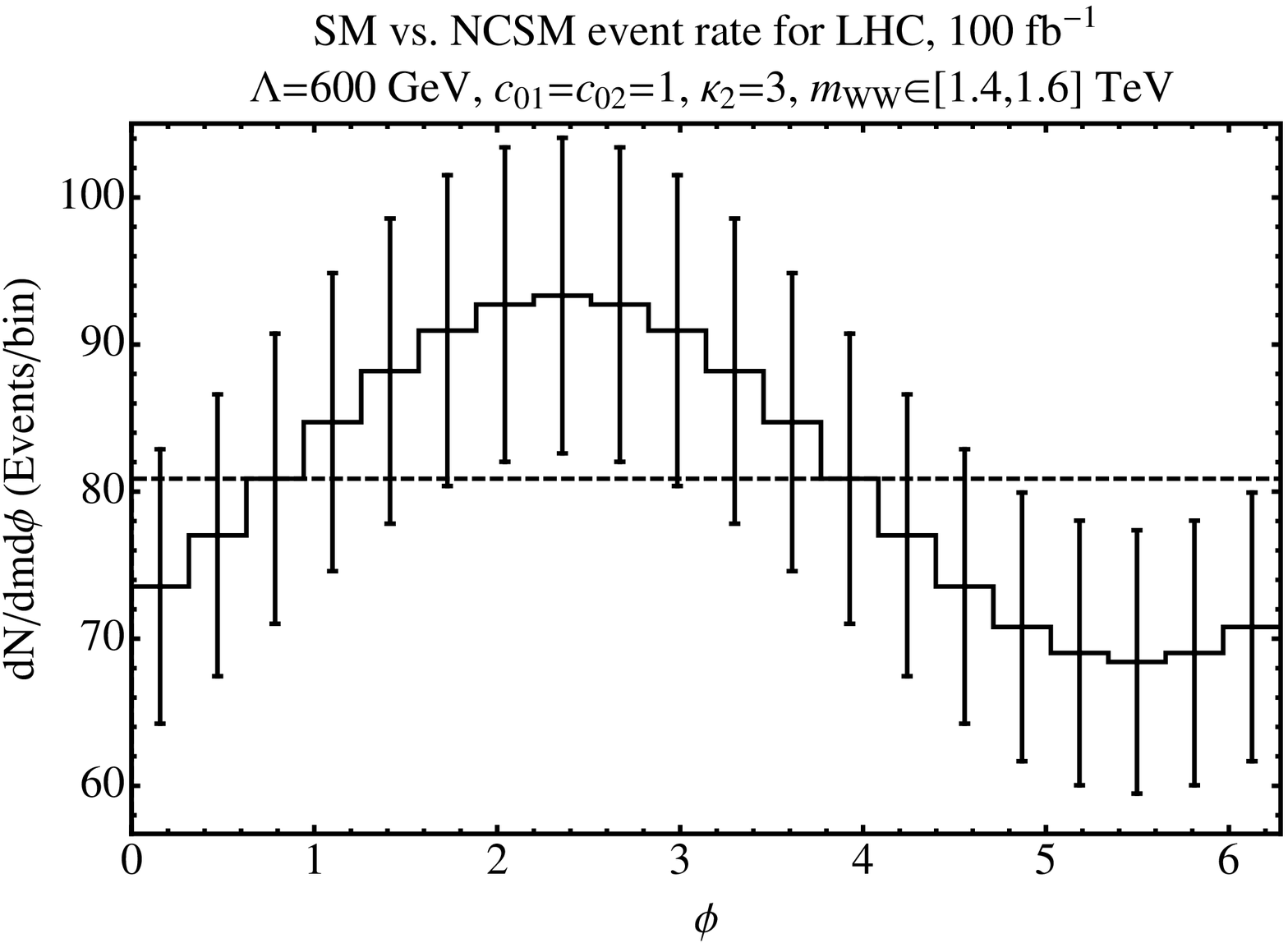}
    \end{center}
    \caption{A comparison of LHC expectations for the SM (dotted) and the NCSM (solid).  Shown is the $1.4\lesssim m_{WW}\lesssim 1.6~\tev$ bin of the
    double differential event rate $dN/dm_{WW}d\phi$.  The error bars are statistical plus certain systematics (see text).}
    \label{fig:LHCcomp}
  \end{figure} 

  The signal should also be visible in the invariant mass distribution, so long as we choose a certain range in $\phi$.  From Fig.~\ref{fig:LHCcomp} we can see that
  integrating over the range from $1\lesssim \phi\lesssim 4$ should maximize the excess of the NCSM over the SM.  Figure~\ref{fig:LHCcomp2} shows the invariant
  mass distribution with the $\phi$ integral taken only over this range.  A systematic excess of the NCSM over the SM is apparent at high values of the invariant mass.
  \begin{figure}[htb!]
    \begin{center}
      \includegraphics[width=10cm]{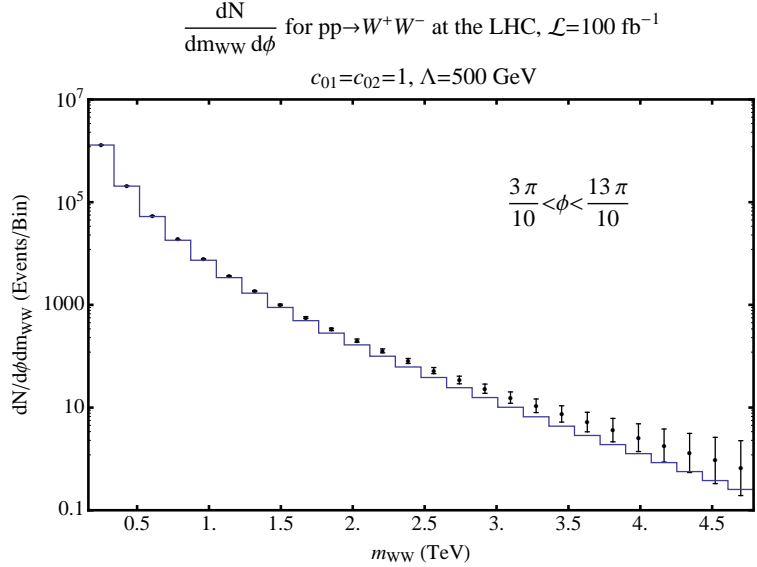}
    \end{center}
    \caption{A comparison of LHC expectations for the SM (solid) and the NCSM (data points) for the 
      differential event rate $dN/dm_{WW}$.  }
    \label{fig:LHCcomp2}
  \end{figure} 

  \subsection{ILC phenomenology: \epem\to\wpwm}

  Our analysis for the ILC is similar to that for LEP except that the center-of-mass energy is higher (500~\gev~and 1~\tev), the luminosity we consider 
  is greater (500~\ifb), and the luminosity uncertainty is smaller ($10^{-4}$).  Also, the ILC will have polarized beam capability, so in addition to 
  the differential cross section $d\sigma/d\cos\theta\,d\phi$, we can consider the differential left-right asymmetry
  $dA_{LR}/d\cos\theta\,d\phi$, where the left-right asymmetry $A_{LR}$ is defined as the asymmetry between cross sections with fully
  polarized electron and positron beams, that is
  \begin{equation}
    A_{LR}=\frac{\sigma(e^-_Le^+_R\to\wpwm)-\sigma(e^-_Re^+_L\to\wpwm)}{\sigma(e^-_Le^+_R\to\wpwm)+\sigma(e^-_Re^+_L\to\wpwm)}\;.
  \end{equation}
  Of course, at a real collider, the beam polarization is imperfect, but is straightforward to relate the asymmetry measured
  with partially polarized beams to the ``theoretical'' value of $A_{LR}$ defined above.  The beam polarizations and the 
  accuracy to which they are known of course affect the errors on the measurement of $A_{LR}$ and thus the bounds obtainable
  from it.
  We assume that the electron and positron beams have polarizations $P_{e^-}=0.9$ and $P_{e^+}=0.6$, respectively, and
  that the polarization uncertainty is $\Delta P/P=0.25\%$.  
  It should be noted that $A_{LR}$ shows the same dependence on $\theta_{\mu\nu}$ and
  $\phi$ as the unpolarized differential cross section.  As we will see in the next section, distinguishing contributions from 
  the other components of $\theta$ requires measuring the $W$ polarization.
  
  Using these two observables, we again perform a $\chi^2$ analysis and determine the exclusion contours for the ILC.
  Figure \ref{fig:ILCex} shows the discovery reach for the ILC assuming 500~\ifb~of operation at $\sqrt{s}=500~\gev$, taking $c_{01}=c_{02}=1$.  Shown are
  contours both with and without the inclusion of $dA_{LR}/d\cos\theta\,d\phi$, along with the unitarity bound for $\sqrt{s}=500~\gev$.  Unlike at the LHC, at the ILC
  the search reach extends well beyond the unitarity bound.  Also evident is a dip in the no-$A_{LR}$ exclusion contours
  around $\kappa_2\sim 2$ where the unpolarized observables
  are relatively insensitive to the NC effects.  This is due to a partial cancellation in the amplitude between the minimal and non-minimal NCSM contributions.
  As can be seen from the other set of contours, the addition of $A_{LR}$ as an observable untangles this cancellation and improves the search reach greatly.
  \begin{figure}[htb!]
    \begin{center}
      \includegraphics[width=10cm]{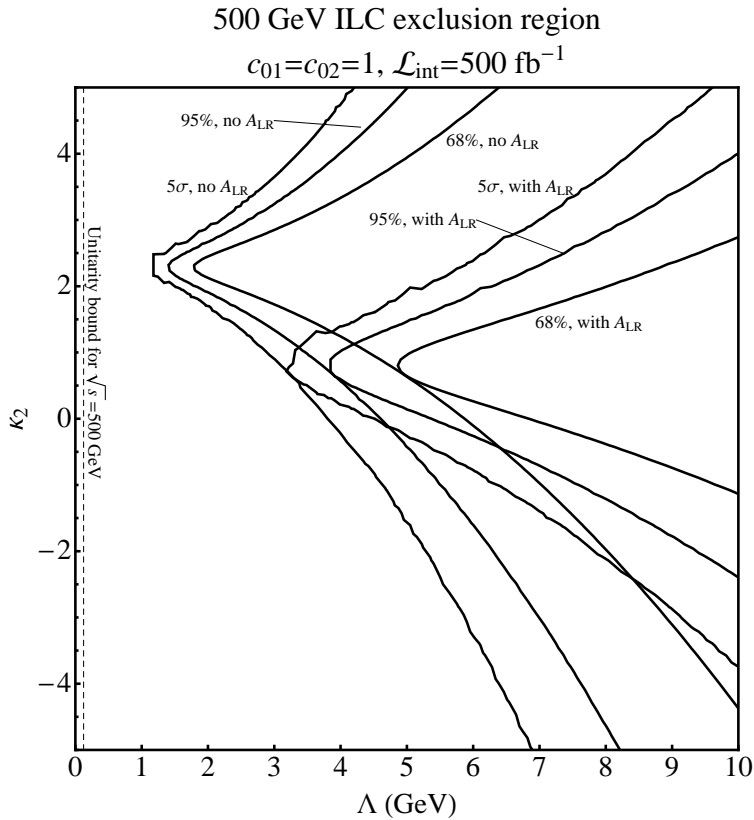}
    \end{center}
    \caption{From left to right, The five-$\sigma$, 95\%, and 68\% confidence level discovery reach for the NCSM from 
      \epem\to\wpwm at the $500~\gev$ ILC, with and without
      $A_{LR}$.}
    \label{fig:ILCex}
  \end{figure} 

  Figure~\ref{fig:ILCcomp} shows a typical example of the unpolarized signal in the discovery region.  We display the $-0.9<\cos\theta<-0.8$ bin of
  $dN/d\cos\theta d\phi$ for $\Lambda=1~\tev$ and $\kappa_2=1$.  The sinusoidal deviation from the flat SM distribution is apparent.
  \begin{figure}[htb!]
    \begin{center}
      \includegraphics[width=10cm]{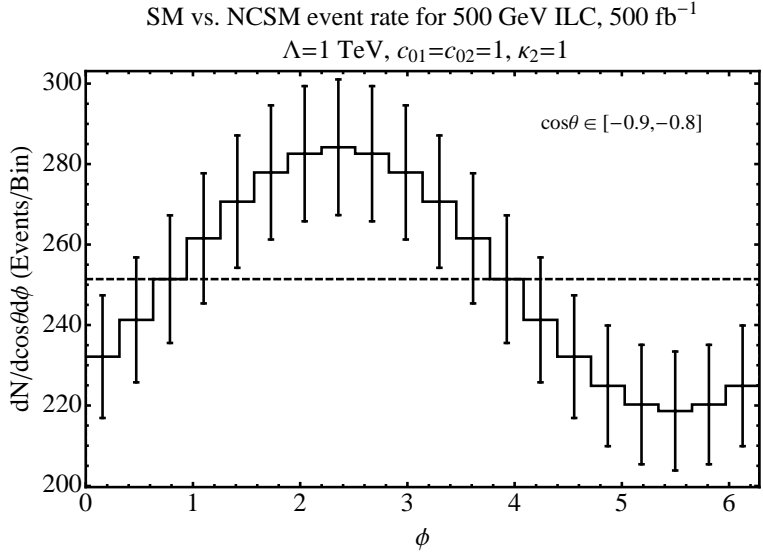}
    \end{center}
    \caption{A comparison of $500~\gev$ ILC expectations for the SM (dotted) and the NCSM (solid).  Shown is the the $-0.9<\cos\theta<-0.8$ bin of 
      the double differential cross section $dN/d\cos\theta d\phi$ for the process \epem\to\wpwm.  The error bars are statistical plus certain systematics (see text).}
    \label{fig:ILCcomp}
  \end{figure} 

  In Fig.~\ref{fig:ILCcompALR}, the $A_{LR}$ signal is presented for a point in parameter space ($\Lambda=3~\tev$, $\kappa_2=2.4$) where the unpolarized differential
  cross section has negligible deviation from the SM.  The differential $A_{LR}$ shows, however, a clear sinusoidal deviation characteristic of the NCSM.
  \begin{figure}[htb!]
    \begin{center}
      \includegraphics[width=10cm]{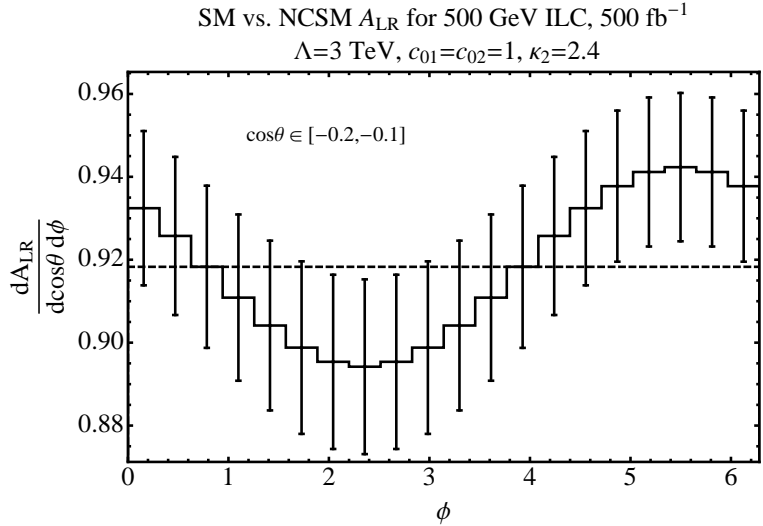}
    \end{center}
    \caption{A comparison of $500~\gev$ ILC $A_{LR}$ data for the SM (dotted) and the NCSM (solid) for \epem\to\wpwm.  
      Shown is the the $-0.2<\cos\theta<-0.1$ bin of 
      the double differential left-right asymmetry $dA_{LR}/d\cos\theta d\phi$.  The error bars are statistical plus certain systematics (see text).}
    \label{fig:ILCcompALR}
  \end{figure} 

  We also consider the process \epem\to\wpwm at a $\sqrt{s}=1~\tev$ ILC with an integrated luminosity of $500~\ifb$.  
  As can be seen in Figs.~\ref{fig:ILCTex},~\ref{fig:ILCTcomp}, and~\ref{fig:ILCTcompALR},
  the phenomenology is qualitatively similar to the $\sqrt{s}=500~\gev$ case but the search reach is significantly higher.  In both cases, the ILC, especially
  when utilizing polarized beams, is sensitive to values of $\Lambda$ many times greater than $\sqrt{s}$.
  \begin{figure}[htb!]
    \begin{center}
      \includegraphics[width=10cm]{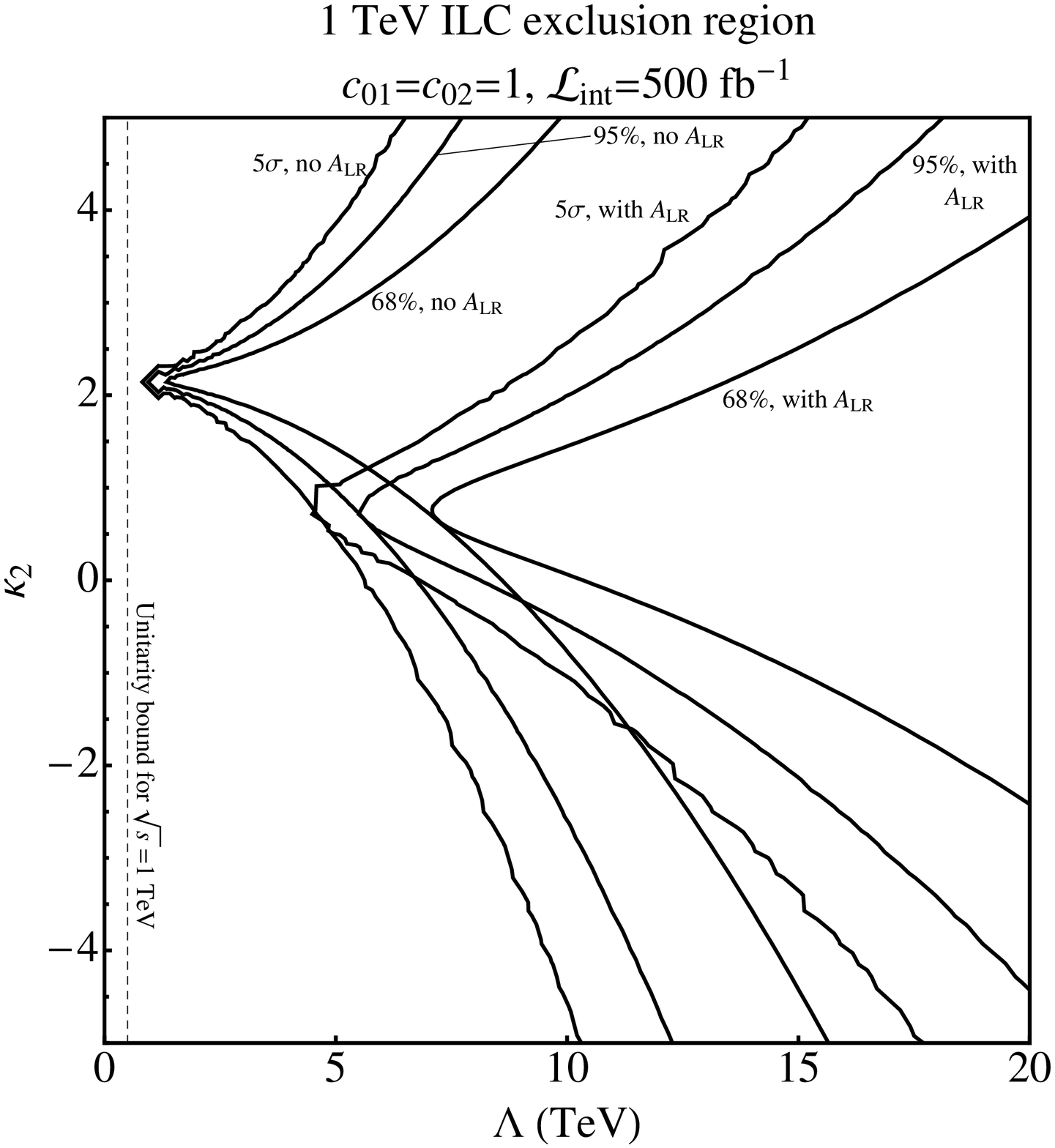}
    \end{center}
    \caption{Same as Fig.~\ref{fig:ILCex}, except for $\sqrt{s}=1~\tev$.}
    \label{fig:ILCTex}
  \end{figure} 
  \begin{figure}[htb!]
    \begin{center}
      \includegraphics[width=10cm]{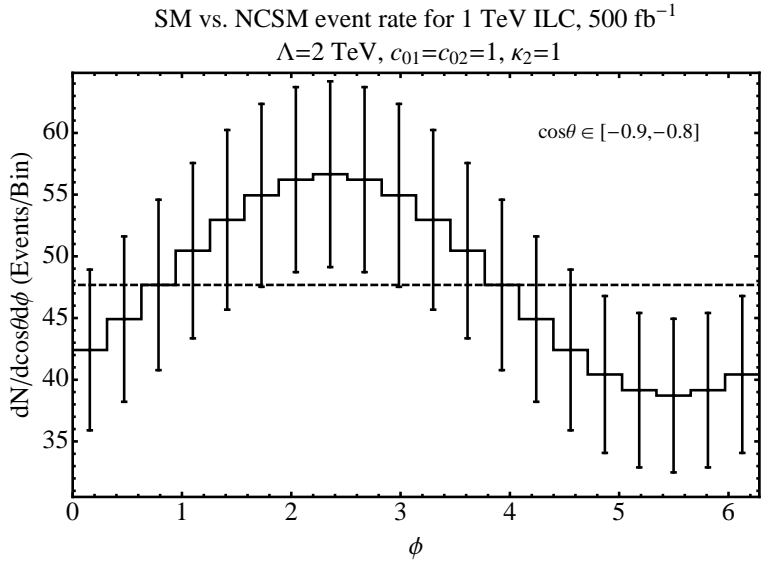}
    \end{center}
    \caption{Same as Fig.~\ref{fig:ILCcomp}, except for $\sqrt{s}=1~\tev$, $\Lambda=2~\tev$, and $-0.9<\cos\theta<-0.8$.}
    \label{fig:ILCTcomp}
  \end{figure} 
  \begin{figure}[htb!]
    \begin{center}
      \includegraphics[width=10cm]{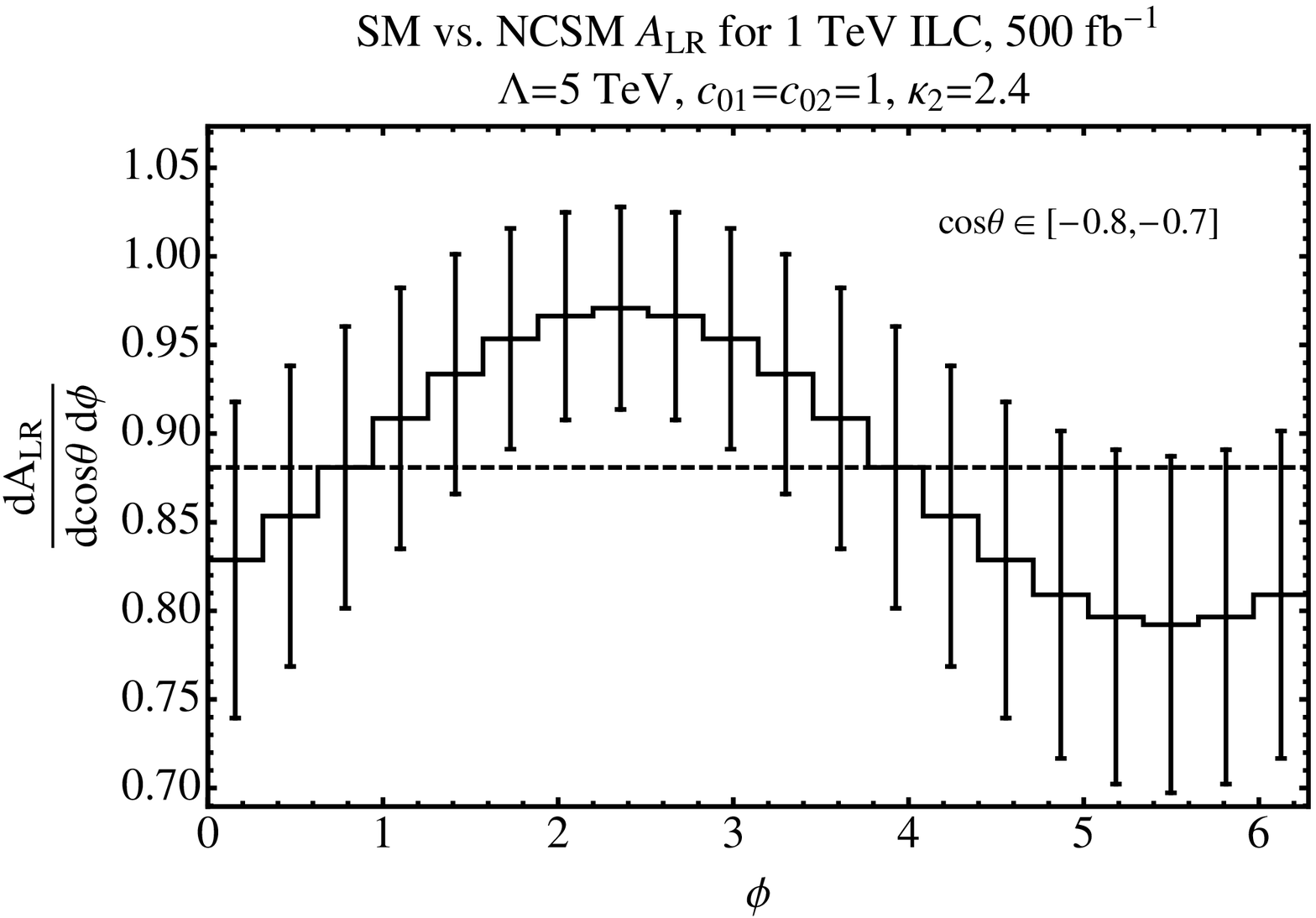}
    \end{center}
    \caption{Same as Fig.~\ref{fig:ILCcompALR}, except for $\sqrt{s}=1~\tev$, $\Lambda=5~\tev$, and $-0.7<\cos\theta<-0.6$.}
    \label{fig:ILCTcompALR}
  \end{figure} 

  \subsection{$W$ polarization at the ILC}

  All the observables discussed above have the same dependence on $\theta_{\mu\nu}$ and $\phi$, that is, 
  $\propto\theta_{01}\sin\phi-\theta_{02}\cos\phi$.  This is true of any observable that does not distinguish
  between {\it final} state polarizations.  At the ILC, however, $W$ polarization can be determined with high
  efficiency \cite{Yehudai:1991gc}.  By comparing the production of different combinations of $W$ polarizations, we can
  disentangle the contributions from different elements of $\theta_{\mu\nu}$ and from $\kappa_2$.  In this way, we can
  measure these different parameters at the ILC.

  In this section, we will compare differential event rates for $W^+W^-$ production at a $\sqrt{s}=1~\tev$~ILC with a
  500~\ifb~integrated luminosity for different $W^+$ and $W^-$ helicities.  To compare different NCSM contributions, 
  each NC distribution we present corresponds to one component of $c_{\mu\nu}=1$ with all others are set to zero.  The 
  analytical expressions for the amplitude squared for the various helicity choices are given in the appendices.

  \subsubsection{$dN/d\phi$}
  First we examine the differential event rate $dN/d\phi$, integrating over the full range in $\cos\theta$. 
  Figure~\ref{fig:pltR00L} compares $W^+W^-$ production at a 1~\tev~ILC for the SM and the NCSM for both \{R,0\}
  and \{0,L\}
  helicity choices (for \{$W^+$,$W^-$\}), and for both $\kappa_2=0$ and $\kappa_2=1$.  
  Here, 0 refers to longitudinal polarization, and R and L to right and left circular or transverse polarizations, 
    sometimes denoted by + and -.
  As is clear from the plot, with $\Lambda=3~\tev$, 
  the $c_{13}=1$ and $c_{23}=1$ distributions are easily distinguishable from each other and from the SM.  The $\kappa_2=1$ NCSM distributions
  are also clearly distinguishable from the $\kappa_2=0$ distributions.
  \begin{figure}[htb!]
    \begin{center}
      \includegraphics[width=15cm]{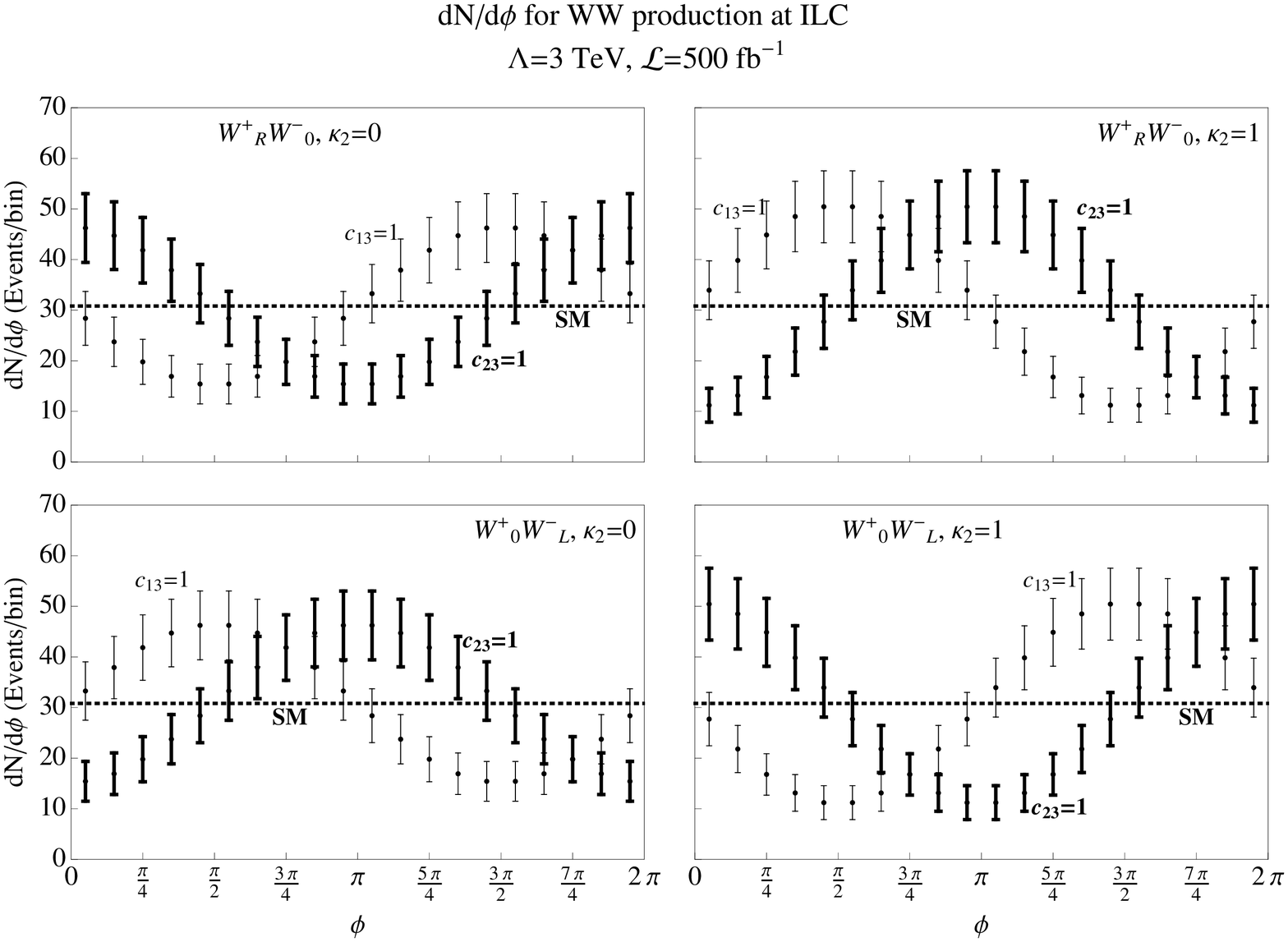}
    \end{center}
    \caption{The differential event rate $dN/d\phi$, integrated over $\cos\theta$ as a function of $\phi$,
      for the SM and various mNCSM ($\kappa_2=0$) contributions with $\Lambda=3~\tev$ for $W^+W^-$ production 
      at a 1~\tev~ILC with 500~\ifb~integrated luminosity.
      Each set of NCSM data points is for $c_{\mu\nu}=1$
      for one choice of $\mu\nu$, and all others being zero.  The left column of plots is for $\kappa_2=0$, while the right column is for 
      $\kappa_2=1$.  The top row of plots is for $W^+_RW^-_0$ production, while the bottom row is for $W^+_0W^-_L$.  The error bars, as before,
      include statistical error and luminosity uncertainty.}
    \label{fig:pltR00L}
  \end{figure} 

  As can also be seen, the NCSM
  contribution from $c_{13}=1$ is opposite in phase for the \{R,0\} and \{0,L\} helicity choices.  When summing over these two helicity choices,
  the $c_{13}=1$ contribution vanishes.  The same is true for the $c_{23}=1$ contribution.  In order to determine the values of $c_{13}$ and
  $c_{23},$ therefore, it is essential to measure the $W$ polarizations.
  
  Different information can be extracted if we look at the $dN/d\phi$ distribution in a particular $\cos\theta$ bin, instead of integrating over all values
  of $\cos\theta$ as we did in Fig.~\ref{fig:pltR00L} above.  
  Figure~\ref{fig:pltR0c} shows the $W^+_RW^-_0$ distribution for
  only the $-1<\cos\theta<-0.9$ bin.  Here, we see that 
  the $c_{12}$ contribution is distinguishable from the SM and other mNCSM ($\kappa_2=0$) contributions (which are negligibly
  different from the SM in this case).  This distribution is also sensitive to the value of $\kappa_2$; the $\kappa_2=0$, distribution is significantly 
  different from the $\kappa_2=1$ distribution.  Interestingly,
  this contribution is flat in $\phi$ rather than having the sinusoidal shape that has characterized the NCSM distributions we have looked at so far.
  \begin{figure}[htb!]
    \begin{center}
      \includegraphics[width=10cm]{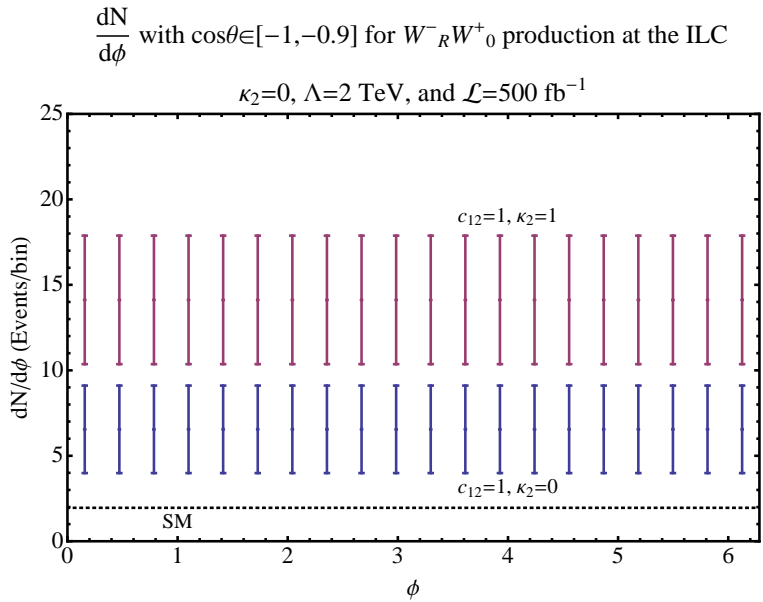}
    \end{center}
    \caption{$dN/d\phi$ for $W^+_RW^-_0$ for the bin $-1<\cos\theta<-0.9$.}
    \label{fig:pltR0c}
  \end{figure} 

  \subsubsection{$dN/d\cos\theta$}
  Instead of integrating over $\cos\theta$ and looking at $dN/d\phi$, we can also integrate over $\phi$ and look at $dN/d\cos\theta$ for different
  $W$ helicity choices as a way of determining the parameters of the NCSM.  In general, integrating over the full range of $\phi$ will remove any sensitivity
  to the NCSM contributions, since they typically vary sinusoidally with $\phi$.  Here we integrate from $\phi=0$ to $\phi=\pi$ and see that the $\cos\theta$
  distributions in this half of the detector are sensitive to the NCSM parameters.

  For example, Fig.~\ref{fig:pltLLcsk1t} shows $dN/d\cos\theta$ for $W^+_LW^-_L$ production for the half-cylinder defined above.  In this case,
  the NCSM distribution with $c_{13}$ and $\kappa_2=1$ is clearly distinguishable from those of the SM and the other 11 NCSM parameter choices,
  all of which have negligible event rate for this helicity choice.    As shown in Fig.~\ref{fig:plt0Rcst}, on the other hand, for $W^+_0W^-_R$ production 
  the $c_{12}=1$ and $c_{13}=1$ with $\kappa_2=0$ distributions show deviations from each other, as well as the SM.  
  
  \begin{figure}[htb!]
    \begin{center}
      \includegraphics[width=10cm]{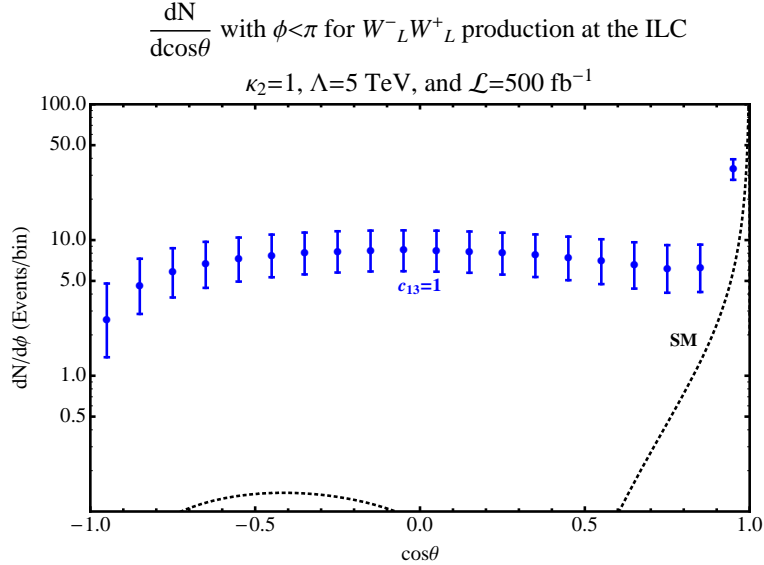}
    \end{center}
    \caption{The differential event rate $dN/d\cos\theta$ integrated over the half-cylinder from $\phi=0$ to $\phi=\pi$
      for the SM and the NCSM ($\kappa_2=1$) contributions for $W^+_LW^-_L$ production at a 1~\tev~ILC with 500~\ifb~integrated luminosity.  
      The data points 
      represent the NCSM event rate for $c_{13}=1$ and all other $c_{\mu\nu}$ set to zero, and the dashed curve is the SM event rate.
      The error bars show the one-sigma statistical errors on the NCSM event rate.}
    \label{fig:pltLLcsk1t}
  \end{figure} 
  \begin{figure}[htb!]
    \begin{center}
      \includegraphics[width=10cm]{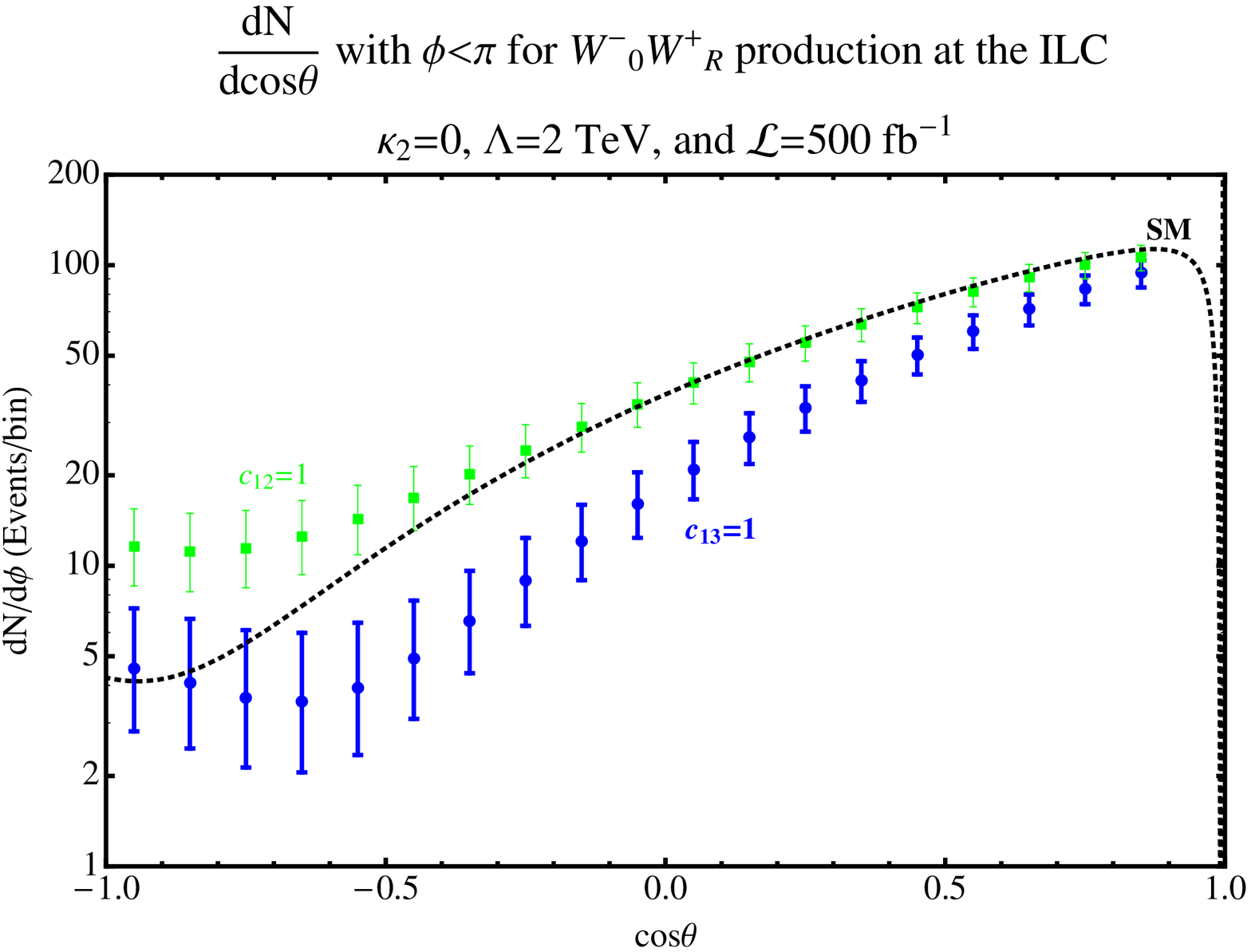}
    \end{center}
    \caption{Like Fig.~\ref{fig:pltLLcsk1t}, except for $W^+_0W^-_R$ and $\kappa_2=0$.  The data points with thick (thin) error bars are for  $c_{13}=1$
      ($c_{12}=1$), with all other $c_{\mu\nu}$ set to zero.  The dashed curve is the SM event rate.}
    \label{fig:plt0Rcst}
  \end{figure} 

  \section{Conclusions}

  The Noncommutative Standard Model is a model of physics with a rich history.  Motivation for its study comes both from theoretical 
  considerations from string theory, and generic properties of quantum gravity, as well as from the NCSM's unique and rich phenomenology.

  In this paper we have contributed to the large body of work on the phenomenology of the NCSM by investigating $WW$ production in this model.
  We have found that in both $\wpwm\to\wpwm$ and $\epem\to\wpwm$ in the NCSM, unlike in the SM, 
  the cancellation of terms in the amplitude that grow with $s$ and $s^2$, respectively, does not occur.  At some value of $s$, these terms violate partial-wave
  unitarity, and we have estimated that value.  

  For an interesting range in NCSM parameter space (for high enough $\Lambda$ and low enough $s$), however, the model satisfies partial
  wave unitarity.  We have seen
  that for much of this range, the model has observable signatures at LEP, the LHC, and the ILC.  We have determined the region of parameter
  space excluded by observations of the $WW$ production differential cross section at LEP, and found that it extends to much higher values
  of the NC scale $\Lambda$ then are ruled out by partial-wave unitarity at LEP energies.  For the LHC, we have found that the search reach obtained
  from deviations from the $WW$ production differential cross section and invariant mass distribution does not extend beyond the region 
  that is ruled out by unitarity.  This shows that the LHC is not an optimal tool (at least using this observable) to probe this model.

  For the ILC, the search reach, using $WW$ production, extends to $\Lambda\sim$~tens of~\tev, well beyond the unitarity bound on $\Lambda$ 
  at ILC energies.  We have shown that the polarization capabilities of the ILC are essential in probing regions of parameter space 
  (via the left-right asymmetry) that are unobservable in the unpolarized cross section because of cancellations in the amplitude.
  Furthermore, measurement of the final state $W$ polarizations provides sensitivity to certain parameters of the model, namely
  $c_{12}$, $c_{13}$, $c_{23}$, and $c_{03}$, whose contributions to the cross section cancel in the sum over final state helicities.  Assuming
  efficient determination of the $W$ helicities is possible, we have given examples where these NCSM parameters at may be measured at
  the ILC for values of $\Lambda$ up to several~\tev.

  There is much more work that can be done in this direction.  One could go beyond calculating differential cross sections and instead 
  incorporate the amplitudes into an event generator.  This would allow the effects of parton showering, detector simulation, and Standard
  Model backgrounds to be taken into account.  This has been undertaken for other NCSM processes at the LHC by Alboteanu et al. \cite{Alboteanu:2006hh}

  \appendix

  \section{Amplitude squared for~\epem\to\wpwm~with $\kappa_2=0$}

  In this and the next appendix we present the helicity amplitudes for~\epem\to\wpwm~production in the NCSM with $\kappa_2$=0.  
  The next section we will present the $\kappa_2$ coefficients in the contribution to the helicity amplitudes from the NCSM with $\kappa_2\neq0$. 
  The amplitude for $\epem\to\wpwm$ in the
  NCSM has contributions from the Feynman diagrams given in Eqs.~\ref{bosons},~\ref{neutrinos}, and~\ref{contact}, so that (for down-type fermions)
  \begin{equation}
    M_{\epem\to\wpwm}=M_{\gamma_s}^{SM}+M_{Z_s}^{SM}+M_{f_t}^{SM}+M_{\gamma_s}^{NC}+M_{Z_s}^{NC}+M_{f_t}^{NC}+M_{c}^{NC},
  \end{equation}
  where $M_{\gamma_s}$, $M_{Z_s}$, $M_{f_t}$, and $M_{c}$ refer to the s-channel photon exchange, s-channel Z boson exchange, t-channel neutrino exchange (for
  down-type fermions this would be u-channel neutrino exchange) and contact interaction diagrams, respectively.  $M^{SM}$ refers to the standard model
  contribution of a diagram, whereas $M^{NC}$ refers to the $\O(\theta)$ NCSM contribution.
  
  To calculate the cross section, we must square the amplitude.  Doing this yields, to $\O(\theta)$,
  \begin{dmath}
    |M_{\epem\to\wpwm}|^2=|M_{\gamma_s}^{SM}+M_{Z_s}^{SM}+M_{f_t}^{SM}|^2+
                        \hat{M}_{\gamma_s\gamma_s}+\hat{M}_{\gamma_sZ_s}+\hat{M}_{\gamma_sf_t}+\hat{M}_{\gamma_sc}+
                        \hat{M}_{Z_sZ_s}+\hat{M}_{Z_sf_t}+\hat{M}_{Z_sc}+
                        \hat{M}_{f_tf_t}+\hat{M}_{f_tc}\;,
  \end{dmath}
  where for a pair of diagrams $a$, $b$, $\hat{M}$ is defined as
  \begin{equation}
    \hat{M}_{a\;b}\equiv 2\Re(M_a^{SM}M_b^{NC}+M_b^{SM}M_a^{NC})\;.
  \end{equation}
  In this appendix we give the values of $\hat{M}$ for every diagram pair and every helicity choice, for 
  $\kappa_2=0$.  The helicities are specified in the order $\hat{M}(h_{e^-}h_{e^+}h_{W^+}h_{W^-})$.

  It is convenient to group the expressions by diagram pair and by initial state helicities.  So, for example, Section A.1 gives the values of
  $\hat{M}_{\gamma_s\gamma_s}=2\Re(M_{\gamma_s}^{SM}M_{\gamma_s}^{NC}+M_{\gamma_s}^{SM}M_{\gamma_s}^{NC})$ with $\kappa_2=0$ for a left-handed electron
  and a right-handed positron--the values are then listed by the helicities of the final-state $W$ bosons.  The prefactor that all the expressions
  share is given first, while the remaining factor for each $W$ helicity choice follows.  So the first line in Section A.1,
  \begin{equation}
    00=\quad  s_\theta ( -c_\phi \theta_{02} + s_\phi \theta_{01} ) (3+\beta ^4-4 \beta ^2 ) s,
  \end{equation}
  gives the value $\hat{M}_{\gamma_s\gamma_s}(LR00)$ if multiplied by the prefactor given at the beginning of Section A.1.  
  If any polarization choice is not shown, it is because
  the corresponding contribution is zero.

  { 
    \setlength{\abovedisplayskip}{0pt} 
    \setlength{\belowdisplayskip}{0pt} 
    \setlength{\abovedisplayshortskip}{0pt} 
    \setlength{\belowdisplayshortskip}{0pt} 
    \subsection{Values of $\hat{M}_{\gamma_s\gamma_s}(LRh_{W^+}h_{W^-})$, with $\kappa_2=0$}
{\bf Prefactor}: $ g^4 Q_f^2 s_W^4 \beta  s /\left( 8 m_W^2 \right)$
\begin{dgroup*}

\begin{dmath*}
00=\quad  s_\theta ( -c_\phi \theta_{02} + s_\phi \theta_{01} ) (3+\beta ^4-4 \beta ^2 ) s 
\end{dmath*}

\begin{dmath*}
0L=\quad  4 m_W^2 ( (-2+s_\theta^2+ 2 c_\theta ) \theta_{12} \beta  + (1-c_\theta) s_\theta s_\phi (\theta_{01}-\theta_{13} \beta  ) + (-1+c_\theta) s_\theta c_\phi (\theta_{02}-\theta_{23} \beta  ) ) 
\end{dmath*}

\begin{dmath*}
0R=\quad  4 m_W^2 ( (-2+s_\theta^2-2 c_\theta ) \theta_{12} \beta  + (-1-c_\theta) s_\theta c_\phi (\theta_{02}+ \theta_{23} \beta  ) + (1+c_\theta) s_\theta s_\phi (\theta_{01}+ \theta_{13} \beta  ) ) 
\end{dmath*}

\begin{dmath*}
L0=\quad  4 m_W^2 ( (2+ 2 c_\theta -s_\theta^2 ) \theta_{12} \beta  + (1+c_\theta) s_\theta c_\phi ( -\theta_{02} + \theta_{23} \beta  ) + (1+c_\theta) s_\theta s_\phi (\theta_{01}-\theta_{13} \beta  ) ) 
\end{dmath*}

\begin{dmath*}
LL=\quad  -4 m_W^2 ( (-2+s_\theta^2+ 2 c_\theta ) \theta_{12} \beta  + (1-c_\theta) s_\theta c_\phi (\theta_{02}+ \theta_{23} \beta  ) + (-1+c_\theta) s_\theta s_\phi (\theta_{01}+ \theta_{13} \beta  ) ) 
\end{dmath*}
\end{dgroup*}

    \subsection{Values of $\hat{M}_{\gamma_s\gamma_s}(RLh_{W^+}h_{W^-})$, with $\kappa_2=0$}
{\bf Prefactor}: $ g^4 Q_f^2 s_W^4 \beta  s /\left( 8 m_W^2 \right)$
\begin{dgroup*}

\begin{dmath*}
00=\quad  -s_\theta ( -c_\phi \theta_{02} + s_\phi \theta_{01} ) (3+\beta ^4-4 \beta ^2 ) s 
\end{dmath*}

\begin{dmath*}
0L=\quad  4 m_W^2 ( (2+ 2 c_\theta -s_\theta^2 ) \theta_{12} \beta  + (-1-c_\theta) s_\theta s_\phi (\theta_{01}+ \theta_{13} \beta  ) + (1+c_\theta) s_\theta c_\phi (\theta_{02}+ \theta_{23} \beta  ) ) 
\end{dmath*}

\begin{dmath*}
0R=\quad  -4 m_W^2 ( (-2+s_\theta^2+ 2 c_\theta ) \theta_{12} \beta  + (1-c_\theta) s_\theta s_\phi (\theta_{01}-\theta_{13} \beta  ) + (-1+c_\theta) s_\theta c_\phi (\theta_{02}-\theta_{23} \beta  ) ) 
\end{dmath*}

\begin{dmath*}
L0=\quad  4 m_W^2 ( (-2+s_\theta^2+ 2 c_\theta ) \theta_{12} \beta  + (1-c_\theta) s_\theta c_\phi (\theta_{02}+ \theta_{23} \beta  ) + (-1+c_\theta) s_\theta s_\phi (\theta_{01}+ \theta_{13} \beta  ) ) 
\end{dmath*}

\begin{dmath*}
LL=\quad  4 m_W^2 ( (-2+s_\theta^2-2 c_\theta ) \theta_{12} \beta  + (1+c_\theta) s_\theta c_\phi (\theta_{02}-\theta_{23} \beta  ) + (1+c_\theta) s_\theta s_\phi ( -\theta_{01} + \theta_{13} \beta  ) ) 
\end{dmath*}
\end{dgroup*}

    \subsection{Values of $\hat{M}_{Z_s\gamma_s}(LRh_{W^+}h_{W^-})$, with $\kappa_2=0$}
{\bf Prefactor}: $ (-g_A^f-g_V^f) g^4 Q_f s_W^2 \beta  s /\left( 32 c_W^2 m_W^2 (s-m_Z^2) \right)$
\begin{dgroup*}

\begin{dmath*}
00=\quad  (-1+2 s_W^2) s_\theta ( -c_\phi \theta_{02} + s_\phi \theta_{01} ) (3+\beta ^4-4 \beta ^2 ) s^2 
\end{dmath*}

\begin{dmath*}
0L=\quad  -2 m_W^2 ( (1-c_\theta) s_\theta c_\phi ( \theta_{02} (-3+4 s_W^2+\beta ^2) + 2 \left(1-2 s_W^2\right) \theta_{23} \beta  ) + (-1+c_\theta) s_\theta s_\phi ( \theta_{01} (-3+4 s_W^2+\beta ^2) + 2 \left(1-2 s_W^2\right) \theta_{13} \beta  ) -2 \left(2 s_W^2-1\right) (-2+s_\theta^2+ 2 c_\theta ) \theta_{12} \beta  ) s 
\end{dmath*}

\begin{dmath*}
0R=\quad  2 m_W^2 ( (-1-c_\theta) s_\theta c_\phi ( -3 \theta_{02} + 4 s_W^2 (\theta_{02}+ \theta_{23} \beta  ) + \theta_{02} \beta ^2 -2 \theta_{23} \beta  ) + (1+c_\theta) s_\theta s_\phi ( -3 \theta_{01} + 4 s_W^2 (\theta_{01}+ \theta_{13} \beta  ) + \theta_{01} \beta ^2 -2 \theta_{13} \beta  ) -2 \left(2 s_W^2-1\right) (2+ 2 c_\theta -s_\theta^2 ) \theta_{12} \beta  ) s 
\end{dmath*}

\begin{dmath*}
L0=\quad  2 m_W^2 ( (-1-c_\theta) s_\theta c_\phi ( \theta_{02} (-3+4 s_W^2+\beta ^2) + 2 \left(1-2 s_W^2\right) \theta_{23} \beta  ) + (1+c_\theta) s_\theta s_\phi ( \theta_{01} (-3+4 s_W^2+\beta ^2) + 2 \left(1-2 s_W^2\right) \theta_{13} \beta  ) + 2 \left(2 s_W^2-1\right) (2+ 2 c_\theta -s_\theta^2 ) \theta_{12} \beta  ) s 
\end{dmath*}

\begin{dmath*}
LL=\quad  8 s_\theta m_W^4 ( c_\phi \theta_{02} -s_\phi \theta_{01} -c_\phi \theta_{23} \beta  + s_\phi \theta_{13} \beta  ) 
\end{dmath*}

\begin{dmath*}
LR=\quad  -2 m_W^2 ( (1-c_\theta) s_\theta c_\phi ( -3 \theta_{02} + 4 s_W^2 (\theta_{02}+ \theta_{23} \beta  ) + \theta_{02} \beta ^2 -2 \theta_{23} \beta  ) + (-1+c_\theta) s_\theta s_\phi ( -3 \theta_{01} + 4 s_W^2 (\theta_{01}+ \theta_{13} \beta  ) + \theta_{01} \beta ^2 -2 \theta_{13} \beta  ) + 2 \left(2 s_W^2-1\right) (-2+s_\theta^2+ 2 c_\theta ) \theta_{12} \beta  ) s 
\end{dmath*}

\begin{dmath*}
R0=\quad  8 s_\theta m_W^4 ( c_\phi (\theta_{02}+ \theta_{23} \beta  ) -s_\phi (\theta_{01}+ \theta_{13} \beta  ) ) 
\end{dmath*}
\end{dgroup*}

    \subsection{Values of $\hat{M}_{Z_s\gamma_s}(RLh_{W^+}h_{W^-})$, with $\kappa_2=0$}
{\bf Prefactor}: $ (g_V^f-g_A^f) g^4 Q_f s_W^2 \beta  s /\left( 32 c_W^2 m_W^2 (s-m_Z^2) \right)$
\begin{dgroup*}

\begin{dmath*}
00=\quad  (-1+2 s_W^2) s_\theta ( -c_\phi \theta_{02} + s_\phi \theta_{01} ) (3+\beta ^4-4 \beta ^2 ) s^2 
\end{dmath*}

\begin{dmath*}
0L=\quad  2 m_W^2 ( (-1-c_\theta) s_\theta c_\phi ( -3 \theta_{02} + 4 s_W^2 (\theta_{02}+ \theta_{23} \beta  ) + \theta_{02} \beta ^2 -2 \theta_{23} \beta  ) + (1+c_\theta) s_\theta s_\phi ( -3 \theta_{01} + 4 s_W^2 (\theta_{01}+ \theta_{13} \beta  ) + \theta_{01} \beta ^2 -2 \theta_{13} \beta  ) -2 \left(2 s_W^2-1\right) (2+ 2 c_\theta -s_\theta^2 ) \theta_{12} \beta  ) s 
\end{dmath*}

\begin{dmath*}
0R=\quad  -2 m_W^2 ( (1-c_\theta) s_\theta c_\phi ( \theta_{02} (-3+4 s_W^2+\beta ^2) + 2 \left(1-2 s_W^2\right) \theta_{23} \beta  ) + (-1+c_\theta) s_\theta s_\phi ( \theta_{01} (-3+4 s_W^2+\beta ^2) + 2 \left(1-2 s_W^2\right) \theta_{13} \beta  ) -2 \left(2 s_W^2-1\right) (-2+s_\theta^2+ 2 c_\theta ) \theta_{12} \beta  ) s 
\end{dmath*}

\begin{dmath*}
L0=\quad  -2 m_W^2 ( (1-c_\theta) s_\theta c_\phi ( -3 \theta_{02} + 4 s_W^2 (\theta_{02}+ \theta_{23} \beta  ) + \theta_{02} \beta ^2 -2 \theta_{23} \beta  ) + (-1+c_\theta) s_\theta s_\phi ( -3 \theta_{01} + 4 s_W^2 (\theta_{01}+ \theta_{13} \beta  ) + \theta_{01} \beta ^2 -2 \theta_{13} \beta  ) + 2 \left(2 s_W^2-1\right) (-2+s_\theta^2+ 2 c_\theta ) \theta_{12} \beta  ) s 
\end{dmath*}

\begin{dmath*}
LL=\quad  8 s_\theta m_W^4 ( c_\phi (\theta_{02}+ \theta_{23} \beta  ) -s_\phi (\theta_{01}+ \theta_{13} \beta  ) ) 
\end{dmath*}

\begin{dmath*}
LR=\quad  2 m_W^2 ( (-1-c_\theta) s_\theta c_\phi ( \theta_{02} (-3+4 s_W^2+\beta ^2) + 2 \left(1-2 s_W^2\right) \theta_{23} \beta  ) + (1+c_\theta) s_\theta s_\phi ( \theta_{01} (-3+4 s_W^2+\beta ^2) + 2 \left(1-2 s_W^2\right) \theta_{13} \beta  ) + 2 \left(2 s_W^2-1\right) (2+ 2 c_\theta -s_\theta^2 ) \theta_{12} \beta  ) s 
\end{dmath*}

\begin{dmath*}
R0=\quad  8 s_\theta m_W^4 ( c_\phi \theta_{02} -s_\phi \theta_{01} -c_\phi \theta_{23} \beta  + s_\phi \theta_{13} \beta  ) 
\end{dmath*}
\end{dgroup*}

    \subsection{Values of $\hat{M}_{Z_sZ_s}(LRh_{W^+}h_{W^-})$, with $\kappa_2=0$}
{\bf Prefactor}: $ -(g_A^f+g_V^f)^2 g^4 \beta  s^2 /\left( 64 c_W^2 m_W^2 (s-m_Z^2)^2 \right)$
\begin{dgroup*}

\begin{dmath*}
00=\quad  s_W^2 s_\theta ( -c_\phi \theta_{02} + s_\phi \theta_{01} ) (3+\beta ^4-4 \beta ^2 ) s^2 
\end{dmath*}

\begin{dmath*}
0L=\quad  2 m_W^2 ( (-1+c_\theta) s_\theta c_\phi ( \theta_{02} (-1+2 s_W^2+\beta ^2) -2 s_W^2 \theta_{23} \beta  ) + (-1+c_\theta) s_\theta s_\phi (\theta_{01}-2 s_W^2 (\theta_{01}-\theta_{13} \beta  ) -\theta_{01} \beta ^2 ) + 2 s_W^2 (-2+s_\theta^2+ 2 c_\theta ) \theta_{12} \beta  ) s 
\end{dmath*}

\begin{dmath*}
0R=\quad  2 m_W^2 ( (-1-c_\theta) s_\theta c_\phi ( -\theta_{02} + 2 s_W^2 (\theta_{02}+ \theta_{23} \beta  ) + \theta_{02} \beta ^2 ) + (1+c_\theta) s_\theta s_\phi ( -\theta_{01} + 2 s_W^2 (\theta_{01}+ \theta_{13} \beta  ) + \theta_{01} \beta ^2 ) + 2 s_W^2 (-2+s_\theta^2-2 c_\theta ) \theta_{12} \beta  ) s 
\end{dmath*}

\begin{dmath*}
L0=\quad  2 m_W^2 ( (1+c_\theta) s_\theta c_\phi (\theta_{02}-2 s_W^2 (\theta_{02}-\theta_{23} \beta  ) -\theta_{02} \beta ^2 ) + (1+c_\theta) s_\theta s_\phi ( \theta_{01} (-1+2 s_W^2+\beta ^2) -2 s_W^2 \theta_{13} \beta  ) + 2 s_W^2 (2+ 2 c_\theta -s_\theta^2 ) \theta_{12} \beta  ) s 
\end{dmath*}

\begin{dmath*}
LL=\quad  8 s_\theta m_W^4 ( c_\phi \theta_{02} -s_\phi \theta_{01} -c_\phi \theta_{23} \beta  + s_\phi \theta_{13} \beta  ) 
\end{dmath*}

\begin{dmath*}
LR=\quad  -2 m_W^2 ( (1-c_\theta) s_\theta c_\phi ( -\theta_{02} + 2 s_W^2 (\theta_{02}+ \theta_{23} \beta  ) + \theta_{02} \beta ^2 ) + (-1+c_\theta) s_\theta s_\phi ( -\theta_{01} + 2 s_W^2 (\theta_{01}+ \theta_{13} \beta  ) + \theta_{01} \beta ^2 ) + 2 s_W^2 (-2+s_\theta^2+ 2 c_\theta ) \theta_{12} \beta  ) s 
\end{dmath*}

\begin{dmath*}
R0=\quad  8 s_\theta m_W^4 ( c_\phi (\theta_{02}+ \theta_{23} \beta  ) -s_\phi (\theta_{01}+ \theta_{13} \beta  ) ) 
\end{dmath*}
\end{dgroup*}

    \subsection{Values of $\hat{M}_{Z_sZ_s}(RLh_{W^+}h_{W^-})$, with $\kappa_2=0$}
{\bf Prefactor}: $ -(g_A^f-g_V^f)^2 g^4 \beta  s^2 /\left( 64 c_W^2 m_W^2 (s-m_Z^2)^2 \right)$
\begin{dgroup*}

\begin{dmath*}
00=\quad  -s_W^2 s_\theta ( -c_\phi \theta_{02} + s_\phi \theta_{01} ) (3+\beta ^4-4 \beta ^2 ) s^2 
\end{dmath*}

\begin{dmath*}
0L=\quad  -2 m_W^2 ( (-1-c_\theta) s_\theta c_\phi ( -\theta_{02} + 2 s_W^2 (\theta_{02}+ \theta_{23} \beta  ) + \theta_{02} \beta ^2 ) + (1+c_\theta) s_\theta s_\phi ( -\theta_{01} + 2 s_W^2 (\theta_{01}+ \theta_{13} \beta  ) + \theta_{01} \beta ^2 ) + 2 s_W^2 (-2+s_\theta^2-2 c_\theta ) \theta_{12} \beta  ) s 
\end{dmath*}

\begin{dmath*}
0R=\quad  -2 m_W^2 ( (-1+c_\theta) s_\theta c_\phi ( \theta_{02} (-1+2 s_W^2+\beta ^2) -2 s_W^2 \theta_{23} \beta  ) + (-1+c_\theta) s_\theta s_\phi (\theta_{01}-2 s_W^2 (\theta_{01}-\theta_{13} \beta  ) -\theta_{01} \beta ^2 ) + 2 s_W^2 (-2+s_\theta^2+ 2 c_\theta ) \theta_{12} \beta  ) s 
\end{dmath*}

\begin{dmath*}
L0=\quad  2 m_W^2 ( (1-c_\theta) s_\theta c_\phi ( -\theta_{02} + 2 s_W^2 (\theta_{02}+ \theta_{23} \beta  ) + \theta_{02} \beta ^2 ) + (-1+c_\theta) s_\theta s_\phi ( -\theta_{01} + 2 s_W^2 (\theta_{01}+ \theta_{13} \beta  ) + \theta_{01} \beta ^2 ) + 2 s_W^2 (-2+s_\theta^2+ 2 c_\theta ) \theta_{12} \beta  ) s 
\end{dmath*}

\begin{dmath*}
LL=\quad  8 s_\theta m_W^4 ( -c_\phi (\theta_{02}+ \theta_{23} \beta  ) + s_\phi (\theta_{01}+ \theta_{13} \beta  ) ) 
\end{dmath*}

\begin{dmath*}
LR=\quad  2 m_W^2 ( (1+c_\theta) s_\theta c_\phi ( \theta_{02} (-1+2 s_W^2+\beta ^2) -2 s_W^2 \theta_{23} \beta  ) + (1+c_\theta) s_\theta s_\phi (\theta_{01}-2 s_W^2 (\theta_{01}-\theta_{13} \beta  ) -\theta_{01} \beta ^2 ) + 2 s_W^2 (-2+s_\theta^2-2 c_\theta ) \theta_{12} \beta  ) s 
\end{dmath*}

\begin{dmath*}
R0=\quad  8 s_\theta m_W^4 ( -c_\phi \theta_{02} + s_\phi \theta_{01} + c_\phi \theta_{23} \beta  -s_\phi \theta_{13} \beta  ) 
\end{dmath*}
\end{dgroup*}

    \subsection{Values of $\hat{M}_{f_t\gamma_s}(LRh_{W^+}h_{W^-})$, with $\kappa_2=0$}
{\bf Prefactor}: $ g^4 Q_f s_W^2 s /\left( 64 t m_W^2 \right)$
\begin{dgroup*}

\begin{dmath*}
00=\quad  -s_\theta ( -c_\phi \theta_{02} + s_\phi \theta_{01} ) (-1+\beta ^2) (\beta ^3+ 2 c_\theta -3 \beta  ) s^2 
\end{dmath*}

\begin{dmath*}
0L=\quad  -s ( -2 m_W^2 ( -\theta_{12} \beta  ( -2 (\beta +1)^2 + 2 c_\theta ((\beta +1)^2-s_\theta^2 ) + s_\theta^2 (3+ \beta  (2+\beta ) ) ) + s_\theta c_\phi ( 2 s_\theta^2 -(\beta +1)^2 + c_\theta (\beta +1)^2 ) ( -\theta_{02} + \theta_{23} \beta  ) + s_\theta s_\phi ( 2 s_\theta^2 -(\beta +1)^2 + c_\theta (\beta +1)^2 ) (\theta_{01}-\theta_{13} \beta  ) ) + s_\theta ( -c_\phi (\theta_{02}+\theta_{23}) + s_\phi (\theta_{01}+\theta_{13}) ) \beta  (-1+c_\theta+ (-1+c_\theta) \beta ^2 -2 (c_\theta-1) c_\theta \beta  ) s ) 
\end{dmath*}

\begin{dmath*}
0R=\quad  s ( -2 m_W^2 ( \theta_{12} \beta  ( 2 (\beta -1)^2 + 2 c_\theta ((\beta -1)^2-s_\theta^2 ) -s_\theta^2 (3+ (-2+\beta ) \beta  ) ) -s_\theta s_\phi ((\beta -1)^2-2 s_\theta^2 + c_\theta (\beta -1)^2 ) (\theta_{01}+ \theta_{13} \beta  ) + s_\theta c_\phi ((\beta -1)^2-2 s_\theta^2 + c_\theta (\beta -1)^2 ) (\theta_{02}+ \theta_{23} \beta  ) ) + s_\theta ( -c_\phi (\theta_{02}+\theta_{23}) + s_\phi (\theta_{01}+\theta_{13}) ) \beta  (1+ c_\theta (\beta -1)^2 + \beta  (-2+\beta + 2 s_\theta^2 ) ) s ) 
\end{dmath*}

\begin{dmath*}
L0=\quad  s ( 2 m_W^2 ( \theta_{12} \beta  ( 2 (\beta -1)^2 + 2 c_\theta ((\beta -1)^2-s_\theta^2 ) -s_\theta^2 (3+ (-2+\beta ) \beta  ) ) + s_\theta c_\phi ((\beta -1)^2-2 s_\theta^2 + c_\theta (\beta -1)^2 ) ( -\theta_{02} + \theta_{23} \beta  ) + s_\theta s_\phi ((\beta -1)^2-2 s_\theta^2 + c_\theta (\beta -1)^2 ) (\theta_{01}-\theta_{13} \beta  ) ) + s_\theta ( c_\phi (\theta_{23}-\theta_{02} ) + s_\phi (\theta_{01}-\theta_{13} ) ) \beta  (1+ c_\theta (\beta -1)^2 + \beta  (-2+\beta + 2 s_\theta^2 ) ) s ) 
\end{dmath*}

\begin{dmath*}
LL=\quad  4 s_\theta ( -c_\phi (\theta_{02}+ c_\theta \theta_{23} ) + s_\phi (\theta_{01}+ c_\theta \theta_{13} ) ) m_W^2 \beta  (1+\beta ^2-2 c_\theta \beta  ) s 
\end{dmath*}

\begin{dmath*}
LR=\quad  16 s_\theta (-2+s_\theta^2-2 c_\theta ) ( -c_\phi \theta_{02} + s_\phi \theta_{01} ) m_W^4 
\end{dmath*}

\begin{dmath*}
R0=\quad  s ( 2 m_W^2 ( \theta_{12} \beta  ( -2 (\beta +1)^2 + 2 c_\theta ((\beta +1)^2-s_\theta^2 ) + s_\theta^2 (3+ \beta  (2+\beta ) ) ) -s_\theta c_\phi ( 2 s_\theta^2 -(\beta +1)^2 + c_\theta (\beta +1)^2 ) (\theta_{02}+ \theta_{23} \beta  ) + s_\theta s_\phi ( 2 s_\theta^2 -(\beta +1)^2 + c_\theta (\beta +1)^2 ) (\theta_{01}+ \theta_{13} \beta  ) ) -s_\theta ( c_\phi (\theta_{23}-\theta_{02} ) + s_\phi (\theta_{01}-\theta_{13} ) ) \beta  (-1+c_\theta+ (-1+c_\theta) \beta ^2 -2 (c_\theta-1) c_\theta \beta  ) s ) 
\end{dmath*}

\begin{dmath*}
RL=\quad  -16 s_\theta (-2+s_\theta^2+ 2 c_\theta ) ( -c_\phi \theta_{02} + s_\phi \theta_{01} ) m_W^4 
\end{dmath*}

\begin{dmath*}
RR=\quad  -4 s_\theta ( c_\phi \theta_{02} -s_\phi \theta_{01} -c_\theta c_\phi \theta_{23} + c_\theta s_\phi \theta_{13} ) m_W^2 \beta  (1+\beta ^2-2 c_\theta \beta  ) s 
\end{dmath*}
\end{dgroup*}

    \subsection{Values of $\hat{M}_{f_tZ_s}(LRh_{W^+}h_{W^-})$, with $\kappa_2=0$}
{\bf Prefactor}: $ (-g_A^f-g_V^f) g^4 s^2 /\left( 128 c_W^2 t m_W^2 (s-m_Z^2) \right)$
\begin{dgroup*}

\begin{dmath*}
00=\quad  -s_W^2 s_\theta ( -c_\phi \theta_{02} + s_\phi \theta_{01} ) (-1+\beta ^2) (\beta ^3+ 2 c_\theta -3 \beta  ) s^2 
\end{dmath*}

\begin{dmath*}
0L=\quad  -s ( m_W^2 ( -s_\theta s_\phi ( 2 s_\theta^2 -(\beta +1)^2 + c_\theta (\beta +1)^2 ) ( \theta_{01} (-1+2 s_W^2+\beta ^2) -2 s_W^2 \theta_{13} \beta  ) + s_\theta c_\phi ( 2 s_\theta^2 -(\beta +1)^2 + c_\theta (\beta +1)^2 ) ( \theta_{02} (-1+2 s_W^2+\beta ^2) -2 s_W^2 \theta_{23} \beta  ) + 2 s_W^2 \theta_{12} \beta  ( -2 (\beta +1)^2 + 2 c_\theta ((\beta +1)^2-s_\theta^2 ) + s_\theta^2 (3+ \beta  (2+\beta ) ) ) ) -c_W^2 s_\theta ( -c_\phi (\theta_{02}+\theta_{23}) + s_\phi (\theta_{01}+\theta_{13}) ) \beta  (-1+c_\theta+ (-1+c_\theta) \beta ^2 -2 (c_\theta-1) c_\theta \beta  ) s ) 
\end{dmath*}

\begin{dmath*}
0R=\quad  s ( m_W^2 ( -s_\theta c_\phi ((\beta -1)^2-2 s_\theta^2 + c_\theta (\beta -1)^2 ) ( -\theta_{02} + 2 s_W^2 (\theta_{02}+ \theta_{23} \beta  ) + \theta_{02} \beta ^2 ) + s_\theta s_\phi ((\beta -1)^2-2 s_\theta^2 + c_\theta (\beta -1)^2 ) ( -\theta_{01} + 2 s_W^2 (\theta_{01}+ \theta_{13} \beta  ) + \theta_{01} \beta ^2 ) + 2 s_W^2 \theta_{12} \beta  ( -2 (\beta -1)^2 + 2 c_\theta (s_\theta^2-(\beta -1)^2 ) + s_\theta^2 (3+ (-2+\beta ) \beta  ) ) ) -c_W^2 s_\theta ( -c_\phi (\theta_{02}+\theta_{23}) + s_\phi (\theta_{01}+\theta_{13}) ) \beta  (1+ c_\theta (\beta -1)^2 + \beta  (-2+\beta + 2 s_\theta^2 ) ) s ) 
\end{dmath*}

\begin{dmath*}
L0=\quad  s ( m_W^2 ( -s_\theta c_\phi ((\beta -1)^2-2 s_\theta^2 + c_\theta (\beta -1)^2 ) ( \theta_{02} (-1+2 s_W^2+\beta ^2) -2 s_W^2 \theta_{23} \beta  ) + s_\theta s_\phi ((\beta -1)^2-2 s_\theta^2 + c_\theta (\beta -1)^2 ) ( \theta_{01} (-1+2 s_W^2+\beta ^2) -2 s_W^2 \theta_{13} \beta  ) + 2 s_W^2 \theta_{12} \beta  ( 2 (\beta -1)^2 + 2 c_\theta ((\beta -1)^2-s_\theta^2 ) -s_\theta^2 (3+ (-2+\beta ) \beta  ) ) ) -c_W^2 s_\theta ( c_\phi (\theta_{23}-\theta_{02} ) + s_\phi (\theta_{01}-\theta_{13} ) ) \beta  (1+ c_\theta (\beta -1)^2 + \beta  (-2+\beta + 2 s_\theta^2 ) ) s ) 
\end{dmath*}

\begin{dmath*}
LL=\quad  4 s_\theta m_W^2 ( c_\theta ( m_W^2 ( 2 c_\phi \theta_{02} -2 s_\phi \theta_{01} -2 c_\phi \theta_{23} \beta  + 2 s_\phi \theta_{13} \beta  ) + c_W^2 \beta  ( c_\phi (\theta_{23}-2 \theta_{02} \beta  + \theta_{23} \beta ^2 ) -s_\phi (\theta_{13}-2 \theta_{01} \beta  + \theta_{13} \beta ^2 ) ) s ) + \beta  ( 2 m_W^2 ( -c_\phi \theta_{02} + s_\phi (\theta_{01}-\theta_{13} \beta  ) + c_\phi \theta_{23} \beta  ) + c_W^2 ( c_\phi (\theta_{02}+ \theta_{02} \beta ^2 -2 c_\theta^2 \theta_{23} \beta  ) -s_\phi (\theta_{01}+ \theta_{01} \beta ^2 -2 c_\theta^2 \theta_{13} \beta  ) ) s ) ) 
\end{dmath*}

\begin{dmath*}
LR=\quad  8 \left(2 s_W^2-1\right) s_\theta (-2+s_\theta^2-2 c_\theta ) ( -c_\phi \theta_{02} + s_\phi \theta_{01} ) m_W^4 
\end{dmath*}

\begin{dmath*}
R0=\quad  s ( m_W^2 ( -s_\theta c_\phi ( 2 s_\theta^2 -(\beta +1)^2 + c_\theta (\beta +1)^2 ) ( -\theta_{02} + 2 s_W^2 (\theta_{02}+ \theta_{23} \beta  ) + \theta_{02} \beta ^2 ) + s_\theta s_\phi ( 2 s_\theta^2 -(\beta +1)^2 + c_\theta (\beta +1)^2 ) ( -\theta_{01} + 2 s_W^2 (\theta_{01}+ \theta_{13} \beta  ) + \theta_{01} \beta ^2 ) + 2 s_W^2 \theta_{12} \beta  ( -2 (\beta +1)^2 + 2 c_\theta ((\beta +1)^2-s_\theta^2 ) + s_\theta^2 (3+ \beta  (2+\beta ) ) ) ) + c_W^2 s_\theta ( c_\phi (\theta_{23}-\theta_{02} ) + s_\phi (\theta_{01}-\theta_{13} ) ) \beta  (-1+c_\theta+ (-1+c_\theta) \beta ^2 -2 (c_\theta-1) c_\theta \beta  ) s ) 
\end{dmath*}

\begin{dmath*}
RL=\quad  -8 \left(2 s_W^2-1\right) s_\theta (-2+s_\theta^2+ 2 c_\theta ) ( -c_\phi \theta_{02} + s_\phi \theta_{01} ) m_W^4 
\end{dmath*}

\begin{dmath*}
RR=\quad  -4 s_\theta m_W^2 ( c_\theta ( 2 m_W^2 ( -c_\phi (\theta_{02}+ \theta_{23} \beta  ) + s_\phi (\theta_{01}+ \theta_{13} \beta  ) ) + c_W^2 \beta  ( c_\phi (\theta_{23}+ 2 \theta_{02} \beta  + \theta_{23} \beta ^2 ) -s_\phi (\theta_{13}+ 2 \theta_{01} \beta  + \theta_{13} \beta ^2 ) ) s ) + \beta  ( 2 m_W^2 ( c_\phi (\theta_{02}+ \theta_{23} \beta  ) -s_\phi (\theta_{01}+ \theta_{13} \beta  ) ) + c_W^2 ( -c_\phi (\theta_{02}+ \theta_{02} \beta ^2 + 2 c_\theta^2 \theta_{23} \beta  ) + s_\phi (\theta_{01}+ \theta_{01} \beta ^2 + 2 c_\theta^2 \theta_{13} \beta  ) ) s ) ) 
\end{dmath*}
\end{dgroup*}

    \subsection{Values of $\hat{M}_{f_u\gamma_s}(LRh_{W^+}h_{W^-})$, with $\kappa_2=0$}
{\bf Prefactor}: $ g^4 Q_f s_W^2 s /\left( 64 u m_W^2 \right)$
\begin{dgroup*}

\begin{dmath*}
00=\quad  s_\theta ( -c_\phi \theta_{02} + s_\phi \theta_{01} ) (-1+\beta ^2) (\beta ^3-2 c_\theta -3 \beta  ) s^2 
\end{dmath*}

\begin{dmath*}
0L=\quad  s ( 2 m_W^2 ( \theta_{12} \beta  ( 2 (\beta -1)^2 + 2 c_\theta (s_\theta^2-(\beta -1)^2 ) -s_\theta^2 (3+ (-2+\beta ) \beta  ) ) + s_\theta c_\phi ( 2 s_\theta^2 -(\beta -1)^2 + c_\theta (\beta -1)^2 ) ( -\theta_{02} + \theta_{23} \beta  ) + s_\theta s_\phi ( 2 s_\theta^2 -(\beta -1)^2 + c_\theta (\beta -1)^2 ) (\theta_{01}-\theta_{13} \beta  ) ) + s_\theta ( c_\phi (\theta_{23}-\theta_{02} ) + s_\phi (\theta_{01}-\theta_{13} ) ) \beta  (1-c_\theta (\beta -1)^2 + \beta  (-2+\beta + 2 s_\theta^2 ) ) s ) 
\end{dmath*}

\begin{dmath*}
0R=\quad  -s ( 2 m_W^2 ( \theta_{12} \beta  ( 2 (\beta +1)^2 + 2 c_\theta ((\beta +1)^2-s_\theta^2 ) -s_\theta^2 (3+ \beta  (2+\beta ) ) ) -s_\theta s_\phi ((\beta +1)^2-2 s_\theta^2 + c_\theta (\beta +1)^2 ) (\theta_{01}+ \theta_{13} \beta  ) + s_\theta c_\phi ((\beta +1)^2-2 s_\theta^2 + c_\theta (\beta +1)^2 ) (\theta_{02}+ \theta_{23} \beta  ) ) -s_\theta ( c_\phi (\theta_{23}-\theta_{02} ) + s_\phi (\theta_{01}-\theta_{13} ) ) \beta  (1+ c_\theta (\beta +1)^2 + \beta  (2+\beta -2 s_\theta^2 ) ) s ) 
\end{dmath*}

\begin{dmath*}
L0=\quad  s ( 2 m_W^2 ( \theta_{12} \beta  ( 2 (\beta +1)^2 + 2 c_\theta ((\beta +1)^2-s_\theta^2 ) -s_\theta^2 (3+ \beta  (2+\beta ) ) ) + s_\theta c_\phi ((\beta +1)^2-2 s_\theta^2 + c_\theta (\beta +1)^2 ) ( -\theta_{02} + \theta_{23} \beta  ) + s_\theta s_\phi ((\beta +1)^2-2 s_\theta^2 + c_\theta (\beta +1)^2 ) (\theta_{01}-\theta_{13} \beta  ) ) + s_\theta ( -c_\phi (\theta_{02}+\theta_{23}) + s_\phi (\theta_{01}+\theta_{13}) ) \beta  (1+ c_\theta (\beta +1)^2 + \beta  (2+\beta -2 s_\theta^2 ) ) s ) 
\end{dmath*}

\begin{dmath*}
LL=\quad  4 s_\theta m_W^2 \beta  ( s_\phi ( \theta_{01} (1+\beta ^2+ 2 c_\theta \beta  ) -\theta_{13} (c_\theta+ 2 \beta  + c_\theta \beta ^2 -2 s_\theta^2 \beta  ) ) -c_\phi \theta_{02} (1+\beta ^2+ 2 c_\theta \beta  ) + c_\phi \theta_{23} (c_\theta+ 2 \beta  + c_\theta \beta ^2 -2 s_\theta^2 \beta  ) ) s 
\end{dmath*}

\begin{dmath*}
LR=\quad  16 s_\theta (-2+s_\theta^2-2 c_\theta ) ( -c_\phi \theta_{02} + s_\phi \theta_{01} ) m_W^4 
\end{dmath*}

\begin{dmath*}
R0=\quad  s ( 2 m_W^2 ( \theta_{12} \beta  ( -2 (\beta -1)^2 + 2 c_\theta ((\beta -1)^2-s_\theta^2 ) + s_\theta^2 (3+ (-2+\beta ) \beta  ) ) -s_\theta c_\phi ( 2 s_\theta^2 -(\beta -1)^2 + c_\theta (\beta -1)^2 ) (\theta_{02}+ \theta_{23} \beta  ) + s_\theta s_\phi ( 2 s_\theta^2 -(\beta -1)^2 + c_\theta (\beta -1)^2 ) (\theta_{01}+ \theta_{13} \beta  ) ) + s_\theta ( -c_\phi (\theta_{02}+\theta_{23}) + s_\phi (\theta_{01}+\theta_{13}) ) \beta  (1-c_\theta (\beta -1)^2 + \beta  (-2+\beta + 2 s_\theta^2 ) ) s ) 
\end{dmath*}

\begin{dmath*}
RL=\quad  -16 s_\theta (-2+s_\theta^2+ 2 c_\theta ) ( -c_\phi \theta_{02} + s_\phi \theta_{01} ) m_W^4 
\end{dmath*}

\begin{dmath*}
RR=\quad  4 s_\theta m_W^2 \beta  ( -c_\phi ( \theta_{02} (1+\beta ^2+ 2 c_\theta \beta  ) + \theta_{23} (c_\theta+ 2 \beta  + c_\theta \beta ^2 -2 s_\theta^2 \beta  ) ) + s_\phi ( \theta_{01} (1+\beta ^2+ 2 c_\theta \beta  ) + \theta_{13} (c_\theta+ 2 \beta  + c_\theta \beta ^2 -2 s_\theta^2 \beta  ) ) ) s 
\end{dmath*}
\end{dgroup*}

    \subsection{Values of $\hat{M}_{f_uZ_s}(LRh_{W^+}h_{W^-})$, with $\kappa_2=0$}
{\bf Prefactor}: $ (-g_A^f-g_V^f) g^4 s^2 /\left( 128 c_W^2 u m_W^2 (s-m_Z^2) \right)$
\begin{dgroup*}

\begin{dmath*}
00=\quad  s_W^2 s_\theta ( -c_\phi \theta_{02} + s_\phi \theta_{01} ) (-1+\beta ^2) (\beta ^3-2 c_\theta -3 \beta  ) s^2 
\end{dmath*}

\begin{dmath*}
0L=\quad  s ( m_W^2 ( -s_\theta c_\phi ( 2 s_\theta^2 -(\beta -1)^2 + c_\theta (\beta -1)^2 ) ( \theta_{02} (-1+2 s_W^2+\beta ^2) -2 s_W^2 \theta_{23} \beta  ) + s_\theta s_\phi ( 2 s_\theta^2 -(\beta -1)^2 + c_\theta (\beta -1)^2 ) ( \theta_{01} (-1+2 s_W^2+\beta ^2) -2 s_W^2 \theta_{13} \beta  ) -2 s_W^2 \theta_{12} \beta  ( -2 (\beta -1)^2 + 2 c_\theta ((\beta -1)^2-s_\theta^2 ) + s_\theta^2 (3+ (-2+\beta ) \beta  ) ) ) -c_W^2 s_\theta ( c_\phi (\theta_{23}-\theta_{02} ) + s_\phi (\theta_{01}-\theta_{13} ) ) \beta  (1-c_\theta (\beta -1)^2 + \beta  (-2+\beta + 2 s_\theta^2 ) ) s ) 
\end{dmath*}

\begin{dmath*}
0R=\quad  -s ( m_W^2 ( -s_\theta s_\phi ((\beta +1)^2-2 s_\theta^2 + c_\theta (\beta +1)^2 ) ( -\theta_{01} + 2 s_W^2 (\theta_{01}+ \theta_{13} \beta  ) + \theta_{01} \beta ^2 ) + s_\theta c_\phi ((\beta +1)^2-2 s_\theta^2 + c_\theta (\beta +1)^2 ) ( -\theta_{02} + 2 s_W^2 (\theta_{02}+ \theta_{23} \beta  ) + \theta_{02} \beta ^2 ) + 2 s_W^2 \theta_{12} \beta  ( 2 (\beta +1)^2 + 2 c_\theta ((\beta +1)^2-s_\theta^2 ) -s_\theta^2 (3+ \beta  (2+\beta ) ) ) ) + c_W^2 s_\theta ( c_\phi (\theta_{23}-\theta_{02} ) + s_\phi (\theta_{01}-\theta_{13} ) ) \beta  (1+ c_\theta (\beta +1)^2 + \beta  (2+\beta -2 s_\theta^2 ) ) s ) 
\end{dmath*}

\begin{dmath*}
L0=\quad  s ( m_W^2 ( -s_\theta c_\phi ((\beta +1)^2-2 s_\theta^2 + c_\theta (\beta +1)^2 ) ( \theta_{02} (-1+2 s_W^2+\beta ^2) -2 s_W^2 \theta_{23} \beta  ) + s_\theta s_\phi ((\beta +1)^2-2 s_\theta^2 + c_\theta (\beta +1)^2 ) ( \theta_{01} (-1+2 s_W^2+\beta ^2) -2 s_W^2 \theta_{13} \beta  ) + 2 s_W^2 \theta_{12} \beta  ( 2 (\beta +1)^2 + 2 c_\theta ((\beta +1)^2-s_\theta^2 ) -s_\theta^2 (3+ \beta  (2+\beta ) ) ) ) -c_W^2 s_\theta ( -c_\phi (\theta_{02}+\theta_{23}) + s_\phi (\theta_{01}+\theta_{13}) ) \beta  (1+ c_\theta (\beta +1)^2 + \beta  (2+\beta -2 s_\theta^2 ) ) s ) 
\end{dmath*}

\begin{dmath*}
LL=\quad  4 s_\theta m_W^2 ( c_\theta ( 2 m_W^2 ( c_\phi \theta_{02} -s_\phi \theta_{01} -c_\phi \theta_{23} \beta  + s_\phi \theta_{13} \beta  ) + c_W^2 \beta  ( -c_\phi (\theta_{23}-2 \theta_{02} \beta  + \theta_{23} \beta ^2 ) + s_\phi (\theta_{13}-2 \theta_{01} \beta  + \theta_{13} \beta ^2 ) ) s ) + \beta  ( m_W^2 ( 2 c_\phi \theta_{02} -2 s_\phi \theta_{01} -2 c_\phi \theta_{23} \beta  + 2 s_\phi \theta_{13} \beta  ) + c_W^2 ( c_\phi (\theta_{02}+ \theta_{02} \beta ^2 -2 c_\theta^2 \theta_{23} \beta  ) -s_\phi (\theta_{01}+ \theta_{01} \beta ^2 -2 c_\theta^2 \theta_{13} \beta  ) ) s ) ) 
\end{dmath*}

\begin{dmath*}
LR=\quad  8 \left(2 s_W^2-1\right) s_\theta (-2+s_\theta^2-2 c_\theta ) ( -c_\phi \theta_{02} + s_\phi \theta_{01} ) m_W^4 
\end{dmath*}

\begin{dmath*}
R0=\quad  s ( m_W^2 ( -s_\theta c_\phi ( 2 s_\theta^2 -(\beta -1)^2 + c_\theta (\beta -1)^2 ) ( -\theta_{02} + 2 s_W^2 (\theta_{02}+ \theta_{23} \beta  ) + \theta_{02} \beta ^2 ) + s_\theta s_\phi ( 2 s_\theta^2 -(\beta -1)^2 + c_\theta (\beta -1)^2 ) ( -\theta_{01} + 2 s_W^2 (\theta_{01}+ \theta_{13} \beta  ) + \theta_{01} \beta ^2 ) + 2 s_W^2 \theta_{12} \beta  ( -2 (\beta -1)^2 + 2 c_\theta ((\beta -1)^2-s_\theta^2 ) + s_\theta^2 (3+ (-2+\beta ) \beta  ) ) ) -c_W^2 s_\theta ( -c_\phi (\theta_{02}+\theta_{23}) + s_\phi (\theta_{01}+\theta_{13}) ) \beta  (1-c_\theta (\beta -1)^2 + \beta  (-2+\beta + 2 s_\theta^2 ) ) s ) 
\end{dmath*}

\begin{dmath*}
RL=\quad  -8 \left(2 s_W^2-1\right) s_\theta (-2+s_\theta^2+ 2 c_\theta ) ( -c_\phi \theta_{02} + s_\phi \theta_{01} ) m_W^4 
\end{dmath*}

\begin{dmath*}
RR=\quad  4 s_\theta m_W^2 ( c_\theta ( 2 m_W^2 ( c_\phi (\theta_{02}+ \theta_{23} \beta  ) -s_\phi (\theta_{01}+ \theta_{13} \beta  ) ) + c_W^2 \beta  ( c_\phi (\theta_{23}+ 2 \theta_{02} \beta  + \theta_{23} \beta ^2 ) -s_\phi (\theta_{13}+ 2 \theta_{01} \beta  + \theta_{13} \beta ^2 ) ) s ) + \beta  ( 2 m_W^2 ( c_\phi (\theta_{02}+ \theta_{23} \beta  ) -s_\phi (\theta_{01}+ \theta_{13} \beta  ) ) + c_W^2 ( c_\phi (\theta_{02}+ \theta_{02} \beta ^2 + 2 c_\theta^2 \theta_{23} \beta  ) -s_\phi (\theta_{01}+ \theta_{01} \beta ^2 + 2 c_\theta^2 \theta_{13} \beta  ) ) s ) ) 
\end{dmath*}
\end{dgroup*}

    \subsection{Values of $\hat{M}_{f_tf_t}(LRh_{W^+}h_{W^-})$, with $\kappa_2=0$}
{\bf Prefactor}: $ g^4 s_\theta s^3 /\left( 512 t^2 m_W^2 \right)$
\begin{dgroup*}

\begin{dmath*}
00=\quad  (- -c_\phi (\theta_{02}+\theta_{23}) - s_\phi (\theta_{01}+\theta_{13}) ) ((\beta +1)^4-c_\theta (1+ \beta  ( -4 s_\theta^2 + (2+\beta ) (2+ \beta  (2+\beta ) ) ) ) -2 s_\theta^2 (1+\beta +\beta ^3+ 3 \beta ^2 ) ) s 
\end{dmath*}

\begin{dmath*}
0L=\quad  ( -c_\phi (\theta_{02}+\theta_{23}) + s_\phi (\theta_{01}+\theta_{13}) ) ((\beta -1)^4+ c_\theta (1+ \beta  ( 4 s_\theta^2 + (-2+\beta ) (2+ (-2+\beta ) \beta  ) ) ) + 2 s_\theta^2 (-1+\beta +\beta ^3-3 \beta ^2 ) ) s 
\end{dmath*}

\begin{dmath*}
0R=\quad  ( c_\phi (\theta_{23}-\theta_{02} ) + s_\phi (\theta_{01}-\theta_{13} ) ) ((\beta -1)^4+ c_\theta (1+ \beta  ( 4 s_\theta^2 + (-2+\beta ) (2+ (-2+\beta ) \beta  ) ) ) + 2 s_\theta^2 (-1+\beta +\beta ^3-3 \beta ^2 ) ) s 
\end{dmath*}

\begin{dmath*}
L0=\quad  8( -c_\phi (\theta_{02}+ c_\theta \theta_{23} ) + s_\phi (\theta_{01}+ c_\theta \theta_{13} ) ) m_W^2 (c_\theta+ 3 c_\theta \beta ^2 -\beta  (3+\beta ^2-2 s_\theta^2 ) ) 
\end{dmath*}

\begin{dmath*}
LL=\quad  -8( -c_\phi \theta_{02} + s_\phi \theta_{01} ) m_W^2 ( 2 (\beta -1)^2 + 2 c_\theta (1+ \beta  (-2+s_\theta^2+\beta ) ) -s_\theta^2 (1+ (-4+\beta ) \beta  ) ) 
\end{dmath*}

\begin{dmath*}
LR=\quad  (- c_\phi (\theta_{23}-\theta_{02} ) - s_\phi (\theta_{01}-\theta_{13} ) ) ((\beta +1)^4-c_\theta (1+ \beta  ( -4 s_\theta^2 + (2+\beta ) (2+ \beta  (2+\beta ) ) ) ) -2 s_\theta^2 (1+\beta +\beta ^3+ 3 \beta ^2 ) ) s 
\end{dmath*}

\begin{dmath*}
R0=\quad  8( -c_\phi \theta_{02} + s_\phi \theta_{01} ) m_W^2 ( 2 (\beta +1)^2 -2 c_\theta (1+ \beta  (2+\beta -s_\theta^2 ) ) -s_\theta^2 (1+ \beta  (4+\beta ) ) ) 
\end{dmath*}

\begin{dmath*}
RL=\quad  -8( c_\phi \theta_{02} -s_\phi \theta_{01} -c_\theta c_\phi \theta_{23} + c_\theta s_\phi \theta_{13} ) m_W^2 (c_\theta+ 3 c_\theta \beta ^2 -\beta  (3+\beta ^2-2 s_\theta^2 ) ) 
\end{dmath*}
\end{dgroup*}

    \subsection{Values of $\hat{M}_{f_uf_u}(LRh_{W^+}h_{W^-})$, with $\kappa_2=0$}
{\bf Prefactor}: $ g^4 s_\theta s^3 /\left( 512 u^2 m_W^2 \right)$
\begin{dgroup*}

\begin{dmath*}
00=\quad  ( c_\phi (\theta_{23}-\theta_{02} ) + s_\phi (\theta_{01}-\theta_{13} ) ) ( -(\beta -1)^4 + c_\theta (1+ \beta  ( 4 s_\theta^2 + (-2+\beta ) (2+ (-2+\beta ) \beta  ) ) ) -2 s_\theta^2 (-1+\beta +\beta ^3-3 \beta ^2 ) ) s 
\end{dmath*}

\begin{dmath*}
0L=\quad  (- c_\phi (\theta_{23}-\theta_{02} ) - s_\phi (\theta_{01}-\theta_{13} ) ) ( -(\beta +1)^4 -c_\theta (1+ \beta  ( -4 s_\theta^2 + (2+\beta ) (2+ \beta  (2+\beta ) ) ) ) + 2 s_\theta^2 (1+\beta +\beta ^3+ 3 \beta ^2 ) ) s 
\end{dmath*}

\begin{dmath*}
0R=\quad  (- -c_\phi (\theta_{02}+\theta_{23}) - s_\phi (\theta_{01}+\theta_{13}) ) ( -(\beta +1)^4 -c_\theta (1+ \beta  ( -4 s_\theta^2 + (2+\beta ) (2+ \beta  (2+\beta ) ) ) ) + 2 s_\theta^2 (1+\beta +\beta ^3+ 3 \beta ^2 ) ) s 
\end{dmath*}

\begin{dmath*}
L0=\quad  -8( c_\phi \theta_{02} -s_\phi \theta_{01} -c_\theta c_\phi \theta_{23} + c_\theta s_\phi \theta_{13} ) m_W^2 (c_\theta+ 3 c_\theta \beta ^2 + \beta  (3+\beta ^2-2 s_\theta^2 ) ) 
\end{dmath*}

\begin{dmath*}
LL=\quad  8( -c_\phi \theta_{02} + s_\phi \theta_{01} ) m_W^2 ( -2 (\beta +1)^2 -2 c_\theta (1+ \beta  (2+\beta -s_\theta^2 ) ) + s_\theta^2 (1+ \beta  (4+\beta ) ) ) 
\end{dmath*}

\begin{dmath*}
LR=\quad  ( -c_\phi (\theta_{02}+\theta_{23}) + s_\phi (\theta_{01}+\theta_{13}) ) ( -(\beta -1)^4 + c_\theta (1+ \beta  ( 4 s_\theta^2 + (-2+\beta ) (2+ (-2+\beta ) \beta  ) ) ) -2 s_\theta^2 (-1+\beta +\beta ^3-3 \beta ^2 ) ) s 
\end{dmath*}

\begin{dmath*}
R0=\quad  -8( -c_\phi \theta_{02} + s_\phi \theta_{01} ) m_W^2 ( -2 (\beta -1)^2 + 2 c_\theta (1+ \beta  (-2+s_\theta^2+\beta ) ) + s_\theta^2 (1+ (-4+\beta ) \beta  ) ) 
\end{dmath*}

\begin{dmath*}
RL=\quad  8( -c_\phi (\theta_{02}+ c_\theta \theta_{23} ) + s_\phi (\theta_{01}+ c_\theta \theta_{13} ) ) m_W^2 (c_\theta+ 3 c_\theta \beta ^2 + \beta  (3+\beta ^2-2 s_\theta^2 ) ) 
\end{dmath*}
\end{dgroup*}

    \subsection{Values of $\hat{M}_{c\gamma_s}(LRh_{W^+}h_{W^-})$, with $\kappa_2=0$}
{\bf Prefactor}: $ g^4 Q_f s_W^2 s_\theta \beta  s /\left( 16 m_W^2 \right)$
\begin{dgroup*}

\begin{dmath*}
00=\quad  (1-c_\theta) ( c_\phi \theta_{23} -s_\phi \theta_{13} -c_\phi \theta_{02} \beta ^2 + s_\phi \theta_{01} \beta ^2 ) s 
\end{dmath*}

\begin{dmath*}
0L=\quad  (1+c_\theta) ( c_\phi \theta_{23} -s_\phi \theta_{13} -c_\phi \theta_{02} \beta ^2 + s_\phi \theta_{01} \beta ^2 ) s 
\end{dmath*}

\begin{dmath*}
0R=\quad  (1+c_\theta) ( -c_\phi (\theta_{23}+ \theta_{02} \beta ^2 ) + s_\phi (\theta_{13}+ \theta_{01} \beta ^2 ) ) s 
\end{dmath*}

\begin{dmath*}
L0=\quad  -4 (c_\theta-1) ( -c_\phi \theta_{23} + s_\phi \theta_{13} ) m_W^2 \beta  
\end{dmath*}

\begin{dmath*}
LL=\quad  (1-c_\theta) ( -c_\phi (\theta_{23}+ \theta_{02} \beta ^2 ) + s_\phi (\theta_{13}+ \theta_{01} \beta ^2 ) ) s 
\end{dmath*}

\begin{dmath*}
LR=\quad  4 (c_\theta-1) ( -c_\phi \theta_{23} + s_\phi \theta_{13} ) m_W^2 \beta  
\end{dmath*}
\end{dgroup*}

    \subsection{Values of $\hat{M}_{cZ_s}(LRh_{W^+}h_{W^-})$, with $\kappa_2=0$}
{\bf Prefactor}: $ (g_A^f+g_V^f) g^4 s_\theta \beta  s^2 /\left( 32 m_W^2 (s-m_Z^2) \right)$
\begin{dgroup*}

\begin{dmath*}
00=\quad  (1-c_\theta) ( c_\phi \theta_{23} -s_\phi \theta_{13} -c_\phi \theta_{02} \beta ^2 + s_\phi \theta_{01} \beta ^2 ) s 
\end{dmath*}

\begin{dmath*}
0L=\quad  (1+c_\theta) ( c_\phi \theta_{23} -s_\phi \theta_{13} -c_\phi \theta_{02} \beta ^2 + s_\phi \theta_{01} \beta ^2 ) s 
\end{dmath*}

\begin{dmath*}
0R=\quad  (1+c_\theta) ( -c_\phi (\theta_{23}+ \theta_{02} \beta ^2 ) + s_\phi (\theta_{13}+ \theta_{01} \beta ^2 ) ) s 
\end{dmath*}

\begin{dmath*}
L0=\quad  -4 (c_\theta-1) ( -c_\phi \theta_{23} + s_\phi \theta_{13} ) m_W^2 \beta  
\end{dmath*}

\begin{dmath*}
LL=\quad  (1-c_\theta) ( -c_\phi (\theta_{23}+ \theta_{02} \beta ^2 ) + s_\phi (\theta_{13}+ \theta_{01} \beta ^2 ) ) s 
\end{dmath*}

\begin{dmath*}
LR=\quad  4 (c_\theta-1) ( -c_\phi \theta_{23} + s_\phi \theta_{13} ) m_W^2 \beta  
\end{dmath*}
\end{dgroup*}

    \subsection{Values of $\hat{M}_{cf_t}(LRh_{W^+}h_{W^-})$, with $\kappa_2=0$}
{\bf Prefactor}: $ g^4 s_\theta s^2 /\left( 128 t m_W^2 \right)$
\begin{dgroup*}

\begin{dmath*}
00=\quad  ( 2 s_\theta^2 -(\beta +1)^2 + c_\theta (\beta +1)^2 ) ( c_\phi \theta_{23} -s_\phi \theta_{13} -c_\phi \theta_{02} \beta ^2 + s_\phi \theta_{01} \beta ^2 ) s 
\end{dmath*}

\begin{dmath*}
0L=\quad  ((\beta -1)^2-2 s_\theta^2 + c_\theta (\beta -1)^2 ) ( c_\phi \theta_{23} -s_\phi \theta_{13} -c_\phi \theta_{02} \beta ^2 + s_\phi \theta_{01} \beta ^2 ) s 
\end{dmath*}

\begin{dmath*}
0R=\quad  ((\beta -1)^2-2 s_\theta^2 + c_\theta (\beta -1)^2 ) ( -c_\phi (\theta_{23}+ \theta_{02} \beta ^2 ) + s_\phi (\theta_{13}+ \theta_{01} \beta ^2 ) ) s 
\end{dmath*}

\begin{dmath*}
L0=\quad  -8( -c_\phi \theta_{23} + s_\phi \theta_{13} ) m_W^2 (c_\theta-\beta  )^2 
\end{dmath*}

\begin{dmath*}
LL=\quad  ( 2 s_\theta^2 -(\beta +1)^2 + c_\theta (\beta +1)^2 ) ( -c_\phi (\theta_{23}+ \theta_{02} \beta ^2 ) + s_\phi (\theta_{13}+ \theta_{01} \beta ^2 ) ) s 
\end{dmath*}

\begin{dmath*}
LR=\quad  8( -c_\phi \theta_{23} + s_\phi \theta_{13} ) m_W^2 (c_\theta-\beta  )^2 
\end{dmath*}
\end{dgroup*}

    \subsection{Values of $\hat{M}_{cf_u}(LRh_{W^+}h_{W^-})$, with $\kappa_2=0$}
{\bf Prefactor}: $ g^4 s_\theta s^2 /\left( 128 u m_W^2 \right)$
\begin{dgroup*}

\begin{dmath*}
00=\quad  ( 2 s_\theta^2 -(\beta -1)^2 + c_\theta (\beta -1)^2 ) ( c_\phi \theta_{23} -s_\phi \theta_{13} -c_\phi \theta_{02} \beta ^2 + s_\phi \theta_{01} \beta ^2 ) s 
\end{dmath*}

\begin{dmath*}
0L=\quad  ((\beta +1)^2-2 s_\theta^2 + c_\theta (\beta +1)^2 ) ( c_\phi \theta_{23} -s_\phi \theta_{13} -c_\phi \theta_{02} \beta ^2 + s_\phi \theta_{01} \beta ^2 ) s 
\end{dmath*}

\begin{dmath*}
0R=\quad  ((\beta +1)^2-2 s_\theta^2 + c_\theta (\beta +1)^2 ) ( -c_\phi (\theta_{23}+ \theta_{02} \beta ^2 ) + s_\phi (\theta_{13}+ \theta_{01} \beta ^2 ) ) s 
\end{dmath*}

\begin{dmath*}
L0=\quad  8( -c_\phi \theta_{23} + s_\phi \theta_{13} ) m_W^2 (\beta ^2-c_\theta^2 ) 
\end{dmath*}

\begin{dmath*}
LL=\quad  ( 2 s_\theta^2 -(\beta -1)^2 + c_\theta (\beta -1)^2 ) ( -c_\phi (\theta_{23}+ \theta_{02} \beta ^2 ) + s_\phi (\theta_{13}+ \theta_{01} \beta ^2 ) ) s 
\end{dmath*}

\begin{dmath*}
LR=\quad  8( -c_\phi \theta_{23} + s_\phi \theta_{13} ) m_W^2 (c_\theta^2-\beta ^2 ) 
\end{dmath*}
\end{dgroup*}

  }

  \section{Amplitude squared for \epem\to\wpwm, $\kappa_2$ terms}

  In this appendix, we give the values of the coefficients of $\kappa_2$ in the NCSM 
  contributions to $\hat{M}$ for every diagram pair and every helicity choice.
  The notation and organization is the same as in the previous appendix.  For example, section B.1 gives the $\O(\kappa_2)$ contribution to
  the quantity $\hat{M}_{\gamma_s\gamma_s}=2\Re(M_{\gamma_s}^{SM}M_{\gamma_s}^{NC}+M_{\gamma_s}^{SM}M_{\gamma_s}^{NC})$ for a left-handed electron and a right-handed
  positron.  The line 
  \begin{equation}
    00=\quad  -s_\theta ( -c_\phi \theta_{02} + s_\phi \theta_{01} ) (-3+\beta ^2) \left(\beta ^2-1\right)^2 s^2
  \end{equation}
  gives the coefficient of $\kappa_2$ in $\hat{M}_{\gamma_s\gamma_s}(LR00)$, if this expression is multiplied by the prefactor given at the beginning of
  Section B.1.  If any polarization choice is not shown, it is because
  the corresponding contribution is zero.

  { 
    \setlength{\abovedisplayskip}{0pt} 
    \setlength{\belowdisplayskip}{0pt} 
    \setlength{\abovedisplayshortskip}{0pt} 
    \setlength{\belowdisplayshortskip}{0pt}
    \subsection{Coefficients of $\kappa_2$ for $\hat{M}_{\gamma_s\gamma_s}(LRh_{W^+}h_{W^-})$}
{\bf Prefactor}: $ g^6 Q_f^2 s_W^4 \beta  s /\left( 8 m_W^4 \right)$
\begin{dgroup*}

\begin{dmath*}
00=\quad  -s_\theta ( -c_\phi \theta_{02} + s_\phi \theta_{01} ) (-3+\beta ^2) \left(\beta ^2-1\right)^2 s^2 
\end{dmath*}

\begin{dmath*}
0L=\quad  -4 m_W^2 (-1+\beta ^2) ( (-2+s_\theta^2+ 2 c_\theta ) \theta_{12} \beta  + (1-c_\theta) s_\theta c_\phi (\theta_{02}+ \theta_{23} \beta  ) + (-1+c_\theta) s_\theta s_\phi (\theta_{01}+ \theta_{13} \beta  ) ) s 
\end{dmath*}

\begin{dmath*}
0R=\quad  -4 m_W^2 (-1+\beta ^2) ( (-2+s_\theta^2-2 c_\theta ) \theta_{12} \beta  + (1+c_\theta) s_\theta c_\phi (\theta_{02}-\theta_{23} \beta  ) + (1+c_\theta) s_\theta s_\phi ( -\theta_{01} + \theta_{13} \beta  ) ) s 
\end{dmath*}

\begin{dmath*}
L0=\quad  4 m_W^2 (-1+\beta ^2) ( (-2+s_\theta^2-2 c_\theta ) \theta_{12} \beta  + (-1-c_\theta) s_\theta c_\phi (\theta_{02}+ \theta_{23} \beta  ) + (1+c_\theta) s_\theta s_\phi (\theta_{01}+ \theta_{13} \beta  ) ) s 
\end{dmath*}

\begin{dmath*}
LL=\quad  32 s_\theta m_W^4 ( c_\phi \theta_{02} -s_\phi \theta_{01} -c_\phi \theta_{23} \beta  + s_\phi \theta_{13} \beta  ) 
\end{dmath*}

\begin{dmath*}
LR=\quad  4 m_W^2 (-1+\beta ^2) ( (-2+s_\theta^2+ 2 c_\theta ) \theta_{12} \beta  + (1-c_\theta) s_\theta s_\phi (\theta_{01}-\theta_{13} \beta  ) + (-1+c_\theta) s_\theta c_\phi (\theta_{02}-\theta_{23} \beta  ) ) s 
\end{dmath*}

\begin{dmath*}
R0=\quad  32 s_\theta m_W^4 ( c_\phi (\theta_{02}+ \theta_{23} \beta  ) -s_\phi (\theta_{01}+ \theta_{13} \beta  ) ) 
\end{dmath*}
\end{dgroup*}

    \subsection{Coefficients of $\kappa_2$ for $\hat{M}_{\gamma_s\gamma_s}(RLh_{W^+}h_{W^-})$}
{\bf Prefactor}: $ g^6 Q_f^2 s_W^4 \beta  s /\left( 8 m_W^4 \right)$
\begin{dgroup*}

\begin{dmath*}
00=\quad  s_\theta ( -c_\phi \theta_{02} + s_\phi \theta_{01} ) (-3+\beta ^2) \left(\beta ^2-1\right)^2 s^2 
\end{dmath*}

\begin{dmath*}
0L=\quad  4 m_W^2 (-1+\beta ^2) ( (-2+s_\theta^2-2 c_\theta ) \theta_{12} \beta  + (1+c_\theta) s_\theta c_\phi (\theta_{02}-\theta_{23} \beta  ) + (1+c_\theta) s_\theta s_\phi ( -\theta_{01} + \theta_{13} \beta  ) ) s 
\end{dmath*}

\begin{dmath*}
0R=\quad  4 m_W^2 (-1+\beta ^2) ( (-2+s_\theta^2+ 2 c_\theta ) \theta_{12} \beta  + (1-c_\theta) s_\theta c_\phi (\theta_{02}+ \theta_{23} \beta  ) + (-1+c_\theta) s_\theta s_\phi (\theta_{01}+ \theta_{13} \beta  ) ) s 
\end{dmath*}

\begin{dmath*}
L0=\quad  -4 m_W^2 (-1+\beta ^2) ( (-2+s_\theta^2+ 2 c_\theta ) \theta_{12} \beta  + (1-c_\theta) s_\theta s_\phi (\theta_{01}-\theta_{13} \beta  ) + (-1+c_\theta) s_\theta c_\phi (\theta_{02}-\theta_{23} \beta  ) ) s 
\end{dmath*}

\begin{dmath*}
LL=\quad  32 s_\theta m_W^4 ( -c_\phi (\theta_{02}+ \theta_{23} \beta  ) + s_\phi (\theta_{01}+ \theta_{13} \beta  ) ) 
\end{dmath*}

\begin{dmath*}
LR=\quad  -4 m_W^2 (-1+\beta ^2) ( (-2+s_\theta^2-2 c_\theta ) \theta_{12} \beta  + (-1-c_\theta) s_\theta c_\phi (\theta_{02}+ \theta_{23} \beta  ) + (1+c_\theta) s_\theta s_\phi (\theta_{01}+ \theta_{13} \beta  ) ) s 
\end{dmath*}

\begin{dmath*}
R0=\quad  32 s_\theta m_W^4 ( -c_\phi \theta_{02} + s_\phi \theta_{01} + c_\phi \theta_{23} \beta  -s_\phi \theta_{13} \beta  ) 
\end{dmath*}
\end{dgroup*}

    \subsection{Coefficients of $\kappa_2$ for $\hat{M}_{Z_s\gamma_s}(LRh_{W^+}h_{W^-})$}
{\bf Prefactor}: $ (g_A^f+g_V^f) g^6 Q_f s_W^2 \left(2 s_W^4-3 s_W^2+1\right) \beta  s^2 /\left( 32 c_W^4 m_W^4 (s-m_Z^2) \right)$
\begin{dgroup*}

\begin{dmath*}
00=\quad  -s_\theta ( -c_\phi \theta_{02} + s_\phi \theta_{01} ) (-3+\beta ^2) \left(\beta ^2-1\right)^2 s^2 
\end{dmath*}

\begin{dmath*}
0L=\quad  -4 m_W^2 (-1+\beta ^2) ( (-2+s_\theta^2+ 2 c_\theta ) \theta_{12} \beta  + (1-c_\theta) s_\theta c_\phi (\theta_{02}+ \theta_{23} \beta  ) + (-1+c_\theta) s_\theta s_\phi (\theta_{01}+ \theta_{13} \beta  ) ) s 
\end{dmath*}

\begin{dmath*}
0R=\quad  4 m_W^2 (-1+\beta ^2) ( (2+ 2 c_\theta -s_\theta^2 ) \theta_{12} \beta  + (1+c_\theta) s_\theta c_\phi ( -\theta_{02} + \theta_{23} \beta  ) + (1+c_\theta) s_\theta s_\phi (\theta_{01}-\theta_{13} \beta  ) ) s 
\end{dmath*}

\begin{dmath*}
L0=\quad  4 m_W^2 (-1+\beta ^2) ( (-2+s_\theta^2-2 c_\theta ) \theta_{12} \beta  + (-1-c_\theta) s_\theta c_\phi (\theta_{02}+ \theta_{23} \beta  ) + (1+c_\theta) s_\theta s_\phi (\theta_{01}+ \theta_{13} \beta  ) ) s 
\end{dmath*}

\begin{dmath*}
LL=\quad  32 s_\theta m_W^4 ( c_\phi \theta_{02} -s_\phi \theta_{01} -c_\phi \theta_{23} \beta  + s_\phi \theta_{13} \beta  ) 
\end{dmath*}

\begin{dmath*}
LR=\quad  4 m_W^2 (-1+\beta ^2) ( (-2+s_\theta^2+ 2 c_\theta ) \theta_{12} \beta  + (1-c_\theta) s_\theta s_\phi (\theta_{01}-\theta_{13} \beta  ) + (-1+c_\theta) s_\theta c_\phi (\theta_{02}-\theta_{23} \beta  ) ) s 
\end{dmath*}

\begin{dmath*}
R0=\quad  32 s_\theta m_W^4 ( c_\phi (\theta_{02}+ \theta_{23} \beta  ) -s_\phi (\theta_{01}+ \theta_{13} \beta  ) ) 
\end{dmath*}
\end{dgroup*}

    \subsection{Coefficients of $\kappa_2$ for $\hat{M}_{Z_s\gamma_s}(RLh_{W^+}h_{W^-})$}
{\bf Prefactor}: $ (g_A^f-g_V^f) g^6 Q_f s_W^2 \left(2 s_W^4-3 s_W^2+1\right) \beta  s^2 /\left( 32 c_W^4 m_W^4 (s-m_Z^2) \right)$
\begin{dgroup*}

\begin{dmath*}
00=\quad  -s_\theta ( -c_\phi \theta_{02} + s_\phi \theta_{01} ) (-3+\beta ^2) \left(\beta ^2-1\right)^2 s^2 
\end{dmath*}

\begin{dmath*}
0L=\quad  4 m_W^2 (-1+\beta ^2) ( (2+ 2 c_\theta -s_\theta^2 ) \theta_{12} \beta  + (1+c_\theta) s_\theta c_\phi ( -\theta_{02} + \theta_{23} \beta  ) + (1+c_\theta) s_\theta s_\phi (\theta_{01}-\theta_{13} \beta  ) ) s 
\end{dmath*}

\begin{dmath*}
0R=\quad  -4 m_W^2 (-1+\beta ^2) ( (-2+s_\theta^2+ 2 c_\theta ) \theta_{12} \beta  + (1-c_\theta) s_\theta c_\phi (\theta_{02}+ \theta_{23} \beta  ) + (-1+c_\theta) s_\theta s_\phi (\theta_{01}+ \theta_{13} \beta  ) ) s 
\end{dmath*}

\begin{dmath*}
L0=\quad  4 m_W^2 (-1+\beta ^2) ( (-2+s_\theta^2+ 2 c_\theta ) \theta_{12} \beta  + (1-c_\theta) s_\theta s_\phi (\theta_{01}-\theta_{13} \beta  ) + (-1+c_\theta) s_\theta c_\phi (\theta_{02}-\theta_{23} \beta  ) ) s 
\end{dmath*}

\begin{dmath*}
LL=\quad  32 s_\theta m_W^4 ( c_\phi (\theta_{02}+ \theta_{23} \beta  ) -s_\phi (\theta_{01}+ \theta_{13} \beta  ) ) 
\end{dmath*}

\begin{dmath*}
LR=\quad  4 m_W^2 (-1+\beta ^2) ( (-2+s_\theta^2-2 c_\theta ) \theta_{12} \beta  + (-1-c_\theta) s_\theta c_\phi (\theta_{02}+ \theta_{23} \beta  ) + (1+c_\theta) s_\theta s_\phi (\theta_{01}+ \theta_{13} \beta  ) ) s 
\end{dmath*}

\begin{dmath*}
R0=\quad  32 s_\theta m_W^4 ( c_\phi \theta_{02} -s_\phi \theta_{01} -c_\phi \theta_{23} \beta  + s_\phi \theta_{13} \beta  ) 
\end{dmath*}
\end{dgroup*}

    \subsection{Coefficients of $\kappa_2$ for $\hat{M}_{Z_sZ_s}(LRh_{W^+}h_{W^-})$}
{\bf Prefactor}: $ (g_A^f+g_V^f)^2 g^6 s_W^2 \beta  s^3 /\left( 64 c_W^2 m_W^4 (s-m_Z^2)^2 \right)$
\begin{dgroup*}

\begin{dmath*}
00=\quad  s_\theta ( -c_\phi \theta_{02} + s_\phi \theta_{01} ) (-3+\beta ^2) \left(\beta ^2-1\right)^2 s^2 
\end{dmath*}

\begin{dmath*}
0L=\quad  4 m_W^2 (-1+\beta ^2) ( (-2+s_\theta^2+ 2 c_\theta ) \theta_{12} \beta  + (1-c_\theta) s_\theta c_\phi (\theta_{02}+ \theta_{23} \beta  ) + (-1+c_\theta) s_\theta s_\phi (\theta_{01}+ \theta_{13} \beta  ) ) s 
\end{dmath*}

\begin{dmath*}
0R=\quad  4 m_W^2 (-1+\beta ^2) ( (-2+s_\theta^2-2 c_\theta ) \theta_{12} \beta  + (1+c_\theta) s_\theta c_\phi (\theta_{02}-\theta_{23} \beta  ) + (1+c_\theta) s_\theta s_\phi ( -\theta_{01} + \theta_{13} \beta  ) ) s 
\end{dmath*}

\begin{dmath*}
L0=\quad  -4 m_W^2 (-1+\beta ^2) ( (-2+s_\theta^2-2 c_\theta ) \theta_{12} \beta  + (-1-c_\theta) s_\theta c_\phi (\theta_{02}+ \theta_{23} \beta  ) + (1+c_\theta) s_\theta s_\phi (\theta_{01}+ \theta_{13} \beta  ) ) s 
\end{dmath*}

\begin{dmath*}
LL=\quad  32 s_\theta m_W^4 ( -c_\phi \theta_{02} + s_\phi \theta_{01} + c_\phi \theta_{23} \beta  -s_\phi \theta_{13} \beta  ) 
\end{dmath*}

\begin{dmath*}
LR=\quad  -4 m_W^2 (-1+\beta ^2) ( (-2+s_\theta^2+ 2 c_\theta ) \theta_{12} \beta  + (1-c_\theta) s_\theta s_\phi (\theta_{01}-\theta_{13} \beta  ) + (-1+c_\theta) s_\theta c_\phi (\theta_{02}-\theta_{23} \beta  ) ) s 
\end{dmath*}

\begin{dmath*}
R0=\quad  32 s_\theta m_W^4 ( -c_\phi (\theta_{02}+ \theta_{23} \beta  ) + s_\phi (\theta_{01}+ \theta_{13} \beta  ) ) 
\end{dmath*}
\end{dgroup*}

    \subsection{Coefficients of $\kappa_2$ for $\hat{M}_{Z_sZ_s}(RLh_{W^+}h_{W^-})$}
{\bf Prefactor}: $ (g_A^f-g_V^f)^2 g^6 s_W^2 \beta  s^3 /\left( 64 c_W^2 m_W^4 (s-m_Z^2)^2 \right)$
\begin{dgroup*}

\begin{dmath*}
00=\quad  -s_\theta ( -c_\phi \theta_{02} + s_\phi \theta_{01} ) (-3+\beta ^2) \left(\beta ^2-1\right)^2 s^2 
\end{dmath*}

\begin{dmath*}
0L=\quad  4 m_W^2 (-1+\beta ^2) ( (2+ 2 c_\theta -s_\theta^2 ) \theta_{12} \beta  + (1+c_\theta) s_\theta c_\phi ( -\theta_{02} + \theta_{23} \beta  ) + (1+c_\theta) s_\theta s_\phi (\theta_{01}-\theta_{13} \beta  ) ) s 
\end{dmath*}

\begin{dmath*}
0R=\quad  -4 m_W^2 (-1+\beta ^2) ( (-2+s_\theta^2+ 2 c_\theta ) \theta_{12} \beta  + (1-c_\theta) s_\theta c_\phi (\theta_{02}+ \theta_{23} \beta  ) + (-1+c_\theta) s_\theta s_\phi (\theta_{01}+ \theta_{13} \beta  ) ) s 
\end{dmath*}

\begin{dmath*}
L0=\quad  4 m_W^2 (-1+\beta ^2) ( (-2+s_\theta^2+ 2 c_\theta ) \theta_{12} \beta  + (1-c_\theta) s_\theta s_\phi (\theta_{01}-\theta_{13} \beta  ) + (-1+c_\theta) s_\theta c_\phi (\theta_{02}-\theta_{23} \beta  ) ) s 
\end{dmath*}

\begin{dmath*}
LL=\quad  32 s_\theta m_W^4 ( c_\phi (\theta_{02}+ \theta_{23} \beta  ) -s_\phi (\theta_{01}+ \theta_{13} \beta  ) ) 
\end{dmath*}

\begin{dmath*}
LR=\quad  4 m_W^2 (-1+\beta ^2) ( (-2+s_\theta^2-2 c_\theta ) \theta_{12} \beta  + (-1-c_\theta) s_\theta c_\phi (\theta_{02}+ \theta_{23} \beta  ) + (1+c_\theta) s_\theta s_\phi (\theta_{01}+ \theta_{13} \beta  ) ) s 
\end{dmath*}

\begin{dmath*}
R0=\quad  32 s_\theta m_W^4 ( c_\phi \theta_{02} -s_\phi \theta_{01} -c_\phi \theta_{23} \beta  + s_\phi \theta_{13} \beta  ) 
\end{dmath*}
\end{dgroup*}

    \subsection{Coefficients of $\kappa_2$ for $\hat{M}_{f_t\gamma_s}(LRh_{W^+}h_{W^-})$}
{\bf Prefactor}: $ g^6 Q_f s_W^2 s^2 /\left( 64 t m_W^4 \right)$
\begin{dgroup*}

\begin{dmath*}
00=\quad  s_\theta ( -c_\phi \theta_{02} + s_\phi \theta_{01} ) \left(\beta ^2-1\right)^2 (\beta ^3+ 2 c_\theta -3 \beta  ) s^2 
\end{dmath*}

\begin{dmath*}
0L=\quad  2 m_W^2 (-1+\beta ^2) ( \theta_{12} \beta  ( -2 (\beta +1)^2 + 2 c_\theta ((\beta +1)^2-s_\theta^2 ) + s_\theta^2 (3+ \beta  (2+\beta ) ) ) -s_\theta c_\phi ( 2 s_\theta^2 -(\beta +1)^2 + c_\theta (\beta +1)^2 ) (\theta_{02}+ \theta_{23} \beta  ) + s_\theta s_\phi ( 2 s_\theta^2 -(\beta +1)^2 + c_\theta (\beta +1)^2 ) (\theta_{01}+ \theta_{13} \beta  ) ) s 
\end{dmath*}

\begin{dmath*}
0R=\quad  -2 m_W^2 (-1+\beta ^2) ( \theta_{12} \beta  ( -2 (\beta -1)^2 + 2 c_\theta (s_\theta^2-(\beta -1)^2 ) + s_\theta^2 (3+ (-2+\beta ) \beta  ) ) + s_\theta c_\phi ((\beta -1)^2-2 s_\theta^2 + c_\theta (\beta -1)^2 ) (\theta_{02}-\theta_{23} \beta  ) + s_\theta s_\phi ((\beta -1)^2-2 s_\theta^2 + c_\theta (\beta -1)^2 ) ( -\theta_{01} + \theta_{13} \beta  ) ) s 
\end{dmath*}

\begin{dmath*}
L0=\quad  -2 m_W^2 (-1+\beta ^2) ( \theta_{12} \beta  ( 2 (\beta -1)^2 + 2 c_\theta ((\beta -1)^2-s_\theta^2 ) -s_\theta^2 (3+ (-2+\beta ) \beta  ) ) -s_\theta s_\phi ((\beta -1)^2-2 s_\theta^2 + c_\theta (\beta -1)^2 ) (\theta_{01}+ \theta_{13} \beta  ) + s_\theta c_\phi ((\beta -1)^2-2 s_\theta^2 + c_\theta (\beta -1)^2 ) (\theta_{02}+ \theta_{23} \beta  ) ) s 
\end{dmath*}

\begin{dmath*}
LL=\quad  32 s_\theta m_W^4 (c_\theta-\beta  ) ( c_\phi \theta_{02} -s_\phi \theta_{01} -c_\phi \theta_{23} \beta  + s_\phi \theta_{13} \beta  ) 
\end{dmath*}

\begin{dmath*}
LR=\quad  -16 s_\theta (-2+s_\theta^2-2 c_\theta ) ( -c_\phi \theta_{02} + s_\phi \theta_{01} ) m_W^4 (1+\beta ^2) 
\end{dmath*}

\begin{dmath*}
R0=\quad  -2 m_W^2 (-1+\beta ^2) ( \theta_{12} \beta  ( -2 (\beta +1)^2 + 2 c_\theta ((\beta +1)^2-s_\theta^2 ) + s_\theta^2 (3+ \beta  (2+\beta ) ) ) + s_\theta c_\phi ( 2 s_\theta^2 -(\beta +1)^2 + c_\theta (\beta +1)^2 ) (\theta_{02}-\theta_{23} \beta  ) + s_\theta s_\phi ( 2 s_\theta^2 -(\beta +1)^2 + c_\theta (\beta +1)^2 ) ( -\theta_{01} + \theta_{13} \beta  ) ) s 
\end{dmath*}

\begin{dmath*}
RL=\quad  16 s_\theta (-2+s_\theta^2+ 2 c_\theta ) ( -c_\phi \theta_{02} + s_\phi \theta_{01} ) m_W^4 (1+\beta ^2) 
\end{dmath*}

\begin{dmath*}
RR=\quad  -32 s_\theta m_W^4 (c_\theta-\beta  ) ( -c_\phi (\theta_{02}+ \theta_{23} \beta  ) + s_\phi (\theta_{01}+ \theta_{13} \beta  ) ) 
\end{dmath*}
\end{dgroup*}

    \subsection{Coefficients of $\kappa_2$ for $\hat{M}_{f_tZ_s}(LRh_{W^+}h_{W^-})$}
{\bf Prefactor}: $ (-g_A^f-g_V^f) g^6 s_W^2 s^3 /\left( 128 c_W^2 t m_W^4 (s-m_Z^2) \right)$
\begin{dgroup*}

\begin{dmath*}
00=\quad  s_\theta ( -c_\phi \theta_{02} + s_\phi \theta_{01} ) \left(\beta ^2-1\right)^2 (\beta ^3+ 2 c_\theta -3 \beta  ) s^2 
\end{dmath*}

\begin{dmath*}
0L=\quad  2 m_W^2 (-1+\beta ^2) ( \theta_{12} \beta  ( -2 (\beta +1)^2 + 2 c_\theta ((\beta +1)^2-s_\theta^2 ) + s_\theta^2 (3+ \beta  (2+\beta ) ) ) -s_\theta c_\phi ( 2 s_\theta^2 -(\beta +1)^2 + c_\theta (\beta +1)^2 ) (\theta_{02}+ \theta_{23} \beta  ) + s_\theta s_\phi ( 2 s_\theta^2 -(\beta +1)^2 + c_\theta (\beta +1)^2 ) (\theta_{01}+ \theta_{13} \beta  ) ) s 
\end{dmath*}

\begin{dmath*}
0R=\quad  -2 m_W^2 (-1+\beta ^2) ( \theta_{12} \beta  ( -2 (\beta -1)^2 + 2 c_\theta (s_\theta^2-(\beta -1)^2 ) + s_\theta^2 (3+ (-2+\beta ) \beta  ) ) + s_\theta c_\phi ((\beta -1)^2-2 s_\theta^2 + c_\theta (\beta -1)^2 ) (\theta_{02}-\theta_{23} \beta  ) + s_\theta s_\phi ((\beta -1)^2-2 s_\theta^2 + c_\theta (\beta -1)^2 ) ( -\theta_{01} + \theta_{13} \beta  ) ) s 
\end{dmath*}

\begin{dmath*}
L0=\quad  -2 m_W^2 (-1+\beta ^2) ( \theta_{12} \beta  ( 2 (\beta -1)^2 + 2 c_\theta ((\beta -1)^2-s_\theta^2 ) -s_\theta^2 (3+ (-2+\beta ) \beta  ) ) -s_\theta s_\phi ((\beta -1)^2-2 s_\theta^2 + c_\theta (\beta -1)^2 ) (\theta_{01}+ \theta_{13} \beta  ) + s_\theta c_\phi ((\beta -1)^2-2 s_\theta^2 + c_\theta (\beta -1)^2 ) (\theta_{02}+ \theta_{23} \beta  ) ) s 
\end{dmath*}

\begin{dmath*}
LL=\quad  32 s_\theta m_W^4 (c_\theta-\beta  ) ( c_\phi \theta_{02} -s_\phi \theta_{01} -c_\phi \theta_{23} \beta  + s_\phi \theta_{13} \beta  ) 
\end{dmath*}

\begin{dmath*}
LR=\quad  -16 s_\theta (-2+s_\theta^2-2 c_\theta ) ( -c_\phi \theta_{02} + s_\phi \theta_{01} ) m_W^4 (1+\beta ^2) 
\end{dmath*}

\begin{dmath*}
R0=\quad  -2 m_W^2 (-1+\beta ^2) ( \theta_{12} \beta  ( -2 (\beta +1)^2 + 2 c_\theta ((\beta +1)^2-s_\theta^2 ) + s_\theta^2 (3+ \beta  (2+\beta ) ) ) + s_\theta c_\phi ( 2 s_\theta^2 -(\beta +1)^2 + c_\theta (\beta +1)^2 ) (\theta_{02}-\theta_{23} \beta  ) + s_\theta s_\phi ( 2 s_\theta^2 -(\beta +1)^2 + c_\theta (\beta +1)^2 ) ( -\theta_{01} + \theta_{13} \beta  ) ) s 
\end{dmath*}

\begin{dmath*}
RL=\quad  16 s_\theta (-2+s_\theta^2+ 2 c_\theta ) ( -c_\phi \theta_{02} + s_\phi \theta_{01} ) m_W^4 (1+\beta ^2) 
\end{dmath*}

\begin{dmath*}
RR=\quad  -32 s_\theta m_W^4 (c_\theta-\beta  ) ( -c_\phi (\theta_{02}+ \theta_{23} \beta  ) + s_\phi (\theta_{01}+ \theta_{13} \beta  ) ) 
\end{dmath*}
\end{dgroup*}

    \subsection{Coefficients of $\kappa_2$ for $\hat{M}_{f_u\gamma_s}(LRh_{W^+}h_{W^-})$}
{\bf Prefactor}: $ g^6 Q_f s_W^2 s^2 /\left( 64 u m_W^4 \right)$
\begin{dgroup*}

\begin{dmath*}
00=\quad  s_\theta ( -c_\phi \theta_{02} + s_\phi \theta_{01} ) \left(\beta ^2-1\right)^2 ( 2 c_\theta + 3 \beta  -\beta ^3 ) s^2 
\end{dmath*}

\begin{dmath*}
0L=\quad  2 m_W^2 (-1+\beta ^2) ( \theta_{12} \beta  ( -2 (\beta -1)^2 + 2 c_\theta ((\beta -1)^2-s_\theta^2 ) + s_\theta^2 (3+ (-2+\beta ) \beta  ) ) -s_\theta c_\phi ( 2 s_\theta^2 -(\beta -1)^2 + c_\theta (\beta -1)^2 ) (\theta_{02}+ \theta_{23} \beta  ) + s_\theta s_\phi ( 2 s_\theta^2 -(\beta -1)^2 + c_\theta (\beta -1)^2 ) (\theta_{01}+ \theta_{13} \beta  ) ) s 
\end{dmath*}

\begin{dmath*}
0R=\quad  -2 m_W^2 (-1+\beta ^2) ( \theta_{12} \beta  ( -2 (\beta +1)^2 + 2 c_\theta (s_\theta^2-(\beta +1)^2 ) + s_\theta^2 (3+ \beta  (2+\beta ) ) ) + s_\theta c_\phi ((\beta +1)^2-2 s_\theta^2 + c_\theta (\beta +1)^2 ) (\theta_{02}-\theta_{23} \beta  ) + s_\theta s_\phi ((\beta +1)^2-2 s_\theta^2 + c_\theta (\beta +1)^2 ) ( -\theta_{01} + \theta_{13} \beta  ) ) s 
\end{dmath*}

\begin{dmath*}
L0=\quad  -2 m_W^2 (-1+\beta ^2) ( \theta_{12} \beta  ( 2 (\beta +1)^2 + 2 c_\theta ((\beta +1)^2-s_\theta^2 ) -s_\theta^2 (3+ \beta  (2+\beta ) ) ) -s_\theta s_\phi ((\beta +1)^2-2 s_\theta^2 + c_\theta (\beta +1)^2 ) (\theta_{01}+ \theta_{13} \beta  ) + s_\theta c_\phi ((\beta +1)^2-2 s_\theta^2 + c_\theta (\beta +1)^2 ) (\theta_{02}+ \theta_{23} \beta  ) ) s 
\end{dmath*}

\begin{dmath*}
LL=\quad  32 s_\theta m_W^4 (c_\theta+\beta ) ( c_\phi \theta_{02} -s_\phi \theta_{01} -c_\phi \theta_{23} \beta  + s_\phi \theta_{13} \beta  ) 
\end{dmath*}

\begin{dmath*}
LR=\quad  -16 s_\theta (-2+s_\theta^2-2 c_\theta ) ( -c_\phi \theta_{02} + s_\phi \theta_{01} ) m_W^4 (1+\beta ^2) 
\end{dmath*}

\begin{dmath*}
R0=\quad  2 m_W^2 (-1+\beta ^2) ( \theta_{12} \beta  ( 2 (\beta -1)^2 + 2 c_\theta (s_\theta^2-(\beta -1)^2 ) -s_\theta^2 (3+ (-2+\beta ) \beta  ) ) + s_\theta c_\phi ( 2 s_\theta^2 -(\beta -1)^2 + c_\theta (\beta -1)^2 ) ( -\theta_{02} + \theta_{23} \beta  ) + s_\theta s_\phi ( 2 s_\theta^2 -(\beta -1)^2 + c_\theta (\beta -1)^2 ) (\theta_{01}-\theta_{13} \beta  ) ) s 
\end{dmath*}

\begin{dmath*}
RL=\quad  16 s_\theta (-2+s_\theta^2+ 2 c_\theta ) ( -c_\phi \theta_{02} + s_\phi \theta_{01} ) m_W^4 (1+\beta ^2) 
\end{dmath*}

\begin{dmath*}
RR=\quad  -32 s_\theta m_W^4 (c_\theta+\beta ) ( -c_\phi (\theta_{02}+ \theta_{23} \beta  ) + s_\phi (\theta_{01}+ \theta_{13} \beta  ) ) 
\end{dmath*}
\end{dgroup*}

    \subsection{Coefficients of $\kappa_2$ for $\hat{M}_{f_uZ_s}(LRh_{W^+}h_{W^-})$}
{\bf Prefactor}: $ (g_A^f+g_V^f) g^6 s_W^2 s^3 /\left( 128 c_W^2 u m_W^4 (s-m_Z^2) \right)$
\begin{dgroup*}

\begin{dmath*}
00=\quad  s_\theta ( -c_\phi \theta_{02} + s_\phi \theta_{01} ) \left(\beta ^2-1\right)^2 (\beta ^3-2 c_\theta -3 \beta  ) s^2 
\end{dmath*}

\begin{dmath*}
0L=\quad  -2 m_W^2 (-1+\beta ^2) ( \theta_{12} \beta  ( -2 (\beta -1)^2 + 2 c_\theta ((\beta -1)^2-s_\theta^2 ) + s_\theta^2 (3+ (-2+\beta ) \beta  ) ) -s_\theta c_\phi ( 2 s_\theta^2 -(\beta -1)^2 + c_\theta (\beta -1)^2 ) (\theta_{02}+ \theta_{23} \beta  ) + s_\theta s_\phi ( 2 s_\theta^2 -(\beta -1)^2 + c_\theta (\beta -1)^2 ) (\theta_{01}+ \theta_{13} \beta  ) ) s 
\end{dmath*}

\begin{dmath*}
0R=\quad  2 m_W^2 (-1+\beta ^2) ( \theta_{12} \beta  ( -2 (\beta +1)^2 + 2 c_\theta (s_\theta^2-(\beta +1)^2 ) + s_\theta^2 (3+ \beta  (2+\beta ) ) ) + s_\theta c_\phi ((\beta +1)^2-2 s_\theta^2 + c_\theta (\beta +1)^2 ) (\theta_{02}-\theta_{23} \beta  ) + s_\theta s_\phi ((\beta +1)^2-2 s_\theta^2 + c_\theta (\beta +1)^2 ) ( -\theta_{01} + \theta_{13} \beta  ) ) s 
\end{dmath*}

\begin{dmath*}
L0=\quad  2 m_W^2 (-1+\beta ^2) ( \theta_{12} \beta  ( 2 (\beta +1)^2 + 2 c_\theta ((\beta +1)^2-s_\theta^2 ) -s_\theta^2 (3+ \beta  (2+\beta ) ) ) -s_\theta s_\phi ((\beta +1)^2-2 s_\theta^2 + c_\theta (\beta +1)^2 ) (\theta_{01}+ \theta_{13} \beta  ) + s_\theta c_\phi ((\beta +1)^2-2 s_\theta^2 + c_\theta (\beta +1)^2 ) (\theta_{02}+ \theta_{23} \beta  ) ) s 
\end{dmath*}

\begin{dmath*}
LL=\quad  32 s_\theta m_W^4 (c_\theta+\beta ) ( -c_\phi \theta_{02} + s_\phi \theta_{01} + c_\phi \theta_{23} \beta  -s_\phi \theta_{13} \beta  ) 
\end{dmath*}

\begin{dmath*}
LR=\quad  16 s_\theta (-2+s_\theta^2-2 c_\theta ) ( -c_\phi \theta_{02} + s_\phi \theta_{01} ) m_W^4 (1+\beta ^2) 
\end{dmath*}

\begin{dmath*}
R0=\quad  2 m_W^2 (-1+\beta ^2) ( \theta_{12} \beta  ( -2 (\beta -1)^2 + 2 c_\theta ((\beta -1)^2-s_\theta^2 ) + s_\theta^2 (3+ (-2+\beta ) \beta  ) ) + s_\theta c_\phi ( 2 s_\theta^2 -(\beta -1)^2 + c_\theta (\beta -1)^2 ) (\theta_{02}-\theta_{23} \beta  ) + s_\theta s_\phi ( 2 s_\theta^2 -(\beta -1)^2 + c_\theta (\beta -1)^2 ) ( -\theta_{01} + \theta_{13} \beta  ) ) s 
\end{dmath*}

\begin{dmath*}
RL=\quad  -16 s_\theta (-2+s_\theta^2+ 2 c_\theta ) ( -c_\phi \theta_{02} + s_\phi \theta_{01} ) m_W^4 (1+\beta ^2) 
\end{dmath*}

\begin{dmath*}
RR=\quad  32 s_\theta m_W^4 (c_\theta+\beta ) ( -c_\phi (\theta_{02}+ \theta_{23} \beta  ) + s_\phi (\theta_{01}+ \theta_{13} \beta  ) ) 
\end{dmath*}
\end{dgroup*}

  }

  \clearpage

  \end{fmffile}
  
\bibliography{ncsmBib}

\end{document}